\documentclass[]{jfm}

\usepackage{graphicx}
\usepackage{comment}
\usepackage{newtxtext}
\usepackage{newtxmath}
\usepackage{natbib}
\usepackage{hyperref}
\hypersetup{
    colorlinks = true,
    urlcolor   = blue,
    citecolor  = black,
}

\newcommand{\RomanNumeralCaps}[1]
\linenumbers

\usepackage{bm}
\usepackage{caption}
\usepackage{subcaption}
\usepackage{dcolumn}
\usepackage{amssymb}
\usepackage{multirow}
\newcommand{\rey}{$Re_\lambda$}
\newcommand{\re}{$Re_\lambda$~}

\newcommand{\be}{\begin{equation}}
\newcommand{\ee}{\end{equation}}

\title{Data-driven model for Lagrangian evolution of velocity gradients in incompressible turbulent flows}

\author{Rishita Das\aff{1}
  \corresp{\email{rd3154@nyu.edu,rishitadas@tamu.edu}},
 \and Sharath S. Girimaji\aff{2}}

\affiliation{\aff{1}Department of Mechanical and Aerospace Engineering, New York University, Brooklyn, NY 11201, USA
\aff{2}Department of Ocean Engineering, Texas A\&M University, College Station, TX 77843, USA}

\begin{document}
\maketitle

\begin{abstract}
Velocity gradient tensor, $A_{ij}\equiv \partial u_i/\partial x_j$, in a turbulence flow field is modeled by separating the treatment of intermittent magnitude ($A = \sqrt{A_{ij}A_{ij}}$) from that of the more {\em universal} normalized velocity gradient tensor, $b_{ij} \equiv A_{ij}/A$. The boundedness and compactness of the $b_{ij}$-space along with its {\em universal} dynamics allows for the development of models that are reasonably insensitive to Reynolds number. The near-lognormality of the magnitude $A$ is then exploited to derive a model based on a modified Ornstein-Uhlenbeck process. These models are developed using data-driven strategies employing high-fidelity forced isotropic turbulence data sets. {\em A posteriori} model results agree well with direct numerical simulation (DNS) data over a wide range of velocity-gradient features.
\end{abstract}



\section{Introduction\label{sec:Ch7:intro}}


The velocity gradient (VG) evolution in incompressible turbulent flows depends upon four processes - inertial, pressure, viscous, and large-scale forcing if present. 
The inertial contribution arises from the momentum of the fluid and is local. The pressure contribution is given by the pressure Hessian, which is highly nonlocal (as pressure is governed by the elliptic partial differential equation -- Poisson equation) and can be divided into an isotropic part which is local, and an anisotropic part which is nonlocal.
The viscous forces on the evolution of velocity gradients at any point in the flow depend on its immediate neighborhood, thus also representing a nonlocal contribution. 
The large-scale forcing effect stems from the application of external forces that sustain the turbulence in the flow and is a function of the nonlocal forces in the immediate neighborhood.
The interaction between all these effects leads to the complicated behavior of small-scale intermittency and multifractality in turbulence
\citep{yakhot2017emergence,sreenivasan1997phenomenology,meneveau1991multifractal}.
Despite these features, turbulence small scales exhibit a certain degree of universality \citep{kolmogorov1941local,sreenivasan1998update,schumacher2014small}.

Due to its theoretical significance and its practical utility in a variety of applications, there have been numerous attempts at modeling the Lagrangian evolution of the velocity gradient tensor, which is a difficult pursuit due to the inherent challenges of modeling nonlocality and intermittency.
As discussed above, in velocity-gradient evolution, the local inertial and isotropic pressure Hessian contributions are closed, while the remaining non-local processes require closure.
The earliest attempts \citep{cantwell1992exact,girimaji1995modified,martin1998dynamics2} at modeling VG dynamics neglected the non-local terms and modeled only the closed restricted Euler (RE) equations \citep{vieillefosse1982local,cantwell1992exact} which led to finite-time singularity. 
Beginning with the work of \cite{girimaji1990diffusion}, a series of stochastic velocity gradient models followed over the years that used diverse closure techniques for modeling the effects of the non-local pressure and viscous contributions leading to statistically stationary solutions \citep{chertkov1999lagrangian,jeong2003velocity,chevillard2006lagrangian,chevillard2008modeling,wilczek2014pressure}.
Some of the recent modeling efforts, such as the recent deformation of Gaussian field (RDGF) model proposed by \cite{johnson2016closure}, the multifractal process-based stochastic model by \cite{pereira2018multifractal}, the model based on temporal correlation of strain and rotation rate by \cite{leppin2020capturing}, and the data-driven VG model employing tensor-basis neural network for pressure term's closure \citep{tian2021physics}, have shown improvements over previous models. However, all these models are unable to simultaneously capture both the statistical properties of intermittency and the geometric features of small-scale turbulence with high accuracy.
There is a need for a robust VG model that accurately predicts both these essential features of small-scale dynamics and is generalizable to a wide variety of turbulent flows. 


Modeling the nonlinear dynamics of the VG tensor (${A}_{ij} \equiv \partial u_i/\partial x_j$) is challenging due to its multifractal and intermittent nature.
To overcome this complexity, in this work we adopt a different approach: we isolate the universal features of turbulence from intermittency.
Kolmogorov's refined similarity hypothesis \citep{kolmogorov1962refinement} suggests that considering intermittency, using the local average of dissipation rate better encapsulates the universality of turbulence than its global average. Extending the same principle, we use the absolute local pseudo-dissipation rate or VG magnitude ($A=\sqrt{A_{ij}A_{ij}}$) to normalize the VG tensor as follows \citep{girimaji1995modified}:
\begin{equation}
    b_{ij} \equiv \frac{A_{ij}}{A} \;\; \text{where} \;\; A \equiv ||\bm{A}||_F = \sqrt{A_{mn}A_{mn}}.
    \label{eq:Ch7:defbij}
\end{equation}
The normalized velocity gradient tensor, $b_{ij}$, is a mathematically bounded tensor and is statistically nearly universal across different types of turbulent flows at different Reynolds numbers \citep{das2019reynolds,das2022effect}.
The geometric shape features of the turbulence small scales are encoded in the $b_{ij}$ tensor, while its scale and intermittency lie in the VG magnitude $A$ \citep{das2020characterization}. 
The idea here is to develop separate models for $b_{ij}$ and $A$ tailored for capturing the dynamical behavior of each uniquely, resulting in an overall improved prediction of $A_{ij}$ evolution.  
Within the $b_{ij}$ framework, the effects of different turbulence processes including the nonlocal pressure, viscous, and forcing, are nearly universal, well-behaved, and more amenable to modeling  \citep{das2020characterization,das2022effect} than the nonlocal terms in $A_{ij}$ space modeled in previous studies.
Taking advantage of these properties, we model the unclosed pressure and viscous contributions in the mathematically bounded state-space of $b_{ij}$. 
On the other hand, the intermittency is captured in a separate model of magnitude $A$.

To model the nonlocal terms, we employ a simple data-driven approach based on our physical understanding of the small-scale dynamics.
For this, we utilize highly resolved direct numerical simulations (DNS) data and rely on a lookup table approach in a four-dimensional compact space. It provides a more exact representation of the flow physics as compared to other data-driven modeling methods such as neural networks, due to the compactness of the $b_{ij}$ space. This results in a generalizable closure of the non-local pressure and viscous processes at high enough Reynolds numbers within the $b_{ij}$ framework. 

Modeling the VG magnitude $A$ does not require any additional closures. 
We model the evolution of magnitude $A$ (pseudodissipation rate $\sim A^2$) within the framework of the Ornstein-Uhlenbeck (OU) process \citep{uhlenbeck1930theory}, due to its near lognormal probability distribution and exponential decay of its auto-correlation \citep{kolmogorov1962refinement,oboukhov1962some,yeung1989lagrangian}.
Although multifractal formalism \citep{mandelbrot1974intermittent} suggests that pseudodissipation rate is not precisely lognormal, studies have shown support for the lognormal framework of modeling the temporal dynamics of VG magnitude \citep{pope1990velocity,girimaji1990diffusion,huang2014lagrangian}. 
Therefore, we model the VG magnitude as a Reynolds number-dependent modified lognormal process. 
We further incorporate DNS data-based physical modifications within the OU process, with the expectation of capturing the intermittent nature of small-scale turbulence more accurately than a simple lognormal process. 

Overall, this work presents a data-driven Lagrangian model to accurately reproduce all the essential characteristics of VG dynamics in turbulent flows for a broad range of Reynolds numbers with minimal computational effort.
The novelty of our model lies in two primary features of our approach: (1) we model the geometric aspect of VG dynamics separately from its intermittent magnitude, and (2) within the compact space of VG geometry, we use a cleverly articulated, simple but exact lookup table approach for the data-driven closure of the non-local pressure and viscous processes. 
The remaining sections of the paper are arranged as follows. In section \ref{sec:Ch7:gov} we discuss the properties and present the governing differential equations for the normalized VG tensor and VG magnitude in an incompressible turbulent flow. 
The entire modeling methodology is described in section \ref{sec:Ch7:model},
including the philosophy of the modeling approach and its generalizability, formulation of the model equations and closures, and a complete model summary. 
The numerical procedure of the simulations performed using the model is outlined in section \ref{sec:Ch7:num}.
Finally the results of the model are compared with that of DNS and previous models in section \ref{sec:Ch7:results} and the conclusions are presented in section \ref{sec:Ch7:conc}. 

\section{Governing equations\label{sec:Ch7:gov}}

The Navier-Stokes and continuity equations for velocity fluctuations, $u_i$, in an incompressible turbulent flow can be written as
\begin{subequations}
\begin{eqnarray}
 \frac{\partial u_i}{\partial t} + u_k\frac{\partial u_i}{\partial x_k} &=& - \frac{\partial p}{\partial x_i} + \nu \nabla^2 u_i + f_i \label{eq:Ch7:NS}
\\
\frac{\partial u_i}{\partial x_i}&=& 0 \label{eq:Ch7:cont}
\end{eqnarray}
\label{eq:Ch7:NScont}
\end{subequations}
where, $p$ is the pressure fluctuation, $\nu$ is the kinematic viscosity, and $f_i$ represents the large-scale forcing. Pressure and viscous effects are the key nonlocal processes in turbulence. Forcing, which causes the energy production at large scales to compensate for the viscous dissipation at small scales, can be expressed in the following general form for most commonly encountered flows with a mean flow:
\begin{equation}
    f_i = 
    - \langle U_k \rangle \frac{\partial u_i}{\partial x_k} 
    - u_k \frac{\partial \langle U_i \rangle}{\partial x_k}
        + \frac{\partial}{\partial x_k} \langle u_i u_k \rangle
    \label{eq:forc_inhomo}
\end{equation}
where, $U_i = \langle U_i \rangle + u_i$ is the total velocity and $\langle \;  \; \rangle$ indicates ensemble averaging or spatial averaging in homogeneous directions.
The effective forcing $f_i$ varies from one turbulent flow to another depending on the mean flow field as well as the inhomogeneity and anisotropic nature of the flow geometry \citep{rogallo1981}. In homogeneous isotropic turbulence with no mean flow, forcing simply entails injecting energy at the lowest wavenumbers \citep{eswaran1988examination,donzis2010resolution}. 

From equation (\ref{eq:Ch7:NScont}), the governing equation for the velocity gradient tensor can be derived as:
\begin{subequations}
\begin{eqnarray}
    \frac{dA_{ij}}{dt} &=& - A_{ik}A_{kj} + \frac{1}{3}  A_{mk}A_{km} \delta_{ij} + H_{ij} + T_{ij} + G_{ij}, 
    \label{eq:Ch7:dAijdt} \\
     \text{where} \;\; H_{ij} &=& - \frac{\partial^2 p}{\partial x_i \partial x_j} + \frac{\partial^2 p}{\partial x_k \partial x_k} \frac{\delta_{ij}}{3}, \; 
    T_{ij} = \nu \nabla^2 A_{ij}, \;
    G_{ij} = \frac{\partial f_i}{\partial x_j} - \frac{\partial f_k}{\partial x_k}\frac{\delta_{ij}}{3}.
    \label{eq:Ch7:HijTijGij}
\end{eqnarray}
\end{subequations}
Here, $d/dt = \partial/\partial t + u_k \partial/\partial x_k$ is the material or substantial derivative.
The first two terms on the right-hand-side (RHS) of equation (\ref{eq:Ch7:dAijdt}) represent the non-linear effects, but they are local in space. 
The tensor ${H}_{ij}$ is the anisotropic pressure Hessian tensor, ${T}_{ij}$ is the viscous Laplacian tensor, and ${G}_{ij}$ is the anisotropic forcing tensor.
${H}_{ij}$ and ${T}_{ij}$  tensors represent the non-local effects in velocity gradient dynamics and $G_{ij}$ depends on the nature of forcing.

\subsection{Normalized VG tensor \label{sec:Ch7:govbij}}

The evolution equation for $b_{ij}$ in the flow frame of reference, derived from equation (\ref{eq:Ch7:dAijdt}), is
\begin{eqnarray}
    \frac{d b_{ij}}{d t'} = & - & b_{ik}b_{kj} + \frac{1}{3} b_{km} b_{mk} \delta_{ij} 
    + b_{ij} b_{mk} b_{kn} b_{mn} + (h_{ij} - b_{ij} b_{kl} h_{kl}) \nonumber \\
    & + & (\tau_{ij} - b_{ij} b_{kl} \tau_{kl}) + (g_{ij} - b_{ij} b_{kl} g_{kl})
    \label{eq:Ch7:dbijdt}
\end{eqnarray}
where $dt'=Adt$ is the time increment normalized by local VG magnitude, and the timescale $t'$ is referred to as the local timescale. Here,
\begin{eqnarray}
    h_{ij} = \frac{H_{ij}}{A^2} = \frac{1}{A^2} \bigg(- \frac{\partial^2 p}{\partial x_i \partial x_j} + \frac{\partial^2 p}{\partial x_k \partial x_k} \frac{\delta_{ij}}{3} \bigg)
    \; , \;\; 
    \tau_{ij} = \frac{T_{ij}}{A^2} = \frac{\nu}{A^2} \nabla^2 A_{ij}
    \; , \nonumber \\
    g_{ij} = \frac{G_{ij}}{A^2} = \frac{1}{A^2} \bigg(\frac{\partial f_i}{\partial x_j} - \frac{\partial f_k}{\partial x_k}\frac{\delta_{ij}}{3} \bigg).
\end{eqnarray}
are the normalized anisotropic pressure Hessian, viscous Laplacian, and anisotropic forcing tensors, respectively.
In the $b_{ij}$ equation (\ref{eq:Ch7:dbijdt}), the first three terms on the RHS are closed and represent the nonlinear ($N$) - inertial and isotropic pressure Hessian - effects.
The next three terms constitute the non-local pressure ($P$), viscous ($V$), and forcing ($F$) effects on $b_{ij}$ evolution that require closure.
Consideration of the $b_{ij}$ evolution in local timescale is not only consistent with Kolmogorov's refined similarity hypothesis but also leads to the added practical advantage that all the terms on the RHS of equation \ref{eq:Ch7:dbijdt} are normalized tensors.
These normalized pressure Hessian and viscous Laplacian tensors are not necessarily bounded, but are well-behaved in the phase plane of $b_{ij}$ invariants and are considerably more amenable to modeling than the unnormalized tensors in the $A_{ij}$ equation \citep{das2022effect}. 
They further exhibit a nearly universal behavior across turbulent flows of different Reynolds numbers \citep{das2020characterization,das2022effect}.
Therefore, the Lagrangian evolution of $b_{ij}$ can be modeled in the local timescale $t'$ without any explicit dependence on the VG magnitude. 
The magnitude 
dependence comes in only when determining the $b_{ij}$ evolution in real time.

In order to most effectively employ data-driven techniques, we will now establish the minimum number of free parameters required to completely describe the $b_{ij}$ tensor. A detailed derivation can be found in \cite{das2020characterization}; only the main outcomes are summarized below.
Without any loss of generality, we can express $b_{ij}$ in the principal (eigen) reference frame of normalized strain-rate tensor, ${s}_{ij}$, as follows
\begin{eqnarray}
    \bm{\tilde{b}} =
    \left[
    \begin{array}{ccc}
    a_1 & 0 & 0\\
    0 & a_2 & 0\\
    0 & 0 & a_3
    \end{array}\right]
    + 
    \left[
    \begin{array}{ccc}
    0 & -\omega_3 & \omega_2\\
    \omega_3 & 0 & -\omega_1\\
    -\omega_2 & \omega_1 & 0
    \end{array}\right]
	\;\;\; \text{where} \;\; a_1 \geq a_2 \geq a_3
\label{eq:Ch7:b_sw}
\end{eqnarray}
Here, $\tilde{(\;\;)}$ represents tensors in the principal reference frame of ${s}_{ij}$, and 
$a_i$ are the eigenvalues of $s_{ij}$ in decreasing order, such that $a_1 (> 0)$ is the most expansive strain-rate, $a_3 (< 0)$ is the most compressive strain-rate, and the intermediate strain-rate $a_2$ can be positive, negative or zero, in incompressible flows. 
Further, $\omega_i$ are the components of the normalized vorticity vector ($\vec{\omega}$) along the strain-rate eigendirections.
Since the signs of the strain-rate eigenvectors are not uniquely determined by the eigendecomposition, we consider the eigendirections that provide all vorticity components to be of the same sign (either all positive or all negative).

Applying the constraints of incompressibility $(\tilde{b}_{ii}=0)$ and normalization $(\tilde{b}_{ij}\tilde{b}_{ij}=1)$, the $\tilde{b}_{ij}$ state-space can be reduced to a four-dimensional space of only four independent variables (shape-parameters) -- $q$, $r$, $a_2$, $\omega_2$. Here, $q$ and $r$ are the second and third invariants of the tensor, respectively:
\begin{equation}
    q \equiv -\frac{1}{2} b_{ij}b_{ji}=-\frac{1}{2} \tilde{b}_{ij}\tilde{b}_{ji} \;, \;\; r \equiv -\frac{1}{3} b_{ij}b_{jk}b_{ki} = -\frac{1}{3} \tilde{b}_{ij}\tilde{b}_{jk}\tilde{b}_{ki}
\end{equation}
All the remaining elements of $\tilde{b}_{ij}$ can be determined uniquely once these four variables are known, as shown below:
\begin{subequations}
\begin{eqnarray}
    a_1 = \frac{1}{2}(-a_2+\sqrt{1-3a_2^2-2q}) \;,\;\;\; a_3 = \frac{1}{2}(-a_2-\sqrt{1-3a_2^2-2q}),
    \label{eq:10}
\end{eqnarray}
\begin{eqnarray}
    \omega_1 = \pm \frac{1}{2\sqrt{2}} \sqrt{(1+2q-4\omega_2^2) - \frac{8a_2^3 + 8r - a_2(3-2q-12\omega_2^2)}{\sqrt{1-3a_2^2-2q}}},  \label{eq:12a}
\end{eqnarray}
\begin{eqnarray}
    \omega_3 = \pm \frac{1}{2\sqrt{2}} \sqrt{(1+2q-4\omega_2^2) + \frac{8a_2^3 + 8r - a_2(3-2q-12\omega_2^2)}{\sqrt{1-3a_2^2-2q}}}.   \label{eq:12b}
\end{eqnarray}
\end{subequations}
This shows that 
$q$, $r$, $a_2$, and $\omega_2$ completely define the tensor $\tilde{b}_{ij}$ and thence the geometric-shape of the local flow streamlines.
These four variables are also mathematically bounded as follows:
\begin{eqnarray}
    & q & \in \bigg[-\frac{1}{2}, \leq \frac{1}{2} \bigg], \;\;
     r \in \bigg[ -\frac{1+q}{3} \bigg( \frac{1-2q}{3} \bigg)^{1/2}, \frac{1+q}{3} \bigg( \frac{1-2q}{3} \bigg)^{1/2}\bigg] , \;\; \\
    & a_2 & \in \bigg[ -\sqrt{\frac{1-2q}{12}}, \sqrt{\frac{1-2q}{12}} \bigg], \; \text{and}\;\;
    \omega_2 \in \bigg[ -\sqrt{\frac{q}{2}+\frac{1}{4}}, \sqrt{\frac{q}{2}+\frac{1}{4}} \bigg].
    \label{eq:ranges}
\end{eqnarray}

\subsection{VG magnitude \label{sec:Ch7:govths}}

The velocity gradient magnitude or pseudodissipation rate has been shown to have a nearly lognormal distribution \citep{kolmogorov1962refinement,oboukhov1962some,yeung1989lagrangian,monin2013statistical}.
For this reason, we consider the dynamics of the logarithm of VG magnitude:
\begin{equation}
    \theta \equiv \ln{A},
\end{equation}
which is expected to exhibit a near-normal distribution in a turbulent flow field. 
We introduce the scalar variable -- standardized VG magnitude:
\begin{equation}
    \theta^* \equiv \frac{\theta - \langle \theta \rangle}{\sigma_\theta} \;\;\; \text{where} \;\;\;
    \sigma_\theta = \sqrt{ \langle (\theta - \langle \theta \rangle)^2 \rangle } 
    \label{eq:Ch7:ths}
\end{equation}
which exhibits a nearly standard normal distribution, $\mathcal{N}(0,1)$, in a variety of turbulent flows \citep{das2022effect}. 
The evolution equation for $\theta^*$, derived from equation (\ref{eq:Ch7:dAijdt}) 
is given by: 
\begin{equation}
    \frac{d \theta^*}{d t^*} =  \frac{1}{\sigma_\theta \langle A \rangle} (- b_{ik}b_{kj}A_{ij} + h_{ij}A_{ij}
     + \tau_{ij}A_{ij} + g_{ij}A_{ij}) 
    \label{eq:Ch7:dthsdt}
\end{equation}
where $dt^*=\langle A \rangle \; dt$. Here, $t^*$ is referred to as the global timescale and it represents the timescale normalized by the global mean of VG magnitude. This normalization is in essence similar to normalization by the Kolmogorov timescale ($\tau_\eta \sim 1/{\langle A^2 \rangle}^{1/2}$) and is found to be more appropriate for examining VG magnitude than the local timescale used for $b_{ij}$.
The four terms on the RHS of the above equation represent the nonlinear, pressure, viscous, and forcing effects, respectively, on the VG magnitude evolution.

\section{Model Formulation \label{sec:Ch7:model}}

\begin{figure}
    \centering
    \includegraphics[width=0.8\textwidth,trim={0.5cm 0.5cm 0.5cm 0.5cm},clip]{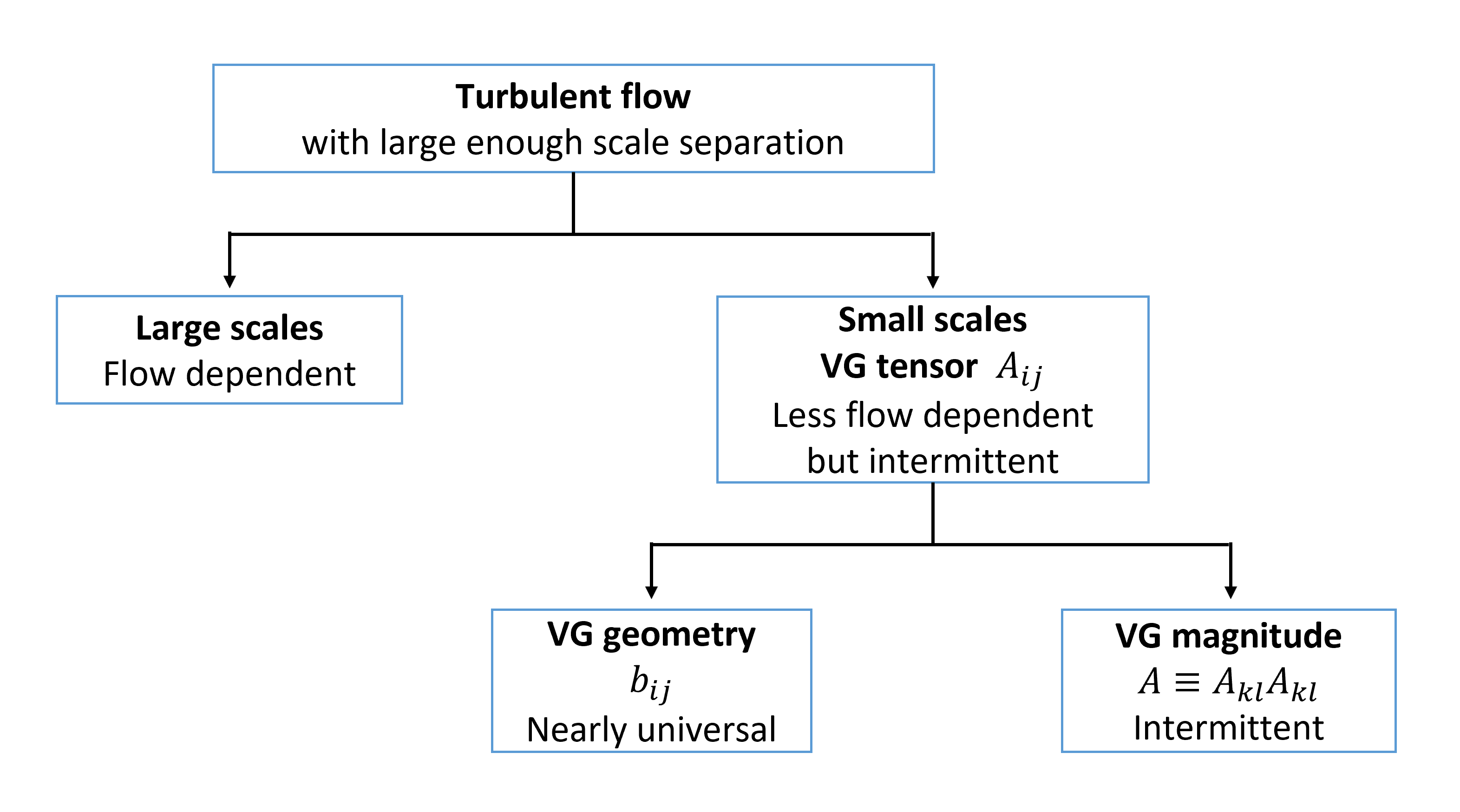}
    \caption{\label{fig:Ch7:flowchart}Flowchart to explain the behavior of velocity gradient tensor and its constituents in turbulence.}
\end{figure}

Instead of considering the evolution of $A_{ij}$ directly, this work isolates the modeling of the normalized velocity gradient tensor ($b_{ij}$) from that of the magnitude ($A$) as shown in the schematic of figure \ref{fig:Ch7:flowchart}.
In this section, the universal features of the dynamics that need to be captured by a velocity-gradient model and that can be used to our advantage in the modeling approach, are first discussed. This is followed by the main modeling strategies and a detailed description of the complete model. Finally, all the model equations and parameters are summarized.

\subsection{Generalizability of modeling VG dynamics\label{sec:Ch7:univ}}

As outlined in figure \ref{fig:Ch7:flowchart}, the large scales of motion in a turbulent flow depend upon the flow geometry and driving mechanism of the flow. It is therefore difficult to develop generalizable models for the large scales that will apply to different turbulent flows.
Models of small-scale dynamics are likely to be more generalizable in comparison since the small scales in turbulent flows (with a large enough scale separation) tend to be isotropic and universal.  
The notion of small-scale universality, which began with the eminent work of \cite{kolmogorov1941local}, has been refined significantly over the years to account for the intermittent nature of small-scale turbulence \citep{kolmogorov1962refinement,oboukhov1962some,sreenivasan1997phenomenology,schumacher2014small}.
The velocity gradient tensor, $A_{ij}$, governs these small-scale motions and exhibits certain universal features across different types of turbulent flows \citep{sreenivasan1998update,schumacher2014small}. However, it also shows a strong dependence on Reynolds number \citep{donzis2005scalar,yeung2018effects}. 
Its multifractal and intermittent nature causes the higher order moments to grow with increasing Reynolds number, deviating far away from Gaussian behavior \citep{yakhot2017emergence}.

In this model, we separate $A_{ij}$ into normalized velocity gradient tensor ($b_{ij}$) and velocity gradient magnitude ($A$), such that the tensor $b_{ij}$ is nearly universal across different turbulent flows while the scalar $A$ reflects all the Reynolds number dependence.
The universality is evident in the PDF and higher order moments of $b_{ij}$ \citep{das2019reynolds} as well as in the mean evolution of $b_{ij}$ invariants \citep{das2020characterization,das2022effect}, that are insensitive to the variation of Taylor Reynolds number (\rey) across different turbulent flows.
Therefore, $b_{ij}$ evolution modeled using DNS data of only one turbulent flow at a given Reynolds number may be considered universal (up to a modeling approximation) and can be applied to reproduce the $b_{ij}$-dynamics of different turbulent flows at different Reynolds numbers.

\begin{figure}
    \centering
    \begin{subfigure}{0.49\textwidth}
        \centering
        \includegraphics[width=\textwidth]{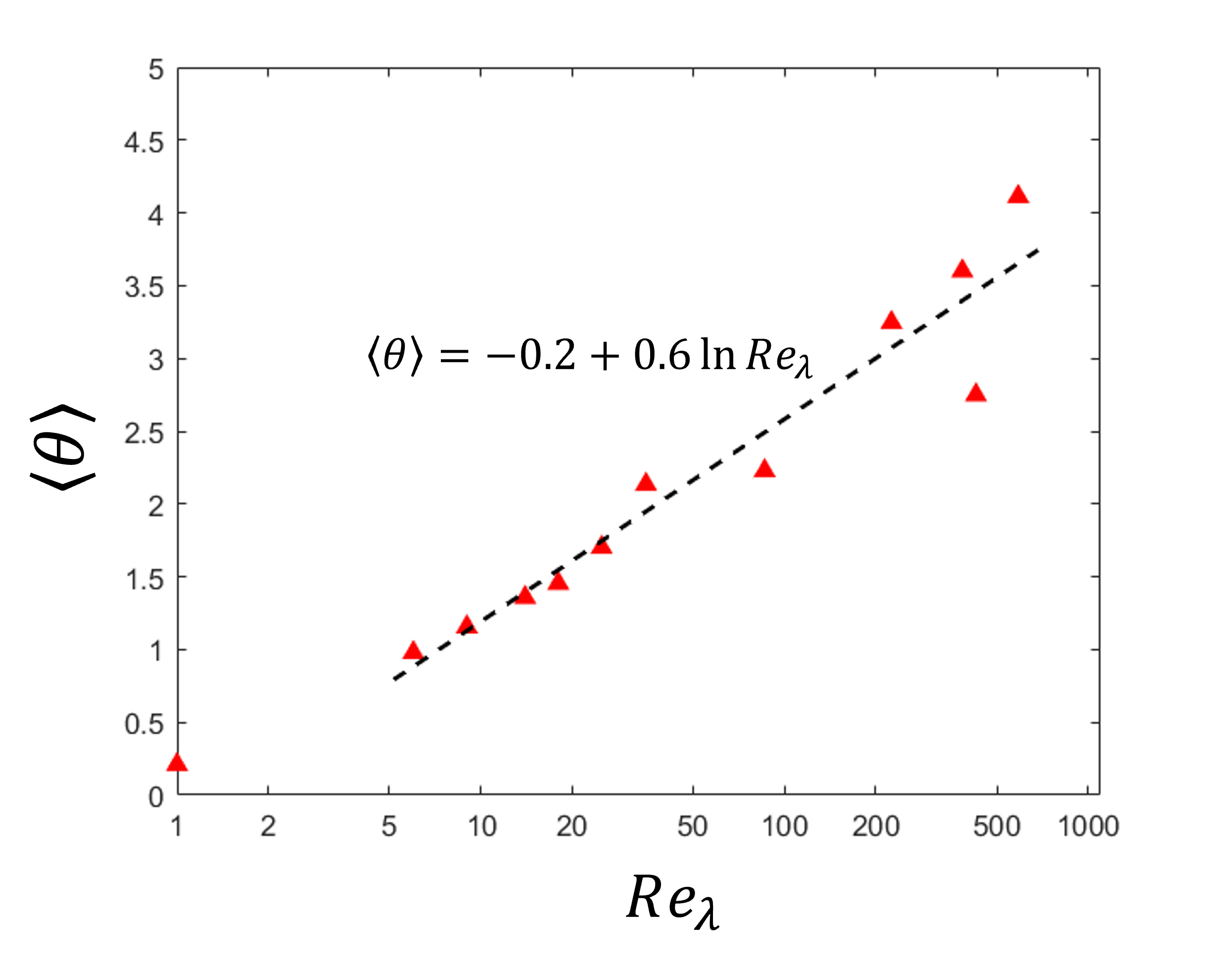}
        \caption{}
    \end{subfigure}
    \hfill
    \begin{subfigure}{0.49\textwidth}
        \centering
        \includegraphics[width=\textwidth]{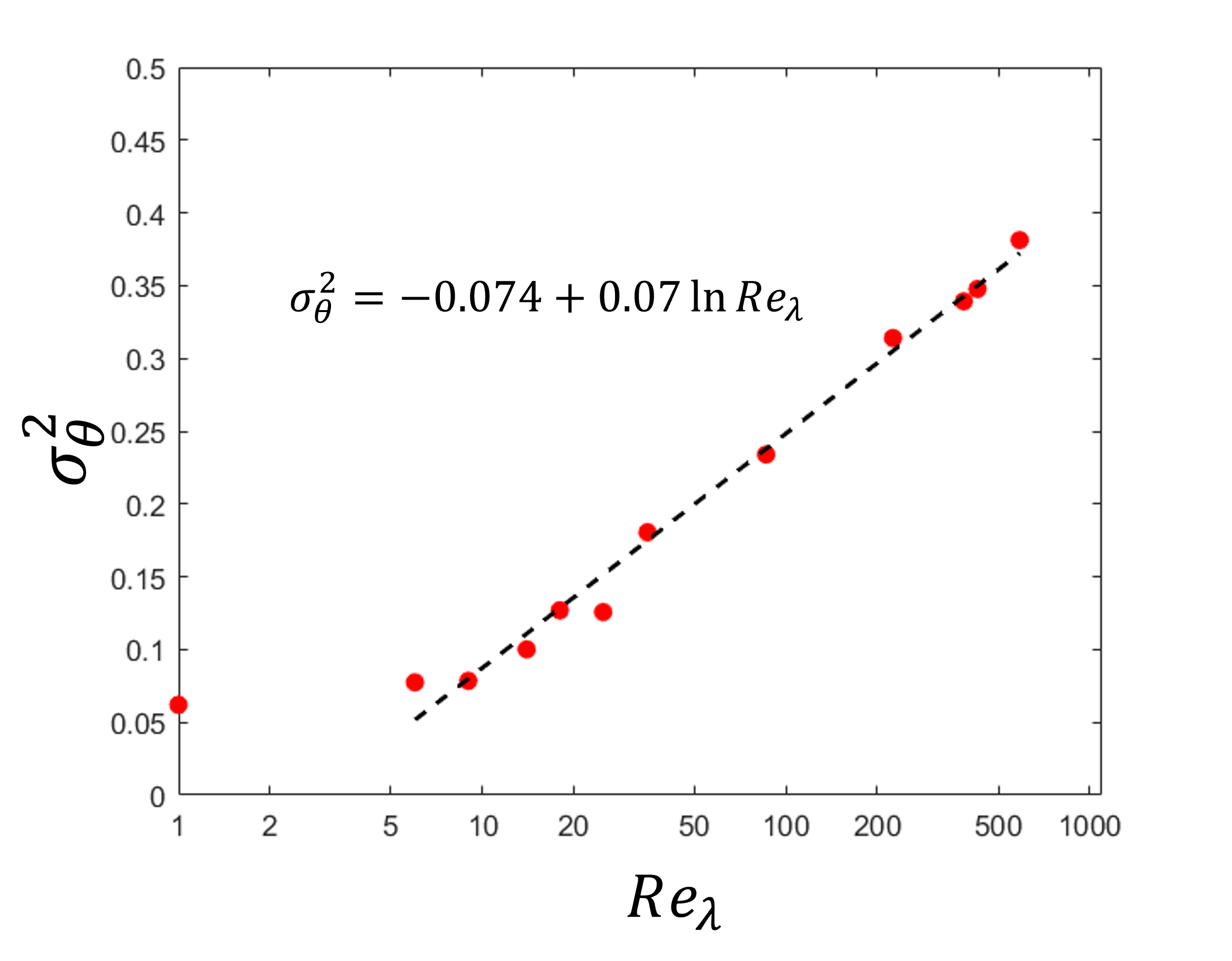}
        \caption{}
    \end{subfigure}
    \caption{\label{fig:Ch7:theta_stats} Statistics of $\theta$ from DNS datasets of forced isotropic turbulent flows at different \re: (\textit{a}) global mean $\langle \theta \rangle$ as a function of $Re_\lambda$ (in natural log scale); dashed line represents a linear least-squares fit of the data ($\langle \theta \rangle = -0.2 + 0.6\ln{Re_\lambda}$); and (\textit{b}) variance $\sigma_\theta^2 = \langle \theta^2 - \langle \theta \rangle^2 \rangle$ as a function of $Re_\lambda$ (in natural log scale); dashed line represents a linear least-squares fit of the data ($\sigma_\theta^2 = -0.074 + 0.07\ln{Re_\lambda}$).}
\end{figure}

The magnitude $A$, on the other hand, exhibits a strong dependence on the Reynolds number of the flow.
The mean and variance of its logarithm ($\theta = \ln{A}$), plotted in figure \ref{fig:Ch7:theta_stats}, clearly increase with increasing \rey.
Preliminary results suggest that $\langle \theta \rangle$ and $\sigma_\theta$ follow approximate scaling laws with \rey, as indicated in the figures.
The scaling law obtained for $\sigma_\theta^2$ is in close agreement with the \rey-scaling of the logarithm of pseudodissipation rate reported by \cite{yeung1989lagrangian}.
But further advanced simulations and analyses are required to develop universal scaling laws for $\langle \theta \rangle$ and $\sigma_\theta^2$. 
In fact, these two quantities are the input parameters of our model for VG magnitude (section \ref{sec:Ch7:modelths}), representing the characteristic Reynolds number dependence of velocity gradients. 

As summarized above, the advantage of this modeling framework is that the nine-components tensorial variable $b_{ij}$ is nearly universal, and one can develop a potentially generalizable $b_{ij}$-model applicable to different types of turbulent flows at wide-ranging Reynolds numbers. Only the scalar $\theta$-model is Reynolds number dependent, which can be represented by scaling laws that are likely generalizable across different types of turbulent flows.

\subsection{Modeling strategy\label{sec:Ch7:strategy}}

The modeling approach of this work constitutes the following:
\begin{enumerate}
    \item \textit{$b_{ij}$-model:} The $b_{ij}$ dynamics in a local timescale (equation \ref{eq:Ch7:dbijdt}) is a function of $b_{ij}$ and other normalized non-local tensors, and it does not explicitly depend on magnitude $A$. Therefore, we formulate a stochastic model for the Lagrangian evolution of $b_{ij}$ in the local timescale ($t'$) without any explicit dependence on $\theta^*$. 
    \item \textit{Closure of nonlocal processes:} As inferred from DNS data in our previous analysis \citep{das2020characterization,das2022effect}, the conditional statistics of the normalized non-local tensors can be reasonably approximated as exclusive functions of $b_{ij}$. Thus, we develop DNS data-driven closure models (generalizable at high $Re_\lambda$) for capturing the conditional mean nonlocal effects of normalized pressure and viscous processes within the four-dimensional bounded state-space of $\tilde{b}_{ij}$. 
    The fluctuations of these nonlocal effects as well as the effect of large-scale forcing are modeled in the stochastic diffusion term using moment constraints.
    \item \textit{$\theta^*$-model:} We model the evolution of VG magnitude in global timescale ($t^*$) within the framework of Ornstein-Uhlenbeck (OU) process \citep{pope1990velocity} in three different ways. The first model is a simple OU model for $\theta^*$ decoupled from $b_{ij}$ dynamics. The second and third models are modified OU models with $b_{ij}$-dependence incorporated into the $\theta^*$ evolution using a DNS data-based diffusion process.
    \item \textit{Timescale:} In addition, an ordinary differential equation provides the relation between the local and the global timescales. 
\end{enumerate}
Finally, the $b_{ij}$ and $\theta^*$ models are combined to form an integrated system of model equations representing the Lagrangian evolution of $A_{ij}$ in global time.

\subsection{Model for normalized VG tensor \label{sec:Ch7:modelbij}}


The Lagrangian dynamics of normalized velocity gradient tensor, $b_{ij}$, is modeled here as a diffusion process \citep{karlin1981second}. A diffusion process is a continuous-time Markov process and is represented by a stochastic differential equation (SDE).
The SDE for $A_{ij}$ commonly used in previously developed models \citep{girimaji1990diffusion,chevillard2006lagrangian,chevillard2008modeling,johnson2016closure} is of the form
\begin{eqnarray}
    & d A_{ij} = M_{ij} dt + K_{ijkl}\; dW_{kl}
    \label{eq:Ch7:AijSDE}
\end{eqnarray}
where $W_{ij}$ is a tensor-valued isotropic Wiener process such that
\begin{equation}
    \langle dW_{ij} \rangle = 0 \;\;\; \text{and} \;\; \langle dW_{ij}dW_{kl} \rangle = \delta_{ik} \delta_{jl} dt.
\end{equation}
The $M_{ij}$ tensor represents the drift coefficient tensor and $K_{ijkl}$ constitutes the diffusion coefficient tensor of the model.
Taking the trace of equation (\ref{eq:Ch7:AijSDE}), one can show that $M_{ii} = K_{iikl} = 0$ satisfies the incompressibility condition $A_{ii}=0$.
Starting from the above equation and using the properties of an It\^{o} process \citep{kloeden1992stochastic}, one can derive the following SDE for $b_{ij}$ in local timescale $t'$ (see appendix \ref{sec:appB} for derivation):
\begin{equation}
    d b_{ij} = (\mu_{ij} + \gamma_{ij}) dt' + D_{ijkl} \;dW'_{kl}
    \label{eq:Ch7:bijSDE}
\end{equation}
where,
\begin{eqnarray}
    & \mu_{ij} = \frac{M_{ij}}{A^2} - b_{ij}b_{kl}\frac{M_{kl}}{A^2}  \;\;, \;\; 
    D_{ijkl} = \frac{K_{ijkl}}{A^{3/2}} - b_{ij}b_{pq}\frac{K_{pqkl}}{A^{3/2}} \;\;, \nonumber \\
    & \gamma_{ij} = -\frac{1}{2}b_{ij}\frac{K_{pqkl}}{A^{3/2}}\frac{K_{pqkl}}{A^{3/2}} 
      - b_{pq}\frac{K_{pqkl}}{A^{3/2}}\frac{K_{ijkl}}{A^{3/2}} + \frac{3}{2}b_{ij}b_{pq}\frac{K_{pqkl}}{A^{3/2}}b_{mn}\frac{K_{mnkl}}{A^{3/2}} \nonumber \\
    & dt' = A dt\;\;, \;\; dW'_{ij} = A^{1/2}dW_{ij}
    \label{eq:Ch7:bijSDEterms}
\end{eqnarray}
and the Wiener process satisfies
\begin{equation}
    \langle dW'_{ij} \rangle = 0 \;\;\; \text{and} \;\;  
    \langle dW'_{ij}dW'_{kl} \rangle = \delta_{ik} \delta_{jl} dt' . 
    \label{eq:Ch7:dWij}
\end{equation}
It is important to note that all the drift and diffusion coefficient tensors of this system of SDEs are dimensionless. 
The drift tensor of the $b_{ij}$ equation has two parts due to the normalization: (i) $\mu_{ij}$ is obtained from the drift tensor of the $A_{ij}$ equation, $M_{ij}$, and (ii) $\gamma_{ij}$ is obtained from the diffusion tensor of the $A_{ij}$ equation, $K_{ijkl}$.
The diffusion tensor of the $b_{ij}$ equation, $D_{ijkl}$, is also obtained from $K_{ijkl}$.
The tensor $\gamma_{ij}$ relates the drift and diffusion processes in the dynamics such that despite the random stochastic forcing term, $b_{ij}$ remains mathematically bounded.
All the coefficient tensors are modeled in the specific functional forms given above.
It can be proved that any system of SDEs for $b_{ij}$, that complies with the above forms of drift and diffusion terms, clearly satisfies the incompressibility constraint:
\begin{equation}
    db_{ii} =  0.
    \label{eq:Ch7:bijconstr1}
\end{equation}
Equation \eqref{eq:Ch7:bijSDE} further satisfies the mathematical constraint of normalization:
\begin{equation}
    d(b_{ij}b_{ij}) = 0
    \label{eq:Ch7:bijconstr2}
\end{equation}
which ensures that the Frobenius norm of the tensor $\bm{b}$ is equal to unity at all times.
The proofs are presented in appendix \ref{sec:appC} and \ref{sec:appD}.


Equation (\ref{eq:Ch7:bijSDE}) leads to a Fokker Planck equation \citep{pope1985pdf} for the joint PDF $\hat{\mathbb{F}}(\bm{b})$ of the tensor $b_{ij}$: 
\begin{equation}
    \frac{d\hat{\mathbb{F}}}{dt'} = - \frac{\partial}{\partial b_{ij}} \big[ \hat{\mathbb{F}} (\mu_{ij} + \gamma_{ij})  \big] + \frac{1}{2} \frac{\partial^2}{\partial b_{ij} \partial b_{pq}}(\hat{\mathbb{F}} D_{ijkl}D_{pqkl})
    \label{eq:Ch7:bijFPE}
\end{equation}
Now, the exact differential equation for the joint PDF of $b_{ij}$, $\mathbb{F}(\bm{b})$, in a turbulent flow can be derived from the $b_{ij}$ governing equation (\ref{eq:Ch7:dbijdt}) as 
\begin{eqnarray}
    \frac{d\mathbb{F}}{dt'} 
    = - \frac{\partial }{\partial b_{ij}} \bigg[ \mathbb{F} \bigg( 
    & - & b_{ik}b_{kj} + \frac{1}{3} b_{km} b_{mk} \delta_{ij} 
    + b_{ij} b_{mk} b_{kn} b_{mn} 
    + \big\langle  h_{ij} - b_{ij}b_{kl}h_{kl} \big| \bm{b} \big\rangle \nonumber \\
    & + & \big\langle  \tau_{ij} - b_{ij}b_{kl}\tau_{kl} \big| \bm{b} \big\rangle
    + \big\langle  g_{ij} - b_{ij}b_{kl}g_{kl} \big| \bm{b} \big\rangle
    \bigg) \bigg]
    \label{eq:Ch7:bijPDFeqn}
\end{eqnarray}
The drift and diffusion coefficient tensors need to be modeled in a way that $\hat{\mathbb{F}}(\bm{b}) \approx {\mathbb{F}}(\bm{b})$.
In data-driven modeling, $\mathbb{F}(\bm{b})$ and its moments are taken from high-fidelity DNS data.
However, requiring the PDFs to be identical is very challenging.
In this work, we constrain the equations of $b_{ij}$-moments up to third order to obtain the parameters of diffusion coefficient tensor along the lines of \cite{girimaji1990diffusion}. 
This modeled diffusion process is, therefore, consistent up to order three, although the numerical results of the model show reasonable agreement of much higher-order moments.

\subsubsection{Drift coefficient tensor \label{sec:Ch7:drift}}

Comparing the terms 
of equations (\ref{eq:Ch7:bijFPE}) and (\ref{eq:Ch7:bijPDFeqn}), the inertial and isotropic pressure Hessian terms are exact, and considering that the role of $D_{ijkl}$ is to model the large-scale forcing effect and that of $\gamma_{ij}$ is to maintain the unit Frobenius norm of $b_{ij}$, the drift coefficient tensor $\mu_{ij}$ takes the form:
\begin{equation}
    \mu_{ij} = - b_{ik} b_{kj} + \frac{1}{3} b_{km}b_{mk} \delta_{ij} + b_{ij} b_{mk} b_{kn} b_{mn} + \big\langle  h_{ij} - b_{ij}b_{kl}h_{kl} \big| \bm{b} \big\rangle 
    + \big\langle \tau_{ij} - b_{ij}b_{kl}\tau_{kl} \big| \bm{b} \big\rangle
    \label{eq:Ch7:muij}
\end{equation}
The term $\mu_{ij}$ represents the dynamics of the inertial and isotropic pressure Hessian contributions as well as the conditional mean of the anisotropic pressure Hessian and viscous contributions. 

The conditional mean normalized anisotropic pressure Hessian and viscous Laplacian tensors,  $\langle {h}_{ij} | \bm{b} \rangle $ and $\langle {\tau}_{ij} | \bm{b} \rangle $, require closure modeling.
As discussed in section \ref{sec:Ch7:govbij}, the tensor $\bm{\tilde{b}}$ in the principal reference frame of the strain-rate tensor can be expressed as a function of only four bounded variables.
Therefore, in order to take advantage of this four-dimensional bounded state-space of $\bm{\tilde{b}}$, the conditional averaging of the normalized pressure Hessian and viscous Laplacian tensors is performed in the $s_{ij}$ principal reference frame.
For a rotation tensor, $\bm{Q}$, with columns constituted by the right eigenvectors of $\bm{s}$, the required tensors in the principal reference frame are
\begin{equation}
    \tilde{b}_{ij} = Q_{ki} b_{kl} Q_{lj} \;\;, \;\; \tilde{h}_{ij} = Q_{ki} h_{kl} Q_{lj} 
    \;\;, \;\; \tilde{\tau}_{ij} = Q_{ki} \tau_{kl} Q_{lj} .
    \label{eq:Ch7:Q}
\end{equation}
Then, the conditional mean pressure Hessian and viscous Laplacian tensors in the flow reference frame can be recovered as follows:
\begin{eqnarray}
    & \langle {h}_{ij} | \bm{b} \rangle =  \langle Q_{ik} \tilde{h}_{kl} Q_{jl} | \bm{b} \rangle 
    = Q_{ik} \langle \tilde{h}_{kl} | \bm{\tilde{b}} \rangle Q_{jl} \;\;, \nonumber \\
    & \langle {\tau}_{ij} | \bm{b} \rangle =  \langle Q_{ik} \tilde{\tau}_{kl} Q_{jl} | \bm{b} \rangle = Q_{ik} \langle \tilde{\tau}_{kl} | \bm{\tilde{b}} \rangle Q_{jl}   
    \label{eq:Ch7:hijtauij}
\end{eqnarray}
since $\bm{Q}$ is a function of $\bm{b}$. 
Therefore, the conditional mean pressure Hessian and viscous Laplacian tensors in the flow reference frame can be obtained if the conditional mean pressure Hessian and viscous Laplacian tensors in the principal frame are known and the local strain-rate eigenvectors are known.

As shown in section \ref{sec:Ch7:govbij}, in the principal reference frame, the tensor $\tilde{b}_{ij}$ can be uniquely defined by a set of only four bounded variables -- $q,r,a_2$ and $\omega_2$.
Therefore, the conditional mean pressure Hessian and viscous Laplacian tensors in the principal frame can be modeled as a function of only four bounded variables, as follows:
\begin{equation}
    \langle \tilde{h}_{ij} | \bm{\tilde{b}} \rangle = \langle \tilde{h}_{ij} | q,r,a_2,\omega_2 \rangle \;\;, \;\; \langle \tilde{\tau}_{ij} | \bm{\tilde{b}} \rangle = \langle \tilde{\tau}_{ij} | q,r,a_2,\omega_2 \rangle 
\end{equation}
The goal is to develop a data-driven model for the above tensors in terms of a four-dimensional input.
Recent studies \citep{parashar2020modeling,tian2021physics} have used tensor-basis neural network to model the unnormalized pressure Hessian ($H_{ij}$) and viscous Laplacian ($T_{ij}$) tensors in the $A_{ij}$-equation as a function of $A_{ij}$. Since $A_{ij}$ constitutes an unbounded space and the behavior of the tensors $H_{ij}$ and $T_{ij}$ is not necessarily invariant across turbulent flows of different Reynolds numbers, network-based modeling becomes essential. 
However, in our case $(q,r,a_2,\omega_2)$ form a bounded state-space and 
the conditional mean dynamics of $\tilde{h}_{ij}$ and $\tilde{\tau}_{ij}$ in the bounded $\tilde{b}_{ij}$ space is nearly unaltered with Reynolds number variation for a broad range of \re \cite{das2020characterization,das2022effect}. Therefore, the simpler and more accurate data-driven approach of direct tabulation based on DNS data is used in this work.

The approach can be summarized as follows:
\begin{enumerate}
    \item The finite space of $q,r,a_2$ and $\omega_2$, as given in equation \eqref{eq:ranges}, is discretized into $(60,60,30,30)$ uniform bins. This discretization strikes the appropriate balance between sampling accuracy in the bins and the desired details of nonlocal flow physics to be captured.
    Other discretizations are tested to show convergence to this combination for the most accurate results.
    \item The conditional expectations of the tensors, $\langle \tilde{h}_{ij} | q,r,a_2,\omega_2 \rangle$ and $\langle \tilde{\tau}_{ij} | q,r,a_2,\omega_2 \rangle $, are computed in this discrete phase-space, using DNS data set of forced isotropic turbulence (see appendix \ref{sec:appG} for details of the dataset). Note that only one lookup table is required to model the mean nonlocal dynamics for turbulent flows of different \rey. 
    Further, this data-driven closure is rotationally invariant since the input and output variables do not depend on the flow reference frame.
    \item This lookup table can then be accessed by an inexpensive array-indexing operation, to determine the conditional mean pressure and viscous dynamics in the principal frame for a given $(q,r,a_2,\omega_2)$ at any point in the flow field or following a fluid particle.
    This is then transformed to the flow reference frame (equation \ref{eq:Ch7:hijtauij}) based on the local eigendirections of strain-rate tensor, to be used in $\mu_{ij}$ for computations.
\end{enumerate}
This completes the modeling of the mean drift tensor $\mu_{ij}$ as a function of the local $b_{ij}$ and it is straightforward to show that our data-driven model for $\mu_{ij}$ is Galilean invariant.
The proof for the same is presented in appendix \ref{sec:appE}.

Our previous work has shown the universality of $b_{ij}$ statistics and associated nonlocal processes across different types of turbulent flows \citep{das2022effect}. 
Therefore, in this work, we restrict ourselves to forced isotropic turbulence. We use DNS data of isotropic turbulence at different Reynolds numbers (appendix \ref{sec:appG}) to illustrate the universality of different modeling components. 
Although the data of different $Re_\lambda$ are considered, since the nonlocal terms requiring closure are insensitive to $Re_\lambda$ variation, only the $Re_\lambda=427$ case is used for final comparison.

\subsubsection{Diffusion coefficient tensor \label{sec:Ch7:diffusion}}

As discussed at the beginning of this section, the interrelationship between the tensors $D_{ijkl}$ and $\gamma_{ij}$ is important in guaranteeing that the mathematical and physical constraints of $b_{ij}$ are satisfied. 
This relationship holds if we use their functional forms as given in equation (\ref{eq:Ch7:bijSDEterms}), in terms of $K_{ijkl}$ from the $A_{ij}$ SDE. 
For this, we assume a general isotropic form of the four-dimensional tensor, $K_{ijkl}$, following previous models \citep{girimaji1990diffusion,chevillard2008modeling,johnson2016closure}:
\begin{equation}
    K_{ijkl} = A^{3/2} ( c_1 \delta_{ij}\delta_{kl} + c_2 \delta_{ik}\delta_{jl} + c_3 \delta_{il}\delta_{jk} )
    \label{eq:Ch7:Kijkl}
\end{equation}
where $c_1,c_2,c_3$ are constant dimensionless diffusion coefficients of the model. 
Only two of these three coefficients are independent subject to the incompressibility condition:
\begin{eqnarray}
    & K_{iikl} = A^{3/2} ( c_1 \delta_{ii}\delta_{kl} + c_2 \delta_{ik}\delta_{il} + c_3 \delta_{il}\delta_{ik} ) = (3 c_1 + c_2 + c_3) \delta_{kl} = 0 \nonumber \\
    & \implies c_1 = - \frac{c_2+c_3}{3}
    \label{eq:Ch7:Kiikl}
\end{eqnarray}
From equations (\ref{eq:Ch7:bijSDEterms}), (\ref{eq:Ch7:Kijkl}) and (\ref{eq:Ch7:Kiikl}), the diffusion coefficient tensor of the $b_{ij}$ equation is
\begin{equation}
    D_{ijkl} = c_2 \bigg( - \frac{1}{3}\delta_{ij}\delta_{kl} + \delta_{ik}\delta_{jl} - b_{ij}b_{kl} \bigg) 
    + c_3 \bigg( - \frac{1}{3}\delta_{ij}\delta_{kl} + \delta_{il}\delta_{jk} - b_{ij}b_{lk} \bigg)
    \label{eq:Ch7:Dijkl}
\end{equation}
and 
\begin{equation}
    \gamma_{ij} = - \bigg( \frac{7}{2}(c_2^2 + c_3^2) + (2+6q)c_2c_3 \bigg) b_{ij}  - 2 c_2 c_3  b_{ji}.
    \label{eq:Ch7:gamij}
\end{equation}
To determine the constant coefficients, $c_2$ and $c_3$, we use a moments constraint method similar to \cite{girimaji1990diffusion}. In this method, the equations of second and third order moments of $b_{ij}$ are constrained to the values obtained from DNS. 
First, the SDEs for the second ($ q $) and third ($ r $) invariants are derived from the $b_{ij}$-SDE (\ref{eq:Ch7:bijSDE}) using It\^{o}'s lemma (appendix \ref{sec:appD}): 
\begin{eqnarray}
    dq &=& - \bigg( b_{ij} \mu_{ji} +  b_{ij}\gamma_{ji} + \frac{1}{2}D_{ijkl}D_{jikl} \bigg) dt'
    - b_{ij} D_{jimn} \; dW'_{mn} \nonumber \\
    dr &=& - \bigg( b_{ik}b_{kj}\mu_{ji} + b_{ik}b_{kj}\gamma_{ji} + b_{ij} D_{jkmn} D_{kimn} \bigg) dt' 
    - b_{ij}b_{jk}D_{kimn} \; dW'_{mn}
    \label{eq:Ch7:qrSDE}
\end{eqnarray}
Taking mean of the above equations and substituting the expressions for $D_{ijkl}$ and $\gamma_{ij}$ from equations (\ref{eq:Ch7:Dijkl}) and (\ref{eq:Ch7:gamij}), yields the following differential equations of the moments -- $\langle q \rangle$ and $\langle r \rangle$:
\begin{eqnarray}
    \frac{d\langle q \rangle}{ dt'} & = &  - \langle b_{ij} \mu_{ji} \rangle - \langle b_{ij}\gamma_{ji} \rangle - \frac{1}{2} \langle D_{ijkl}D_{jikl} \rangle \nonumber \\
     & = &  - \langle b_{ij} \mu_{ji} \rangle 
    - \big( c_2^2 + c_3^2 \big)\big( 8\langle q\rangle + 1 \big) 
    - c_2c_3 \big( 16 \langle q^2\rangle + 4\langle q\rangle + 4 \big)
    \label{eq:Ch7:qm} \\
    \frac{d\langle r \rangle}{ dt'} & = & - \langle b_{ik}b_{kj}\mu_{ji} \rangle - \langle b_{ik}b_{kj}\gamma_{ji} \rangle - \langle b_{ij} D_{jkmn} D_{kimn} \rangle \nonumber \\
     & = &  - \langle b_{ik} b_{kj} \mu_{ji} \rangle  
    - \big( c_2^2 + c_3^2 \big)\bigg( \frac{27}{2} \langle r \rangle \bigg) 
    - c_2c_3 \big( 6\langle r \rangle + 30 \langle qr \rangle - 6 \langle b_{ij}b_{jk}b_{ik} \rangle \big)
    \label{eq:Ch7:rm}
\end{eqnarray}
To model a statistically stationary solution of turbulence, the rate-of-change of moments must be driven to zero while ensuring that the moment values converge to that of DNS. 
For this, we equate the RHS to negative of the error term:
\begin{eqnarray}
    \frac{d\langle q \rangle}{ dt'} & = & - \langle b_{ij} \mu_{ji} \rangle 
    - \big( c_2^2 + c_3^2 \big)\big( 8\langle q\rangle + 1 \big) 
    - c_2c_3 \big( 16 \langle q^2\rangle + 4\langle q\rangle + 4 \big) = - R \big( \langle q \rangle - \overline{q} \big)  \nonumber \\
    \frac{d\langle r \rangle}{ dt'} & = &  - \langle b_{ik} b_{kj} \mu_{ji} \rangle  
    - \big( c_2^2 + c_3^2 \big)\bigg( \frac{27}{2} \langle r \rangle \bigg) 
    - c_2c_3 \big( 6\langle r \rangle + 30 \langle qr \rangle - 6 \langle b_{ij}b_{jk}b_{ik} \rangle \big) \nonumber \\
    & = & - R \big(\langle r \rangle - \overline{r} \big)
\end{eqnarray}
where $\overline{q}, \overline{r}$ are the global mean of $q,r$ obtained from DNS data. 
Here, $R$ represents the rate of convergence of these moments and is set to unity.
An \textit{a priori} simulation of the $b_{ij}$ model equations is run in the normalized timescale $t'$, with an ensemble of $40000$ particles.
At each time step, the above system of nonlinear equations is solved using Newton's method to determine the values of the coefficients $c_2,c_3$. 
In this \textit{a priori} run, the coefficients converge to the following values:
\begin{equation}
    c_2 = 0.0099 \;\; , \;\; c_3 = -0.064
\end{equation}
as the model's moments, $\langle q \rangle$ and $\langle r \rangle$, converge very close to the DNS values of $\overline{q}$ and $\overline{r}$. 
These optimized diffusion coefficient values are used in the stochastic model for $b_{ij}$ and are insensitive to the Reynolds number. 

\subsection{Model for VG magnitude \label{sec:Ch7:modelths}}

The Lagrangian evolution of the scalar $\theta^*$ (equation \ref{eq:Ch7:ths}) is modeled using a modified lognormal approach.
The magnitude $A$ has a nearly lognormal probability distribution and exponential decay of autocorrelation in time \citep{kolmogorov1962refinement,oboukhov1962some,yeung1989lagrangian}.
The exponentiated Ornstein-Uhlenbeck (OU) process is a statistically stationary process that satisfies both these properties \citep{uhlenbeck1930theory,pope1990velocity} and is therefore ideal for modeling $\theta^*$. 
While it has been pointed out that pseudodissipation rate ($A^2$) cannot be precisely lognormal in the context of multifractal formalism \citep{mandelbrot1974intermittent,meneveau1991multifractal}, the OU process models the overall dynamics of $A$ quite accurately \citep{pope1990velocity,girimaji1990diffusion}.
In fact, a recent analysis of Lagrangian trajectories in high Reynolds number turbulence \citep{huang2014lagrangian} has shown evidence that the autocorrelation function of $A$ is consistent with both the exponential decay prescribed by the OU process as well as the logarithmic decay suggested by the multifractal framework, and the two are nearly indistinguishable at such high Reynolds numbers \citep{pereira2018multifractal}.
Since the focus of this work is to accurately reproduce the overall Lagrangian dynamics of the velocity gradients in turbulence, 
we model the velocity gradient magnitude as a Reynolds number-dependent modified lognormal process, without explicitly accounting for multifractal behavior. 

The OU process is a stationary continuous Gaussian Markov process that is often used in modeling systems of finance, mathematics, and physical and biological sciences \citep{pope1990velocity,klebaner2012introduction}.
It further shows the property of mean-reversion. 
The SDE for a general OU process $\theta^*$ evolving in time $t^*$ is given by \citep{girimaji1990diffusion}
\begin{equation}
    d \theta^* = - \alpha (\theta^*-\langle {\theta^*}\rangle) dt^* + \beta \; dW^*
    \label{eq:Ch7:thsSDE1}
\end{equation}
where, $\alpha, \beta > 0$ are parameters of the model, $t^* = \langle A \rangle t$ is the non-dimensional global timescale and $dW^*$ is the increment of a Wiener process or a Gaussian random variable with zero mean and variance $dt^*$.
The parameter $\alpha$ represents the rate of mean-reversion, and without loss of generality it is set to unity since the model propagates in timescale $t^*$ which is already normalized.
The expected value $\langle {\theta^*} \rangle =0$, by construction, in DNS data. 
Therefore, the general form of $\theta^*$-SDE used in this work is
\begin{equation}
    d \theta^* = - \theta^* dt^* + \beta \; dW^*.
    \label{eq:Ch7:thsSDE}
\end{equation}
The diffusion coefficient, $\beta$, is modeled in three different ways as described below.

\subsubsection{Model 1 - simple OU process}
 
Here, we disregard the dependence of $\theta^*$ on $b_{ij}$ and consider the simple OU dynamics that satisfies the global mean and global variance of $\theta^*$. In this case, the diffusion coefficient $\beta$ is taken to be a constant value, which is calculated as follows. 
The equation for the global mean is obtained from equation (\ref{eq:Ch7:thsSDE}),
\begin{equation}
    \frac{d \langle \theta^* \rangle}{dt^*} = - \langle \theta^* \rangle = 0 \;\; \text{since} \;\;
    \langle \theta^* \rangle = 0.
\end{equation}
Thus, the model maintains a stationary mean value of $\theta^*$ once the solution is driven to the zero mean value by the mean-reversion property.
Next, the equation for the global variance is obtained from equation (\ref{eq:Ch7:thsSDE}) using It\^{o}'s product rule,
\begin{equation}
    \frac{d \langle {\theta^*}^2 \rangle}{dt^*} = -2\langle {\theta^*}^2 \rangle + \beta^2
\end{equation}
For a statistically stationary solution, we must have
\begin{equation}
    \frac{d \langle {\theta^*}^2 \rangle}{dt^*} = 0  \;\;\; \implies \;\; \beta = \sqrt{2\langle {\theta^*}^2 \rangle} 
\end{equation}
Therefore, the final form of the $\theta^*$-SDE for Model 1 is given by
\begin{equation}
    d \theta^* = - \theta^* dt^* + \sqrt{2 \langle {\theta^*}^2 \rangle} \; dW^*.
\end{equation}
Here, the value of the variance $\langle {\theta^*}^2 \rangle$ is obtained from DNS data and it can be a function of $Re_\lambda$.

\subsubsection{Model 2 - modified OU process}

\begin{figure}
    \centering
    \begin{subfigure}{0.42\textwidth}
        \centering
        \includegraphics[width=\textwidth]{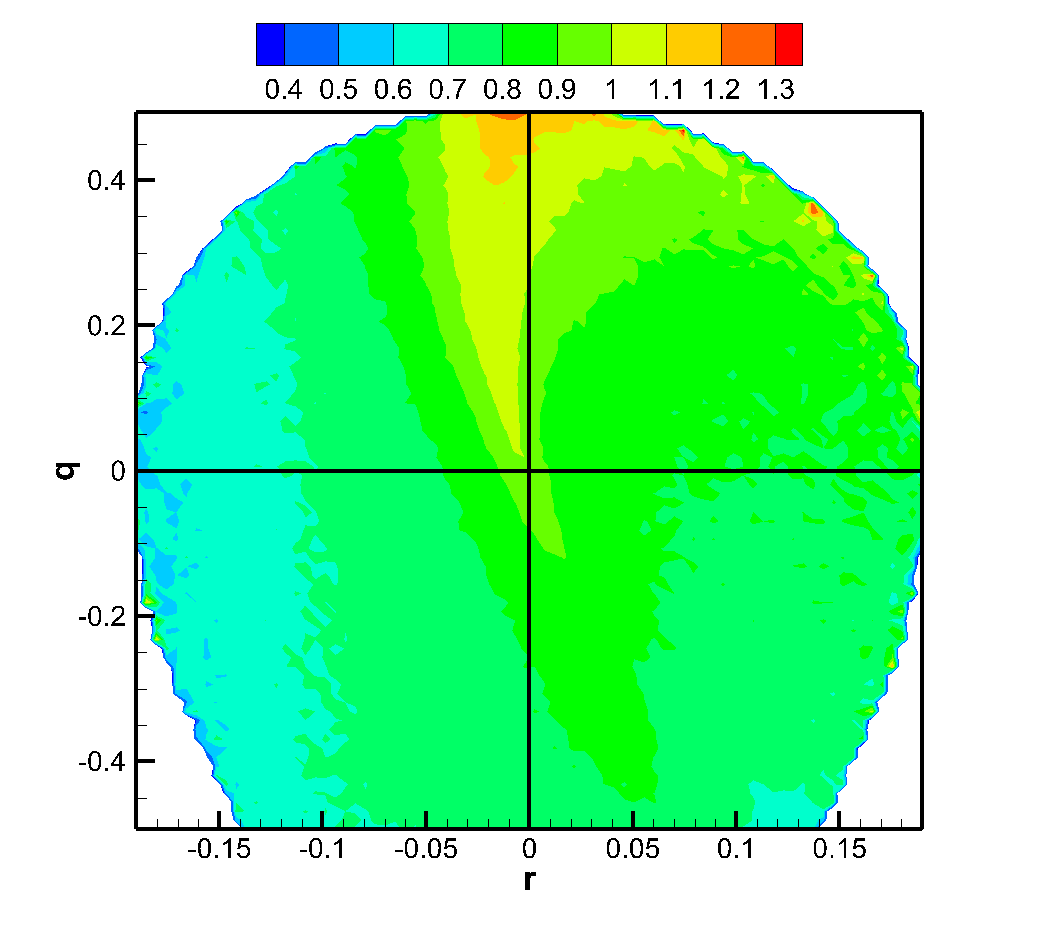}
        \caption{}
    \end{subfigure}
    \begin{subfigure}{0.42\textwidth}
        \centering
        \includegraphics[width=\textwidth]{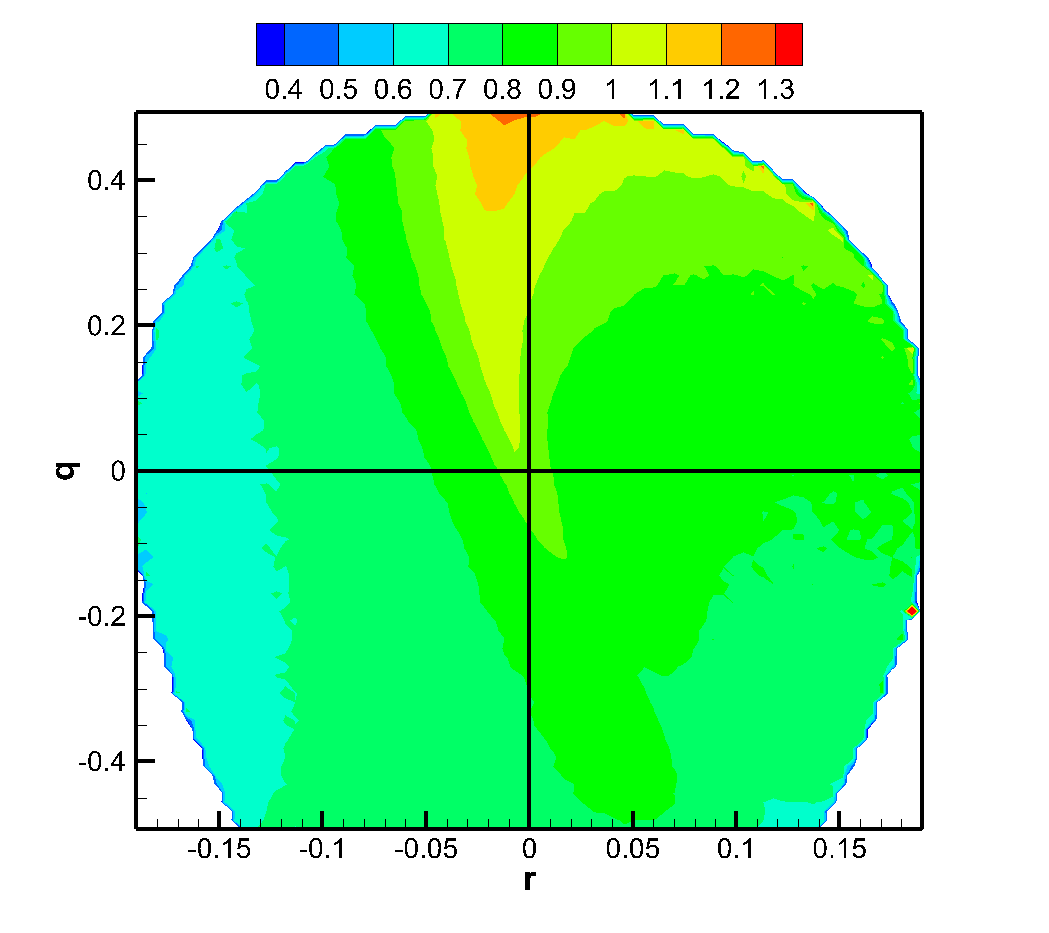}
        \caption{}
    \end{subfigure}
    \begin{subfigure}{0.42\textwidth}
        \centering
        \includegraphics[width=\textwidth]{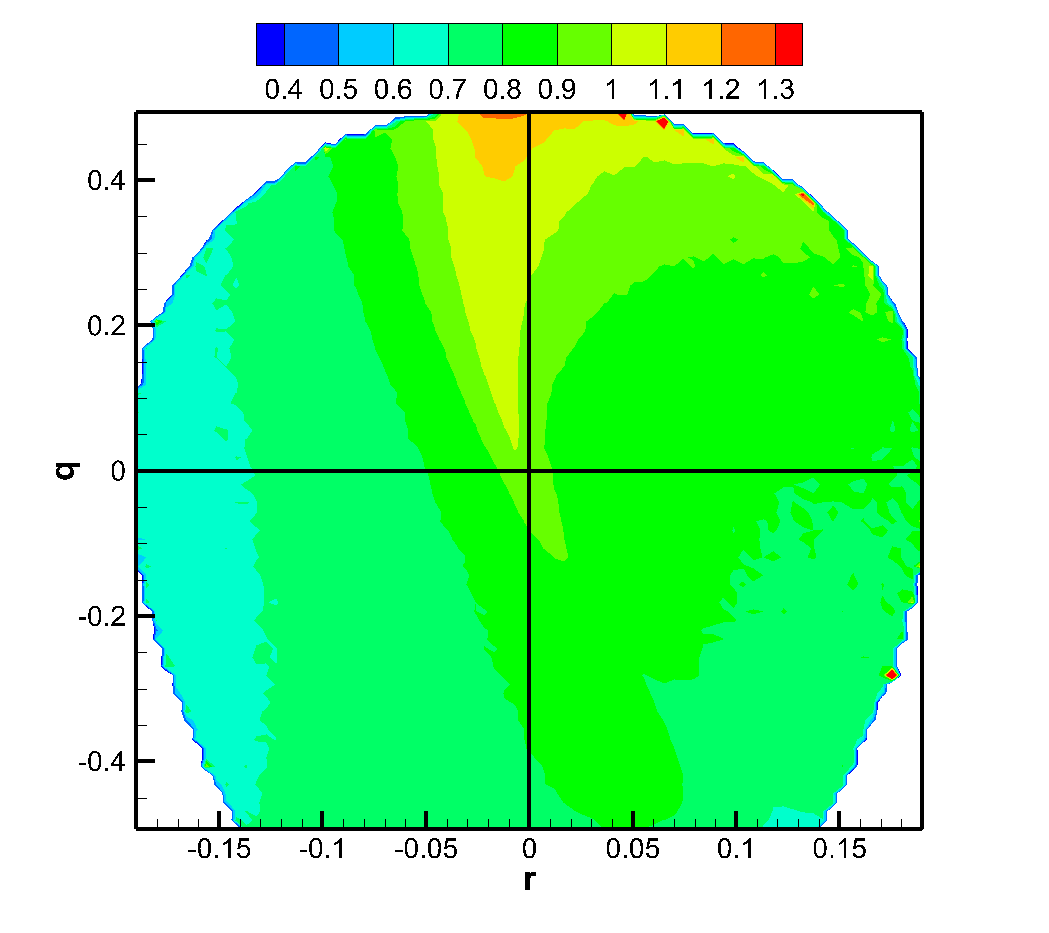}
        \caption{}
    \end{subfigure}
    \begin{subfigure}{0.42\textwidth}
        \centering
        \includegraphics[width=\textwidth]{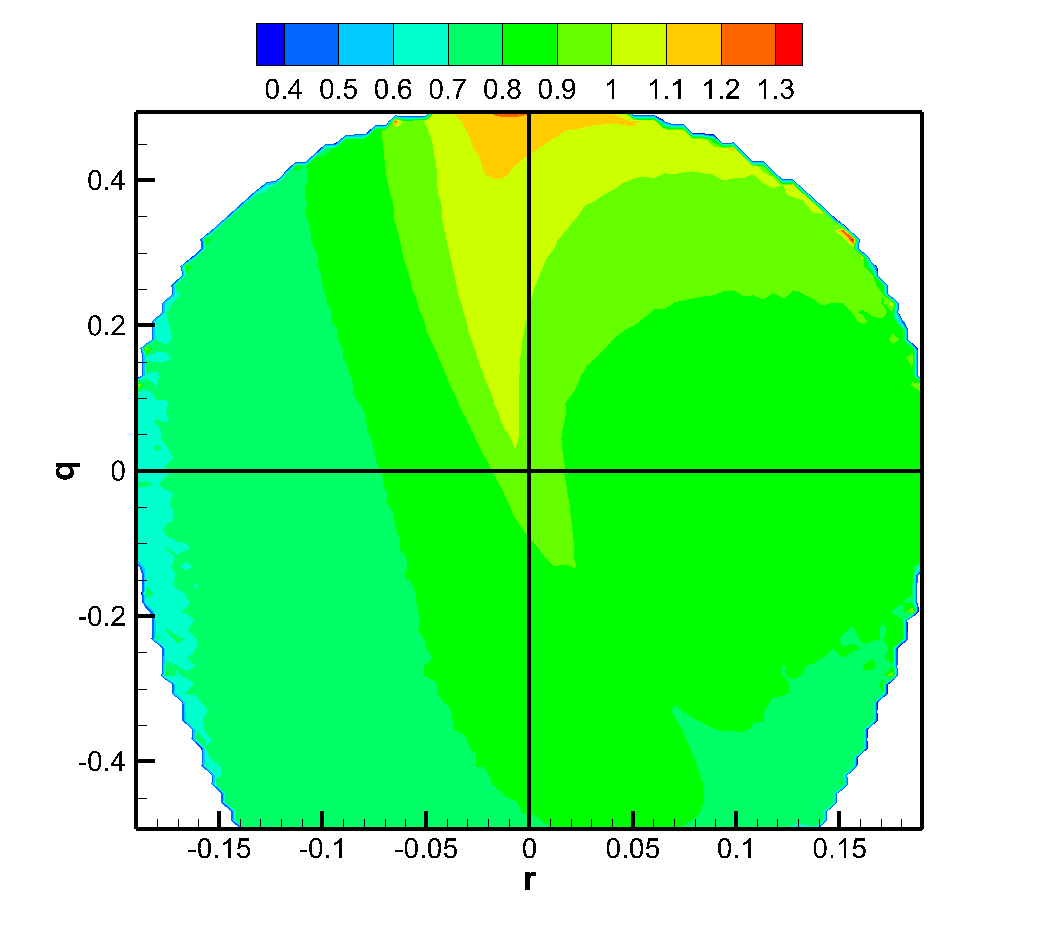}
        \caption{}
    \end{subfigure}
    \caption{\label{fig:Ch7:ths_condVar} Conditional variance of $\theta^*$ conditioned on $q$-$r$, i.e. $\langle (\theta^* - \langle \theta^*| q,r \rangle )^2 | q,r \rangle$, for isotropic turbulent flows of Taylor Reynolds numbers: (a) \re$=225$, (b) \re$=385$, (c) \re$=427$, (d) \re$=588$ .}
\end{figure}

Next, we want to ensure that the value of the variance of $\theta^*$ conditioned on the local streamline geometry ($q,r$) is satisfied. This conditional variance, $\langle (\theta^* - \langle \theta^*| q,r\rangle)^2 | q,r \rangle$, is plotted in figure \ref{fig:Ch7:ths_condVar} for DNS data of forced isotropic turbulence at different Reynolds numbers. It is evident that the conditional variance of $\theta^*$ shows a clear dependence on $q$ and $r$, and is nearly invariant with changing $Re_\lambda$. 
Therefore, we modify the diffusion coefficient as
\begin{equation}
    \beta = \beta(q,r)
\end{equation}
to account for the correct conditional variance of $\theta^*$ for a given $q$-$r$ value.
The equation for the conditional variance can be derived from equation (\ref{eq:Ch7:thsSDE}) as
\begin{equation}
    \frac{d }{dt^*}  \langle (\theta^* - \langle \theta^*| q,r\rangle)^2 | q,r \rangle = - 2 \langle {\theta^*}^2 | q,r \rangle + \langle {\theta^*} | q,r \rangle^2 + (\beta(q,r))^2  
\end{equation}
The conditional variance remains statistically stationary only if
\begin{equation}
    \frac{d }{dt^*}  \langle (\theta^* - \langle \theta^*| q,r\rangle)^2 | q,r \rangle = 0 \;\;\;
    \implies \;\; \beta(q,r) = \sqrt{2(\langle {\theta^*}^2 | q,r \rangle - \langle \theta^* | q,r \rangle^2 )}
\end{equation}
Therefore, the final SDE of $\theta^*$ for Model 2 is given by
\begin{equation}
    d \theta^* = - \theta^* dt^* + \sqrt{2(\langle {\theta^*}^2 | q,r \rangle - \langle \theta^* | q,r \rangle^2 )} \; dW^*.
\end{equation}
Here, the conditional variance values are obtained from DNS data of one \re by discretizing the $q,r$ space into $30\times30$ bins. The same conditional variance table is applicable when modeling turbulent flows of different Reynolds numbers, as evident from figure \ref{fig:Ch7:ths_condVar}.
This $\theta^*$-model is weakly coupled with the $b_{ij}$ dynamics since it depends on $q,r$.

\subsubsection{Model 3 - consistent modified OU process}

Model 1 ensures that the constant diffusion coefficient captures the accurate global variance of $\theta^*$, while model 2 enforces the accurate modeling of conditional variance of $\theta^*$ for a given $q,r$. 
Finally, in Model 3 we propose an adjustment to model 2 such that both the conditional and global variances are satisfied. 
For this, we propose to shift the conditional variance based diffusion coefficient of model 2 by a constant value, $\beta_0$, as follows:
\begin{equation}
    d \theta^* = - \theta^* dt^* + \big( \sqrt{2(\langle {\theta^*}^2 | q,r \rangle - \langle \theta^* | q,r \rangle^2 )} + \beta_0 \big)\; dW^*.
\end{equation}
Here, the value of $\beta_0$ is calculated based on the solutions of models 1 and 2. 
Model 3 also depends upon $b_{ij}$ through its invariants $(q,r)$.


\subsection{Model summary \label{sec:Ch7:modelsumm}}

The resulting model for the Lagrangian evolution of the complete velocity gradient tensor in a turbulent flow is described by a system of stochastic differential equations for the normalized velocity gradient tensor, $b_{ij}$, and a separate stochastic differential equation for the standardized VG magnitude, $\theta^*$.

The final system of equations for $b_{ij}$ in local timescale ($t'$):
\begin{equation}
    d b_{ij} = (\mu_{ij} + \gamma_{ij}) dt' + D_{ijkl} \;dW'_{kl}
    \label{eq:Ch7:bijSDE2}
\end{equation}
\begin{eqnarray}
    \mu_{ij} = &-& b_{ik} b_{kj} + \frac{1}{3} b_{km}b_{mk} \delta_{ij} + b_{ij} b_{mk} b_{kn} b_{mn} + \big\langle  h_{ij} \big| \bm{b} \big\rangle - b_{ij}b_{kl} \big\langle h_{kl} \big| \bm{b} \big\rangle \nonumber \\
    &+& \big\langle \tau_{ij} \big| \bm{b} \big\rangle - b_{ij}b_{kl} \big\langle \tau_{kl} \big| \bm{b} \big\rangle \;, \nonumber \\
    \gamma_{ij} = &-& \bigg( \frac{7}{2}(c_2^2 + c_3^2) + (2+6q)c_2c_3 \bigg) b_{ij}  - 2 c_2 c_3  b_{ji} \;, \nonumber \\
    D_{ijkl} &=& c_2 \bigg( - \frac{1}{3}\delta_{ij}\delta_{kl} + \delta_{ik}\delta_{jl} - b_{ij}b_{kl} \bigg) 
    + c_3 \bigg( - \frac{1}{3}\delta_{ij}\delta_{kl} + \delta_{il}\delta_{jk} - b_{ij}b_{lk} \bigg)
    \label{eq:Ch7:muij2}
\end{eqnarray}
Here, the diffusion coefficient values are
\begin{equation}
    c_2 = 0.009877 \;\; , \;\; c_3 = -0.06402.
\end{equation}
The conditional mean normalized pressure Hessian and viscous Laplacian tensors are obtained from the data-driven closure in the strain-rate ($\bm{s}$) eigen reference frame as a function of the current ($q,r,a_2,\omega_2$), followed by a rotation to the flow reference frame using the local eigenvectors of $\bm{s}$: 
\begin{equation}
    \langle {h}_{ij} | \bm{b} \rangle = Q_{ik} \langle \tilde{h}_{kl} | q,r,a_2,\omega_2 \rangle Q_{jl} \;\; \text{and} \;\; 
    \langle {\tau}_{ij} | \bm{b} \rangle = Q_{ik} \langle \tilde{\tau}_{kl} | q,r,a_2,\omega_2 \rangle Q_{jl} . 
\end{equation}
The above coefficient values and the data-driven closure can be applied to model velocity gradient dynamics of incompressible turbulent flows irrespective of the Taylor Reynolds number.

The final stochastic differential equation for $\theta^*$ in global timescale ($t^* = \langle A \rangle \;t$):
\begin{itemize}
    \item Model 1 -
    \begin{equation}
        d \theta^* = - \theta^* dt^* + \sqrt{2 \langle {\theta^*}^2 \rangle} \; dW^* \;,
        \label{eq:Ch7:Model1ths}
    \end{equation}
    \item Model 2 -
    \begin{equation}
        d \theta^* = - \theta^* dt^* + \sqrt{2(\langle {\theta^*}^2 | q,r \rangle - \langle \theta^* | q,r \rangle^2 )} \; dW^* \;,
        \label{eq:Ch7:Model2ths}
    \end{equation}
    \item Model 3 -
    \begin{equation}
        d \theta^* = - \theta^* dt^* + \big( \sqrt{2(\langle {\theta^*}^2 | q,r \rangle - \langle \theta^* | q,r \rangle^2 )} + \beta_0 \big)\; dW^* \;.
        \label{eq:Ch7:Model3ths}
    \end{equation}
\end{itemize}
where, $\beta_0=0.103$. 
The diffusion coefficients of models 2 and 3 are obtained from the tabulated conditional variance of $\theta^*$ invariant with \re (figure \ref{fig:Ch7:ths_condVar}).
The VG magnitude and the VG tensor are then given by
\begin{equation}
    A = e^{\theta^* \sigma_\theta + \langle \theta \rangle} \;\;, \;\; A_{ij} = A \; b_{ij}
\end{equation}
In general, the parameters in the above equation are Reynolds number dependent as shown in figure \ref{fig:Ch7:theta_stats}. 
For the case of $Re_\lambda=427$, the parameter values are:
\begin{eqnarray}
    \langle \theta \rangle = 2.7493 \;\;,\;\; \sigma_\theta = 0.589655 \;\;,\;\; \langle A \rangle = 18.64173 .
\end{eqnarray} 
The above $\theta^*$ models can be used to simulate the dynamics of VG magnitude for any Reynolds number, provided the corresponding parameter values are known. 

\begin{table}  
  \begin{center}
\def~{\hphantom{0}}
\setlength{\tabcolsep}{1pt}
  \begin{tabular}{lcc}
        Data-based & Discretization & \re \\
        component &  of state-space & dependence \\[1pt]
	\hline \vspace{-0.6cm} \\ 
	\hline 
	    $\langle \bm{\tilde{h}} | q,r,a_2,\omega_2 \rangle $ & (60,60,30,30) & \rey-independent \\[4pt]
	    $\langle \bm{\tilde{\tau}} | q,r,a_2,\omega_2 \rangle $ & (60,60,30,30) & \rey-independent \\[4pt]
	\hline  \vspace{-0.3cm} \\[0.2pt]
	    $\langle (\theta^* - \langle \theta^*|q,r\rangle)^2 | q,r \rangle $ & (30,30) & \rey-independent \\[4pt]
	\hline  \vspace{-0.3cm} \\[0.2pt]
	    $\langle \theta \rangle$ , $\sigma_\theta$ , $\langle A \rangle$ & 1 & \rey-dependent \\[4pt]
  \end{tabular}
  \caption{Components of model based on DNS data.}
  \label{tab:Ch7:data}
  \end{center}
\end{table}

To reconcile between the two different timescales - $t'$ used for $b_{ij}$ evolution and $t^*$ used for $\theta^*$ evolution - a simple, closed ordinary differential equation,
\begin{equation}
    \frac{dt^*}{dt'} = \frac{\langle A \rangle}{A}
    \label{eq:Ch7:timescaleODE}
\end{equation}
is solved to determine $t^*$ for a given $t'$.
No closure is required for this equation.
Finally, the Lagrangian evolution of the velocity gradient tensor, $A_{ij}$, is obtained by multiplying $A$ and $b_{ij}$ at different global time $t^*$. Overall, the velocity gradient model presented here consists of three types of data-based components that are listed in table \ref{tab:Ch7:data}. 
The four-dimensional lookup tables for normalized pressure Hessian and viscous Laplacian tensors and the two-dimensional table for conditional variance of $\theta^*$ can be used in modeling velocity gradient dynamics independent of Reynolds numbers; only the three scalar parameters of the model which are statistics of the VG magnitude are Reynolds number dependent and potentially generalizable in the future with universal scaling laws.

\section{Numerical procedure \label{sec:Ch7:num}}

The numerical solution of the model equations involves numerically propagating the velocity gradient tensor, in terms of the variables $b_{ij}$ and $\theta^*$, of an ensemble of 40000 particles. 
As the initial conditions for the simulations, the particles are picked at random from a randomly generated incompressible isotropic velocity field. 
The trajectories are advanced for a total time period of approximately $1200\tau_\eta$ following these steps at each update:
\begin{enumerate}
    \item The $b_{ij}$ SDEs (equation \ref{eq:Ch7:bijSDE2}) are numerically propagated in the normalized local timescale $t'$, using a second-order weak predictor-corrector scheme (see appendix \ref{sec:appD}) with a constant time increment $dt'$. At each step, the conditional mean nonlocal pressure and viscous contributions are calculated  based on the current $(q,r,a_2,\omega_2)$ values, using the $(60,60,30,30)$ sized lookup-table.
    \item The $\theta^*$ SDE (equation \ref{eq:Ch7:Model1ths}, \ref{eq:Ch7:Model2ths} or \ref{eq:Ch7:Model3ths}) is advanced using the second-order weak predictor-corrector numerical scheme (appendix \ref{sec:appF}) in the global timescale $t^*$, using a first-order approximation of the increment $dt^* = \langle A \rangle dt' / A$ for a fixed value of $dt'$.
    \item The global timescale, $t^*$, is obtained for every particle at each $t'$ by numerically solving the ordinary differential equation \ref{eq:Ch7:timescaleODE} using the implicit second-order Trapezium rule method.
\end{enumerate}
The model solution propagates all particles at a uniform local time increment of $dt'=0.01$, but the global timescale $t^*$ varies from one particle to the other depending upon its current velocity gradient magnitude.
A particle with a smaller magnitude requires fewer steps in $t'$ to reach a certain $t^*$, than a particle with a larger magnitude.
The VG magnitude $\theta^*$ evolves in global time $t^*$, which approximately scales with Kolmogorov timescale. On the other hand, $b_{ij}$ evolves in local timescale $t'$ which varies depending on the local value of $A$.
Issues may arise when $A < \langle A \rangle$, i.e. when $b_{ij}$ evolves faster than $\theta^*$, and appropriate measures should be taken to ensure that the $dt'$ is suitable to propagate the $b_{ij}$ equations. 
However, particles with such low $A$ values do not contribute significantly toward the overall velocity gradient statistics. Convergence of the model's results for $dt'=0.05,0.01$ and $0.002$ suggest that $dt'=0.01$ is sufficient here for accurate statistical modeling.

The incompressibility and normalization constraints are automatically upheld by the model, but are only valid up to the order of numerical error.
Therefore, to avoid the accretion of numerical errors over large periods of time, hard constraints of $b_{ii}=0$ and $||\bm{b}||_F=1$ are enforced after every update.
The computation time is approximately 1.5-2 hours on a single processor for the model's simulations to achieve statistically stationary solutions. 
The results of the model's simulations for the three different $\theta^*$-models are illustrated in the next section as model 1 if equation (\ref{eq:Ch7:Model1ths}) is used, model 2 if equation (\ref{eq:Ch7:Model2ths}) is used, and model 3 if equation (\ref{eq:Ch7:Model3ths}) is used, each along with the $b_{ij}$ equation (\ref{eq:Ch7:bijSDE2}).
The convergence of all the major results have been tested for these models by performing the simulations with $40000$ and $100000$ particles.

\section{Results and comparison with DNS data \label{sec:Ch7:results}}

This section presents a statistical analysis of the solutions of the three models 
and a comparison with the statistics of the corresponding DNS data and some previous models.
First, the statistics of $\theta^*$ are illustrated, followed by the statistics of $b_{ij}$. Finally the complete velocity gradient tensor $A_{ij}$-statistics are shown.
The time evolution of the model's statistics are illustrated as a function of the global normalized time $t^*$.
The time-converged statistical results are plotted by averaging over multiple time realizations of the model's solution, separated by at least $5\tau_\eta$, well after statistical stationarity has been achieved.

\subsection{VG magnitude \label{sec:Ch7:resultsths}}

\begin{figure}
    \centering
    \begin{subfigure}{0.53\textwidth}
        \centering
        \includegraphics[width=\textwidth]{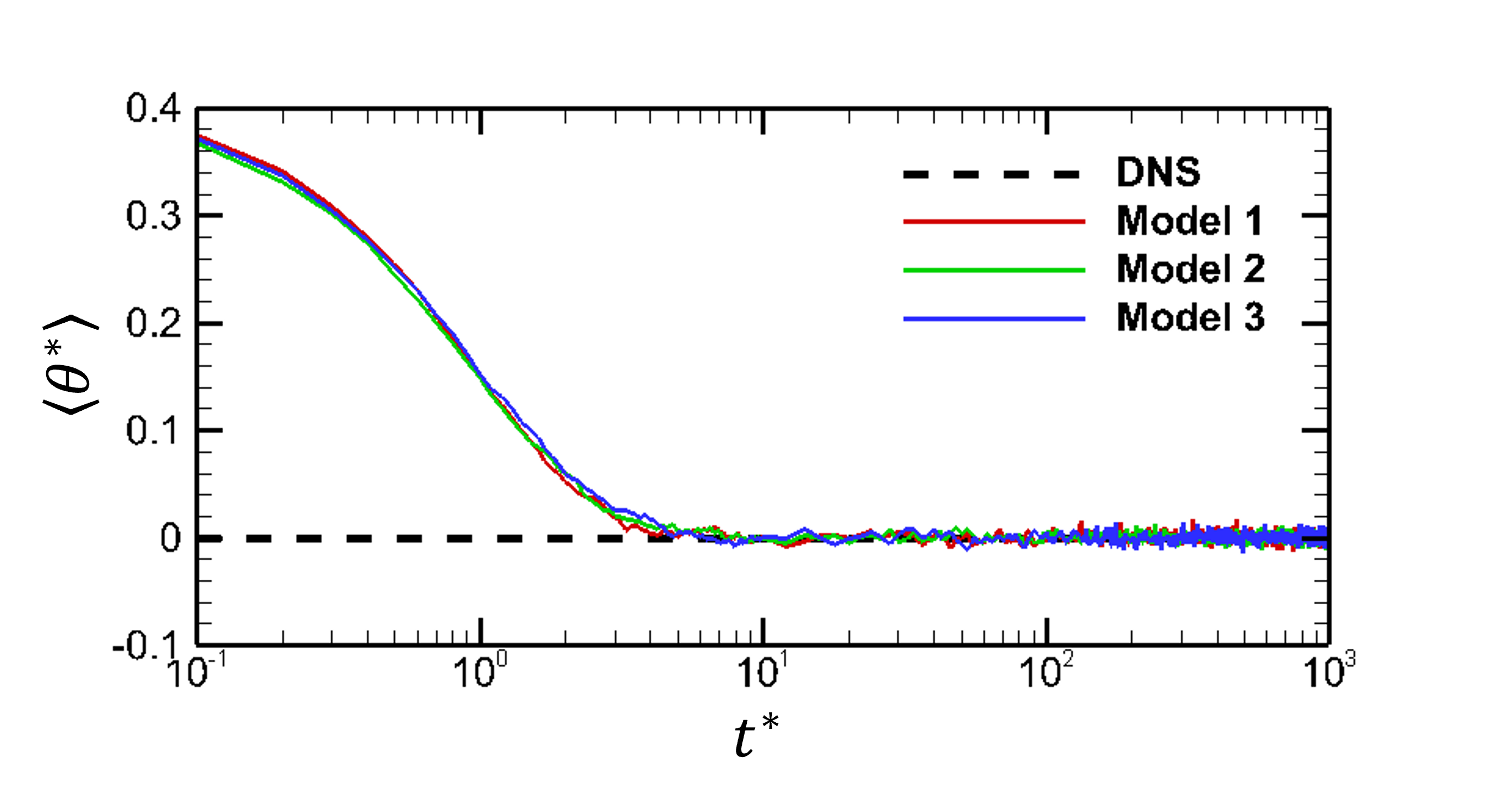}
        \caption{}
    \end{subfigure}
    \hfill 
    \begin{subfigure}{0.53\textwidth}
        \centering
        \includegraphics[width=\textwidth]{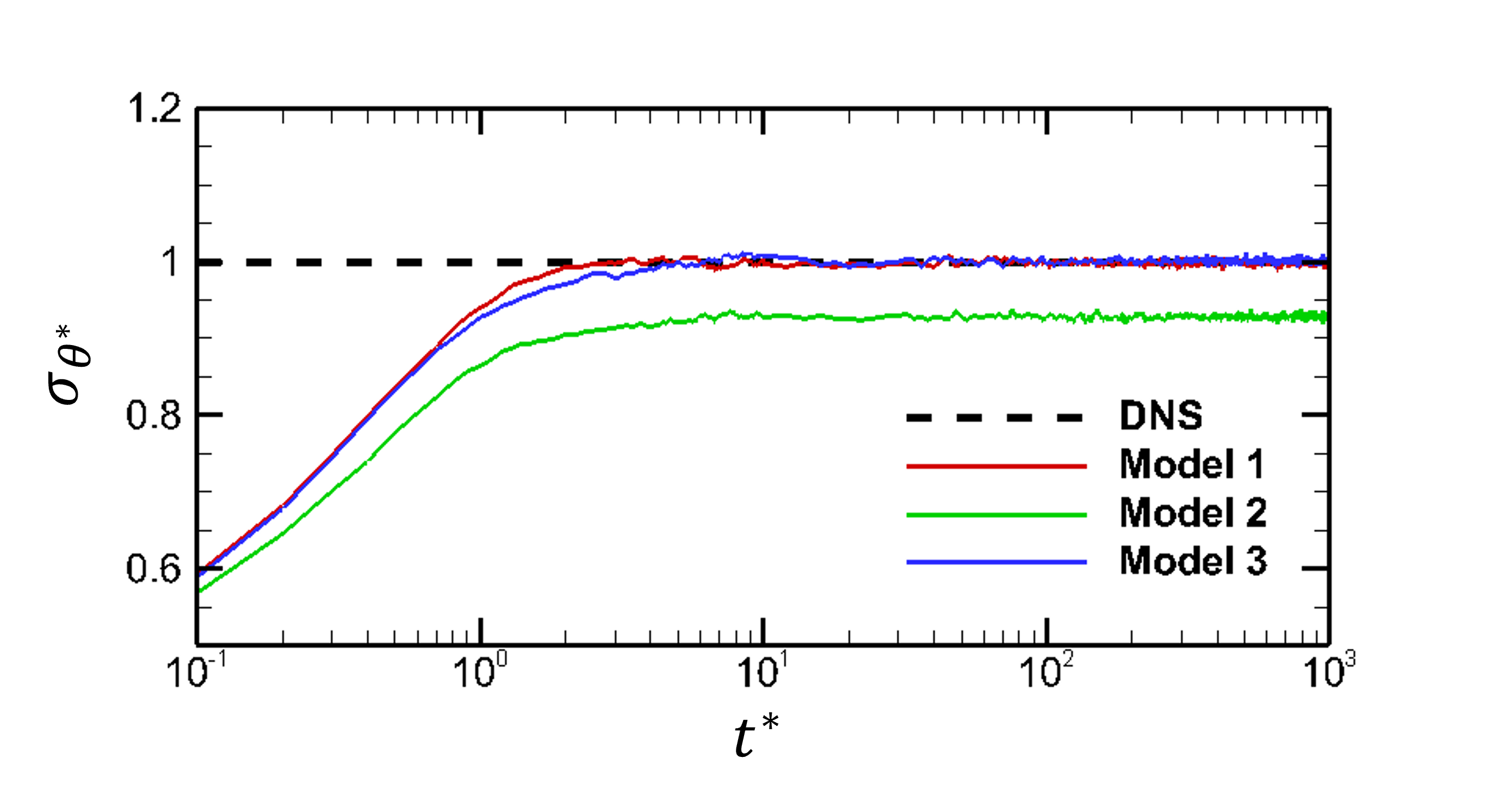}
        \caption{}
    \end{subfigure}
    \caption{\label{fig:Ch7:ths_evol_ts} Evolution of $\theta^*$ statistics: (a) mean, $\langle \theta^* \rangle$ and (b) standard deviation, $\sigma_{\theta^*}$, for the three models with different $\theta^*$ equations. The DNS statistics are marked by dashed lines. The time axis is in logscale.}
\end{figure}

First, the time evolution of the mean and standard deviation of $\theta^*$ 
are plotted for all the three $\theta^*$-model equations in figure \ref{fig:Ch7:ths_evol_ts}. 
Note that the time axes are plotted in logscale to display the transients clearly.
The numerical simulation starts from an initially random field, which is inconsistent with the DNS values of $\langle \theta \rangle$ and $\sigma_\theta$, and therefore the initial values of $\langle \theta^* \rangle$ and $\sigma_\theta^*$ are different from zero and unity, respectively.
Over time, the model's solution evolves toward the DNS value achieving statistically stationary state at about $t \approx 7 \tau_\eta$ ($t^*\approx 6$), where $t$ is the real time. 
As expected, the global mean of $\theta^*$ is captured equally well by all three models due to the mean-reverting property of the OU process.
The global standard deviation of $\theta^*$ is reproduced accurately by both models 1 and 3.
However, it is worse in model 2 compared to the simple OU model (model 1). This indicates that imposing $\theta^*$ to satisfy the variance conditioned on local streamline geometry ($q,r$) in model 2 does not necessarily guarantee that the global variance of $\theta^*$ is automatically satisfied. This justifies the need for the third model to satisfy both the global and conditional standard deviation values.

\begin{figure}
    \centering
    \begin{subfigure}{0.49\textwidth}
        \centering
        \includegraphics[width=\textwidth]{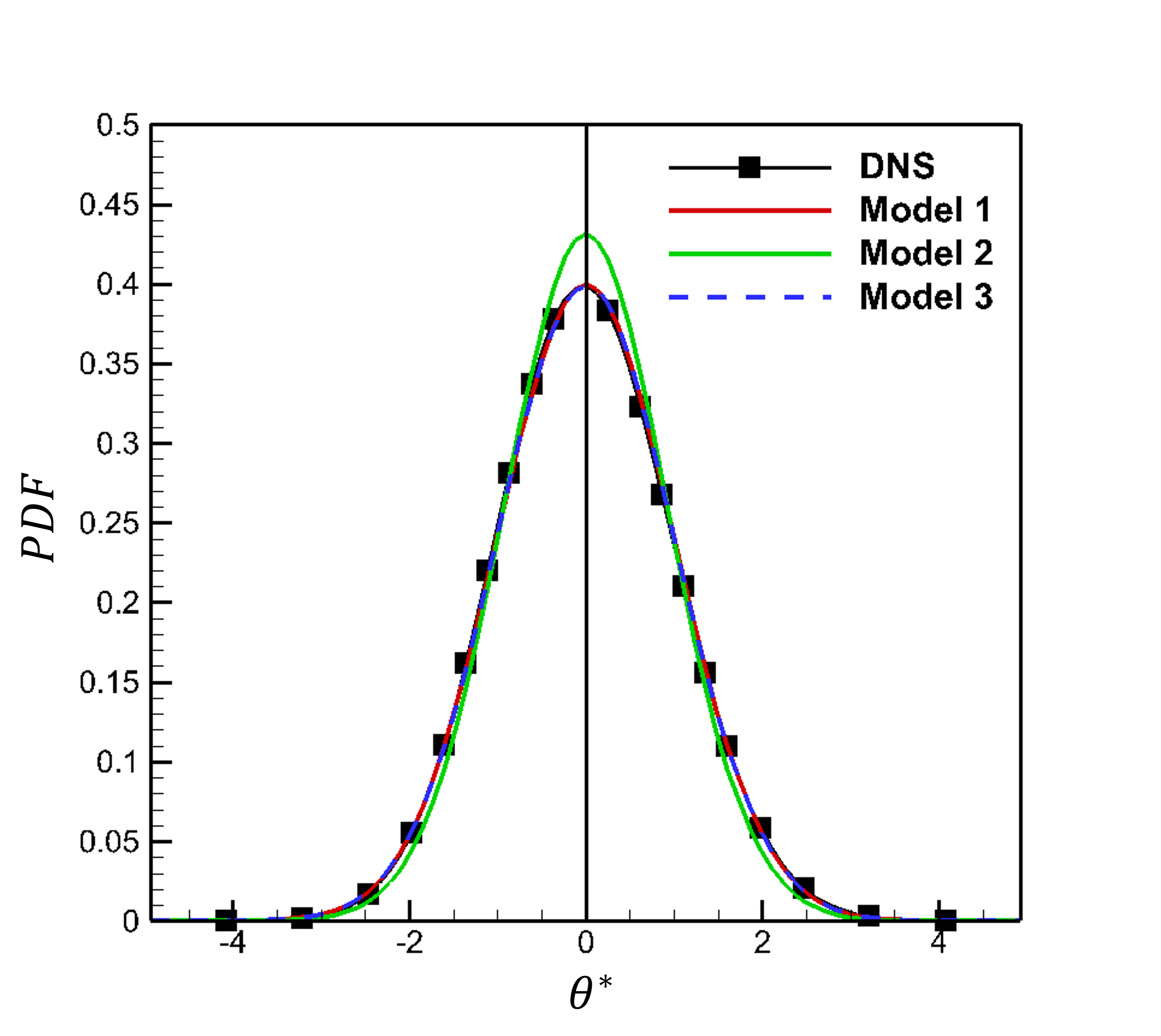}
        \caption{}
    \end{subfigure}
    \hfill 
    \begin{subfigure}{0.49\textwidth}
        \centering
        \includegraphics[width=\textwidth]{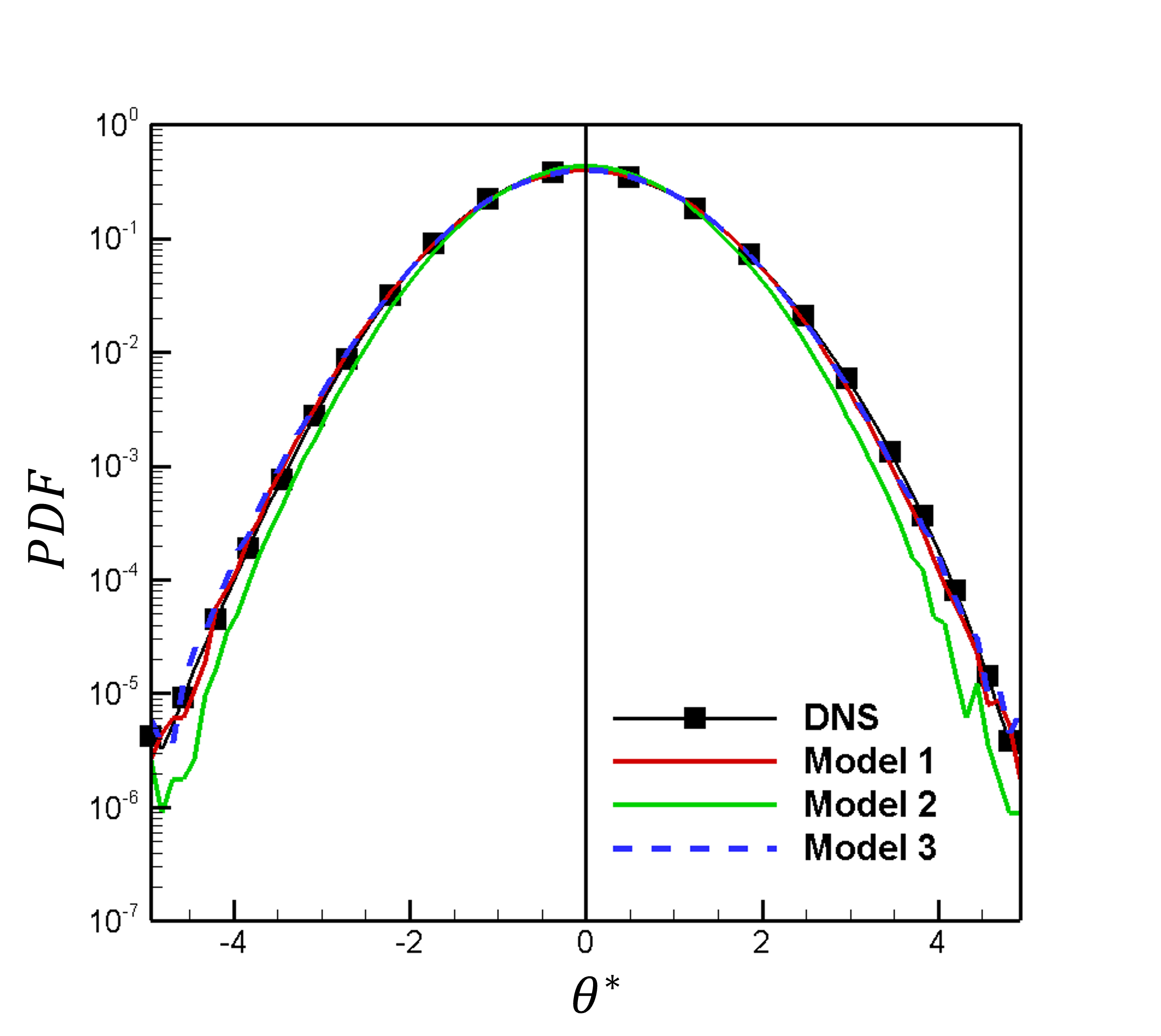}
        \caption{}
    \end{subfigure}
    \caption{\label{fig:Ch7:ths_PDF} PDF of standardized VG magnitude $\theta^*$ in: (a) linear-linear scale and (b) linear-log scale, for the three models. The black solid line with symbols represent the $\theta^*$-PDF from DNS data.}
\end{figure}

\begin{figure}
    \centering
    \includegraphics[width=0.49\textwidth]{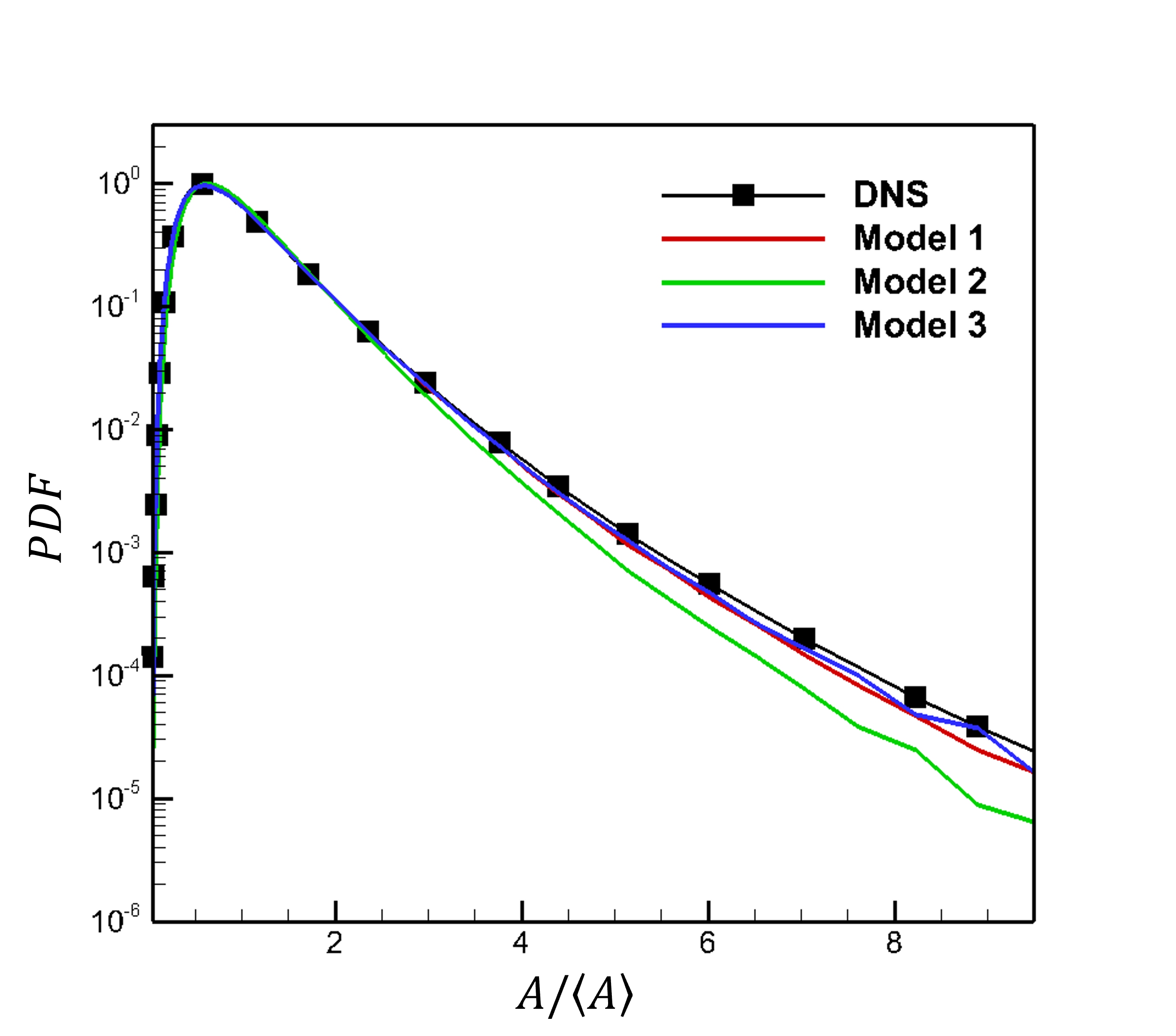}
    \caption{\label{fig:Ch7:A_PDF} PDF of VG magnitude $A/\langle A\rangle$.}
\end{figure}

The converged probability density function (PDF) of the standardized VG magnitude, $\theta^*$, for all the models and DNS data are plotted in figure \ref{fig:Ch7:ths_PDF}.
It is clear that models 1 and 3 are able to reproduce the $\theta^*$ PDF very well, while model 2 shows deviation from the desired DNS result.
The plot in the log-linear scale confirms that the converged PDFs of models 1 and 3 agree well with that of DNS even near the extreme tails of the PDFs.
Next, the converged PDFs of the VG magnitude ($A/\langle A\rangle$) in each of the three models and DNS are plotted in figure \ref{fig:Ch7:A_PDF}. 
All models capture the peak of the PDF reasonably well but model 2 deviates at higher values of magnitude while models 1 and 3 perform better. 
It further shows that model 3 is able to reproduce the tails of the PDF slightly better than model 1. 

\subsection{Normalized VG tensor \label{sec:Ch7:resultsbij}}


\begin{figure}
    \centering
    \begin{subfigure}{0.49\textwidth}
        \centering
        \includegraphics[width=\textwidth]{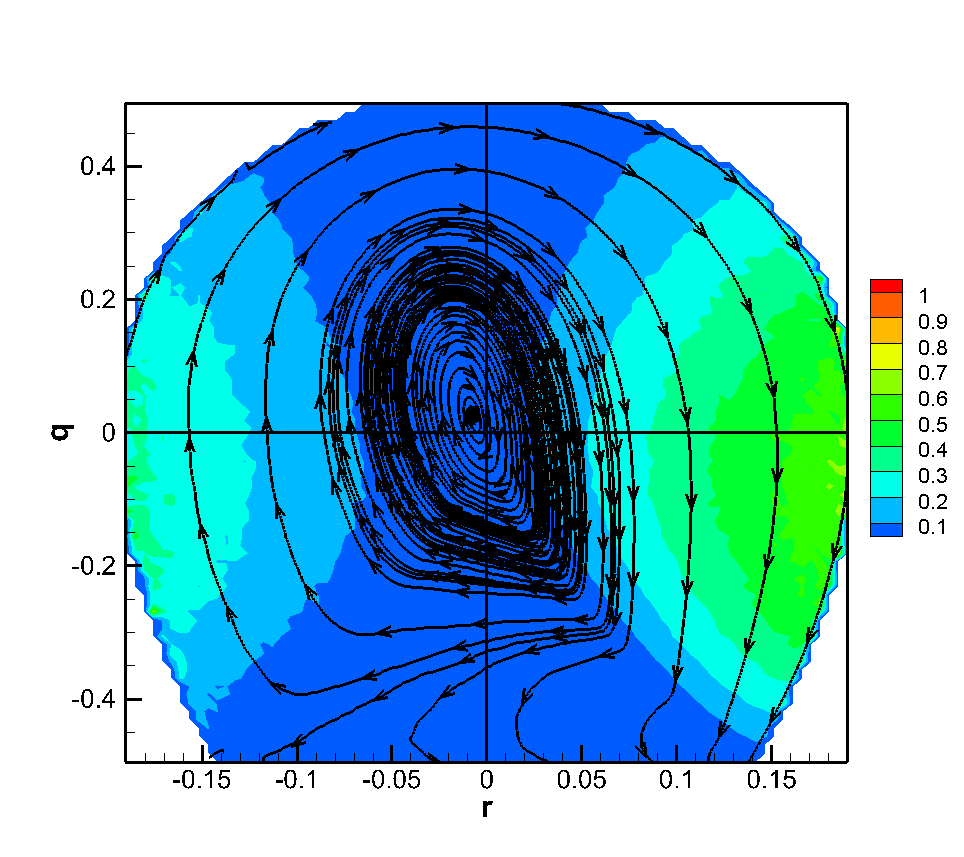}
        \caption{}
    \end{subfigure}
    \hfill 
    \begin{subfigure}{0.49\textwidth}
        \centering
        \includegraphics[width=\textwidth]{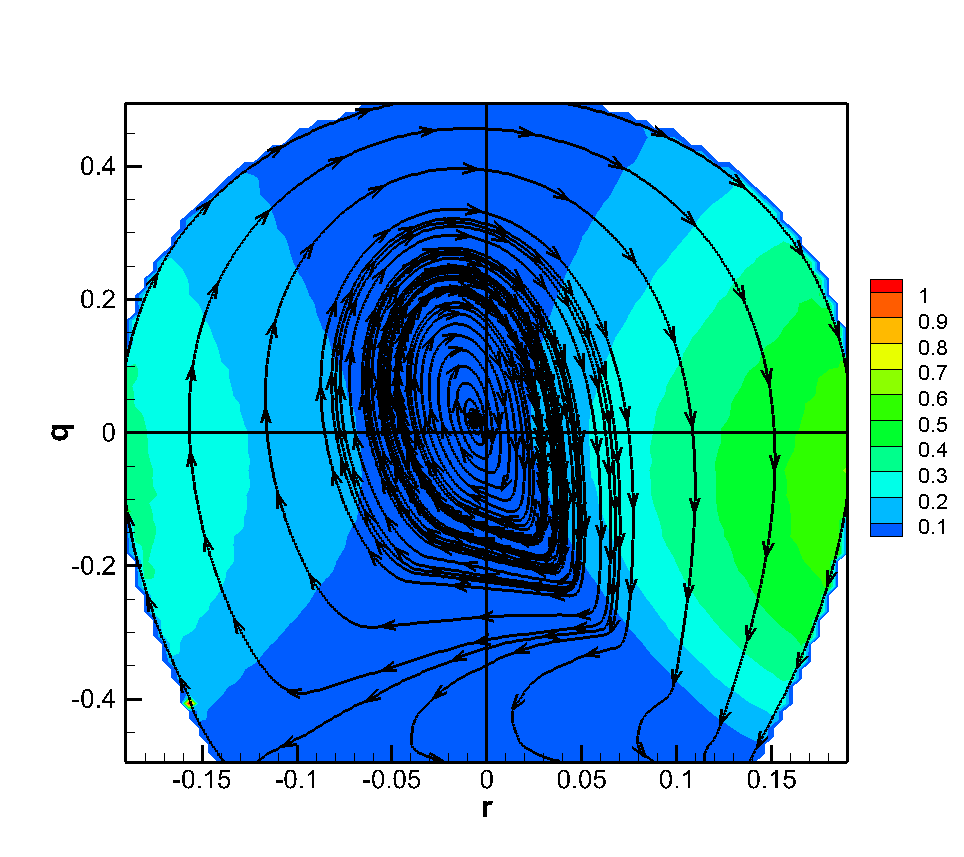}
        \caption{}
    \end{subfigure}
    \caption{\label{fig:Ch7:qrCMT} Conditional mean trajectories in the $q$-$r$ plane due to the inertial, pressure and viscous effects obtained using (a) DNS data and (b) $b_{ij}$ data-driven model. Background contours represent the speed of the trajectory at each point, given by the magnitude of the conditional mean velocity vector, $|\bm{\tilde{v}}|$.}
\end{figure}

The conditional mean trajectories (CMTs) in the phase plane of normalized velocity gradient invariants $(q,r)$ are examined as an \textit{a priori} test of the data-driven closure used to capture the conditional mean nonlocal effects of pressure and viscosity (section \ref{sec:Ch7:modelbij}) on the $b_{ij}$-dynamics.
The $q$-$r$ CMTs are obtained by integrating the vector field of conditional mean velocity ($\bm{\tilde{v}}$) in the $q$-$r$ plane:
\begin{eqnarray}
    \bm{\tilde{v}}
    & = & \begin{pmatrix}
        \tilde{v}_q\\
        \tilde{v}_r
    \end{pmatrix}
    = \Bigg \langle \;
    \begin{pmatrix}
        dq/dt'\\
        dr/dt'
    \end{pmatrix}\;	
    \Bigg | \; q,r \Bigg \rangle  \nonumber \\
     & = & \Bigg \langle \begin{pmatrix}
                        -3r + 2q b_{ij}b_{ik}b_{kj} - h_{ij} (b_{ji} + 2q b_{ij}) - \tau_{ij} (b_{ji} + 2q b_{ij}) \\
                        \frac{2}{3}q^2 + 3rb_{ij}b_{ik}b_{kj} - h_{ij}(b_{jk}b_{ki} + 3rb_{ij}) - \tau_{ij}(b_{jk}b_{ki} + 3rb_{ij})
                        \end{pmatrix} \Bigg | \; q,r \Bigg \rangle .
    \label{eq:Ch7:CMTv}
\end{eqnarray}
due to the inertial, pressure and viscous processes in the turbulent flow.
Note that the effect of the large-scale forcing is not included here because it is not accounted for in the data-driven closure but rather in the stochastic forcing (diffusion) term in the $b_{ij}$-SDE, which can not be tested \textit{a priori}.
The $q$-$r$ CMTs obtained directly from DNS data are plotted in figure \ref{fig:Ch7:qrCMT}(a). 
As discussed in \cite{das2022effect}, trajectories closer to the origin converge toward the attractor near the origin (represents pure-shear streamlines) while trajectories that are outside the separatrix loop are attracted toward the bottom line attractor (represents pure-strain streamlines).
This behavior is almost exactly replicated by the $q$-$r$ CMTs computed using the model's data-driven closure for the conditional mean pressure Hessian and viscous Laplacian tensors (section \ref{sec:Ch7:drift}) in the equation \ref{eq:Ch7:CMTv} (figure \ref{fig:Ch7:qrCMT}b). 
The close resemblance between the two is somewhat expected given the very nature of the lookup table approach for closure. 


\begin{figure}
    \centering
    \begin{subfigure}{0.46\textwidth}
        \centering
        \includegraphics[width=\textwidth]{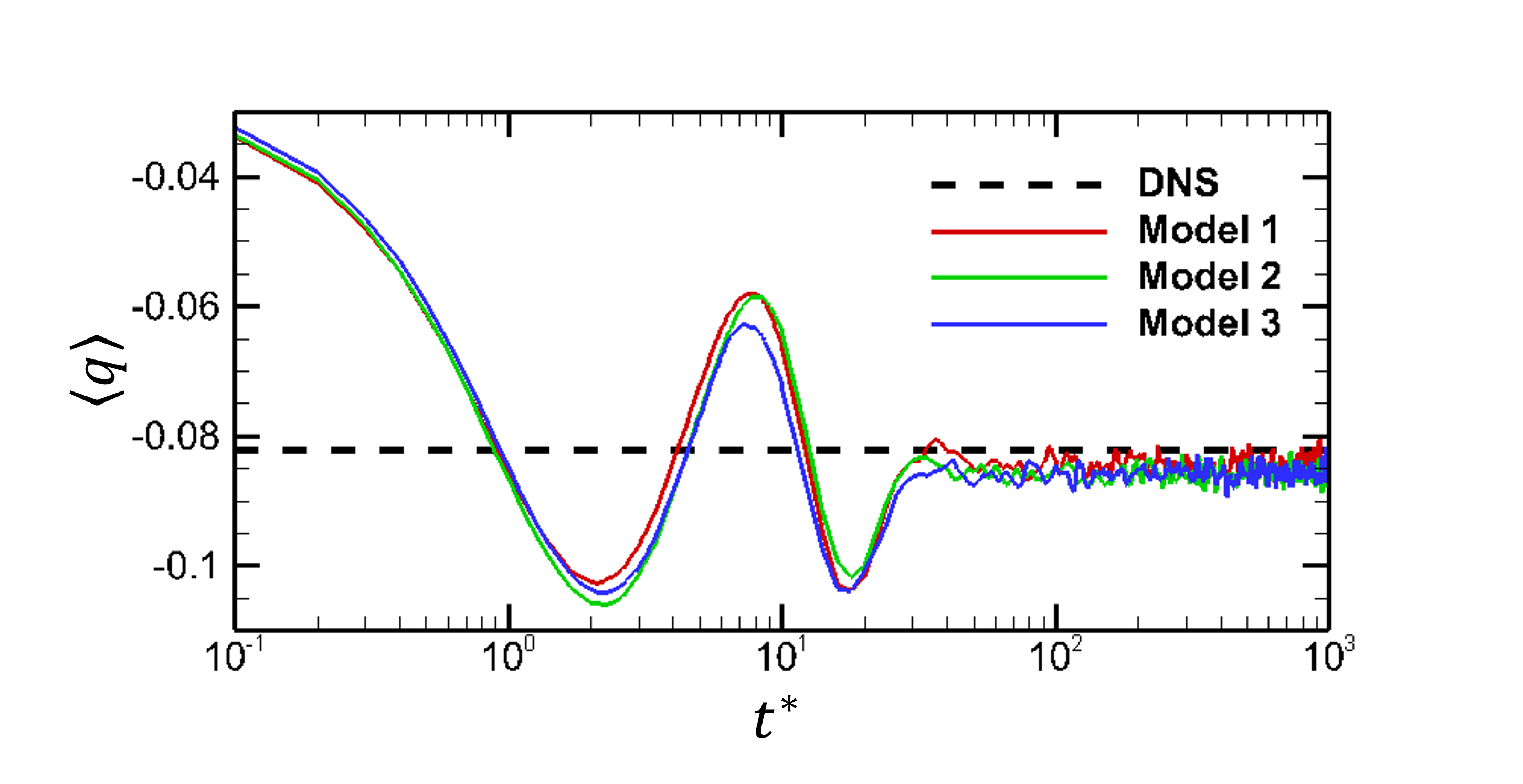}
        \caption{}
    \end{subfigure}
    \hfill 
    \begin{subfigure}{0.46\textwidth}
        \centering
        \includegraphics[width=\textwidth]{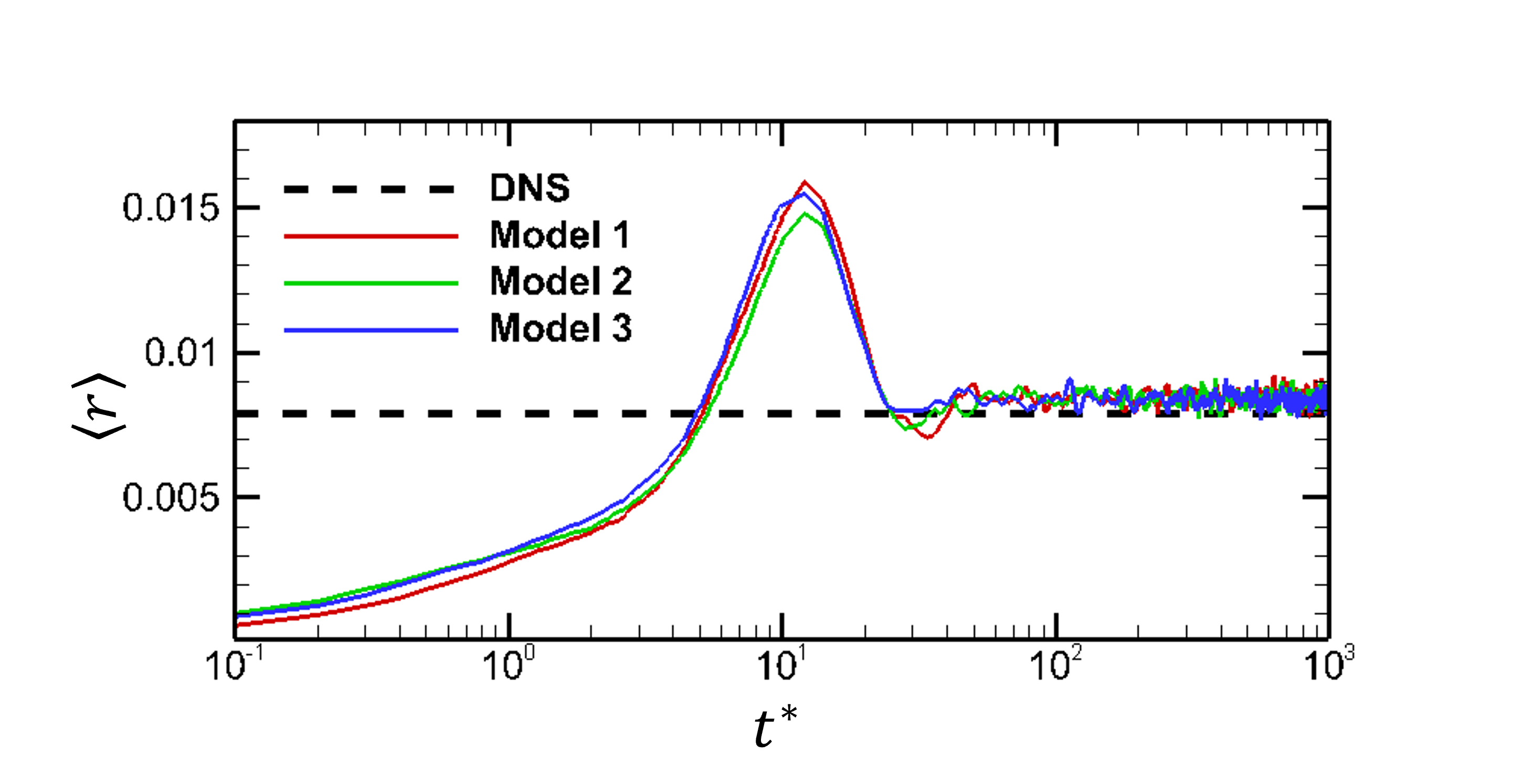}
        \caption{}
    \end{subfigure}
    \begin{subfigure}{0.46\textwidth}
        \centering
        \includegraphics[width=\textwidth]{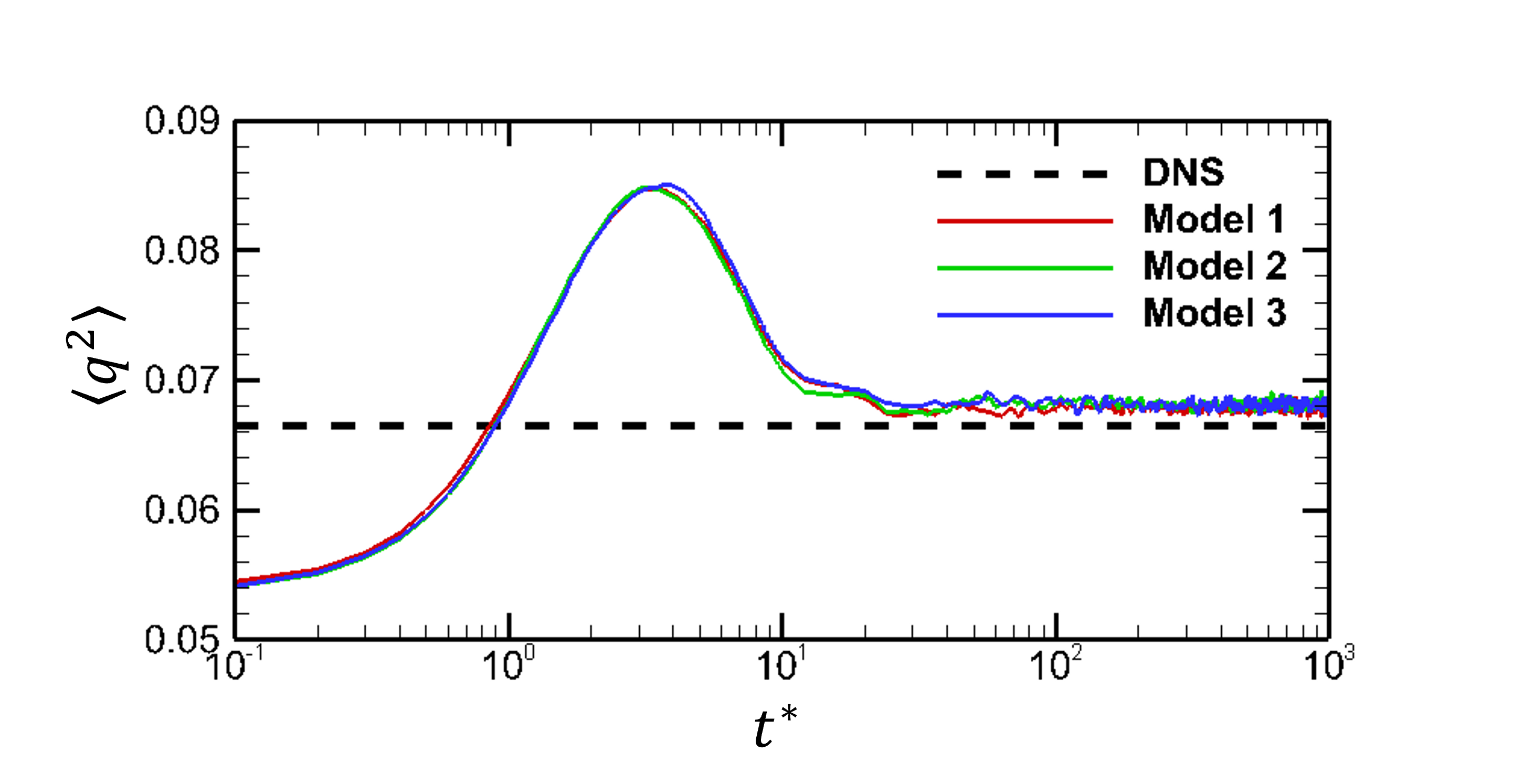}
        \caption{}
    \end{subfigure}
    \hfill 
    \begin{subfigure}{0.46\textwidth}
        \centering
        \includegraphics[width=\textwidth]{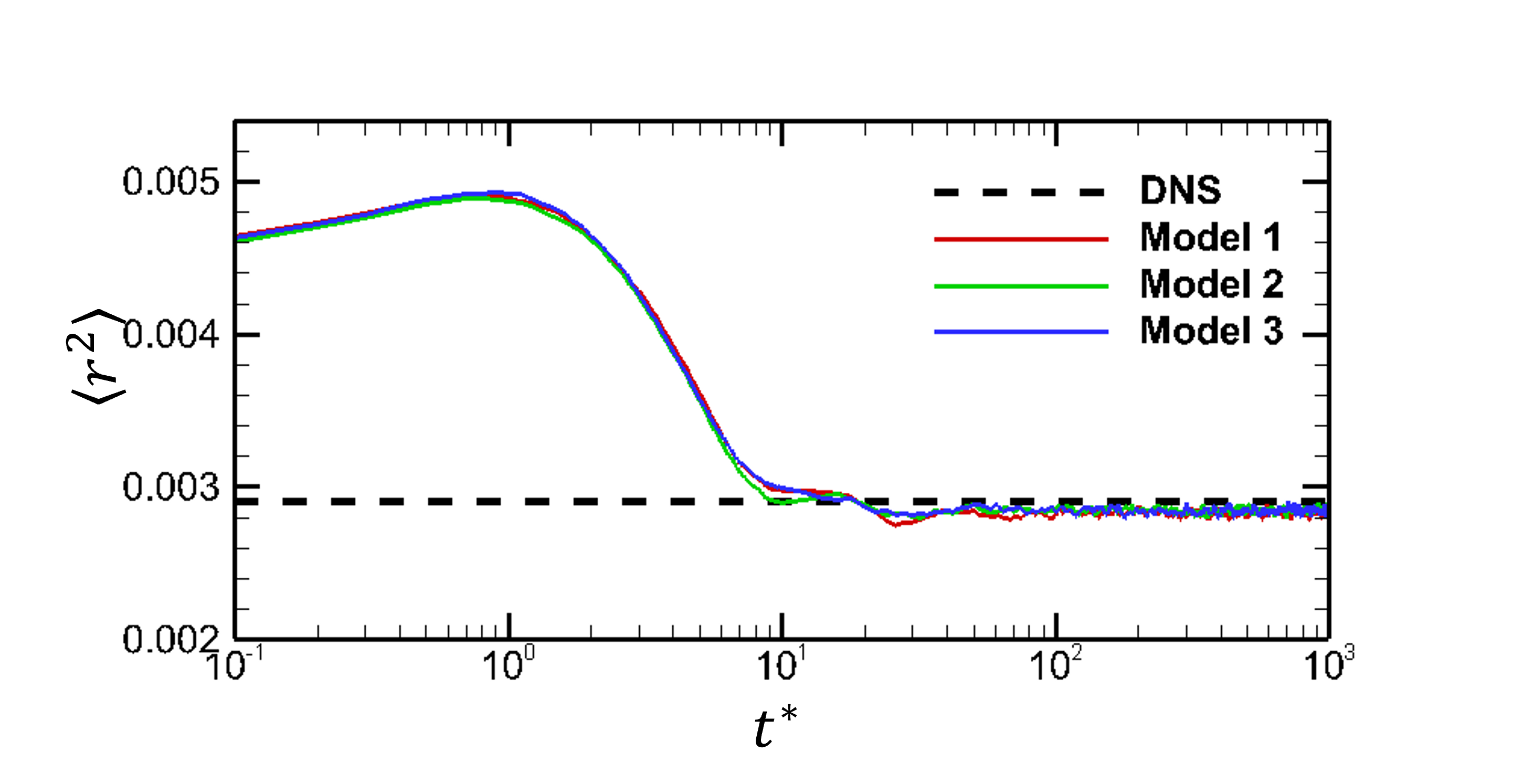}
        \caption{}
    \end{subfigure} 
    \begin{subfigure}{0.46\textwidth}
        \centering
        \includegraphics[width=\textwidth]{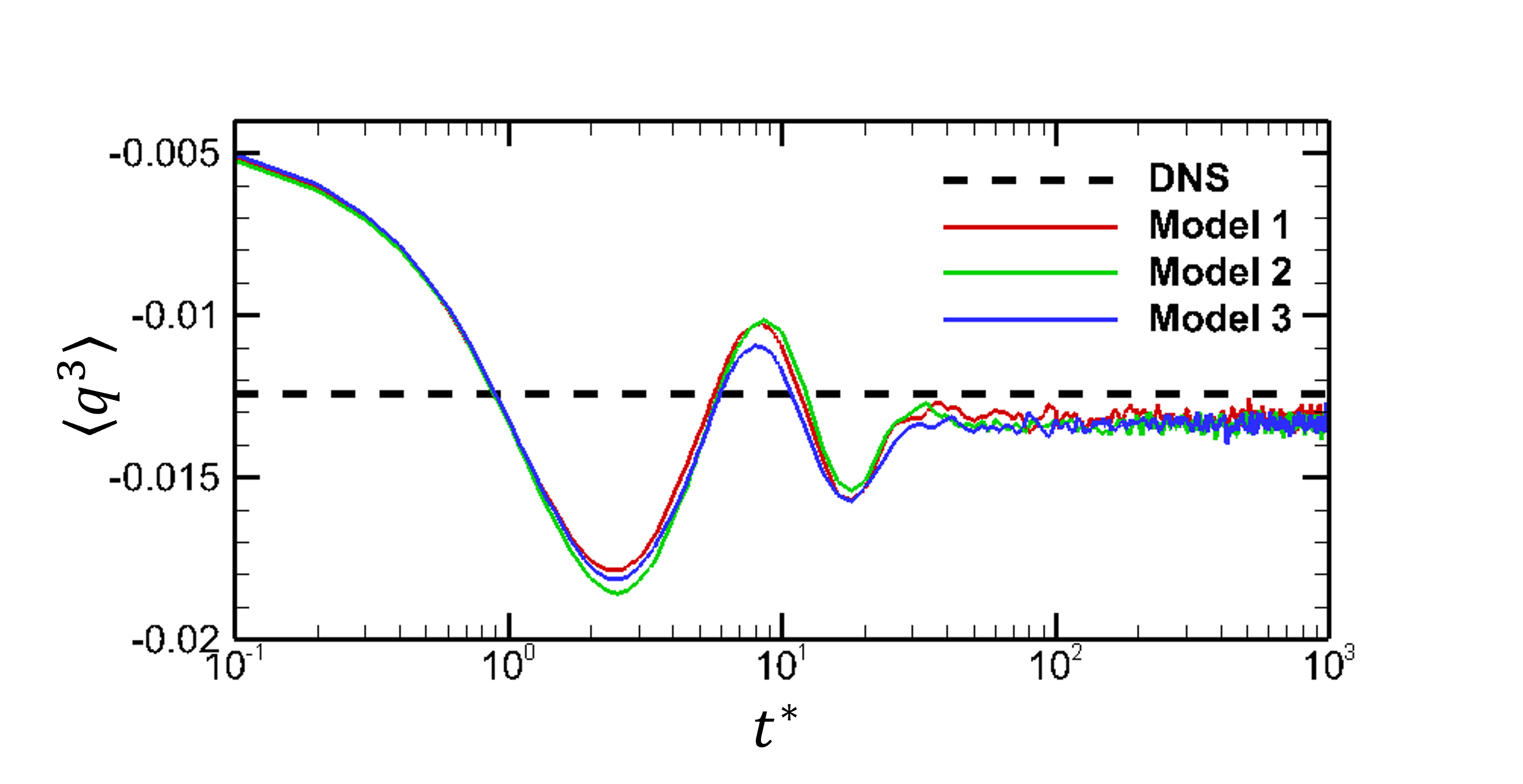}
        \caption{}
    \end{subfigure}
    \hfill 
    \begin{subfigure}{0.46\textwidth}
        \centering
        \includegraphics[width=\textwidth]{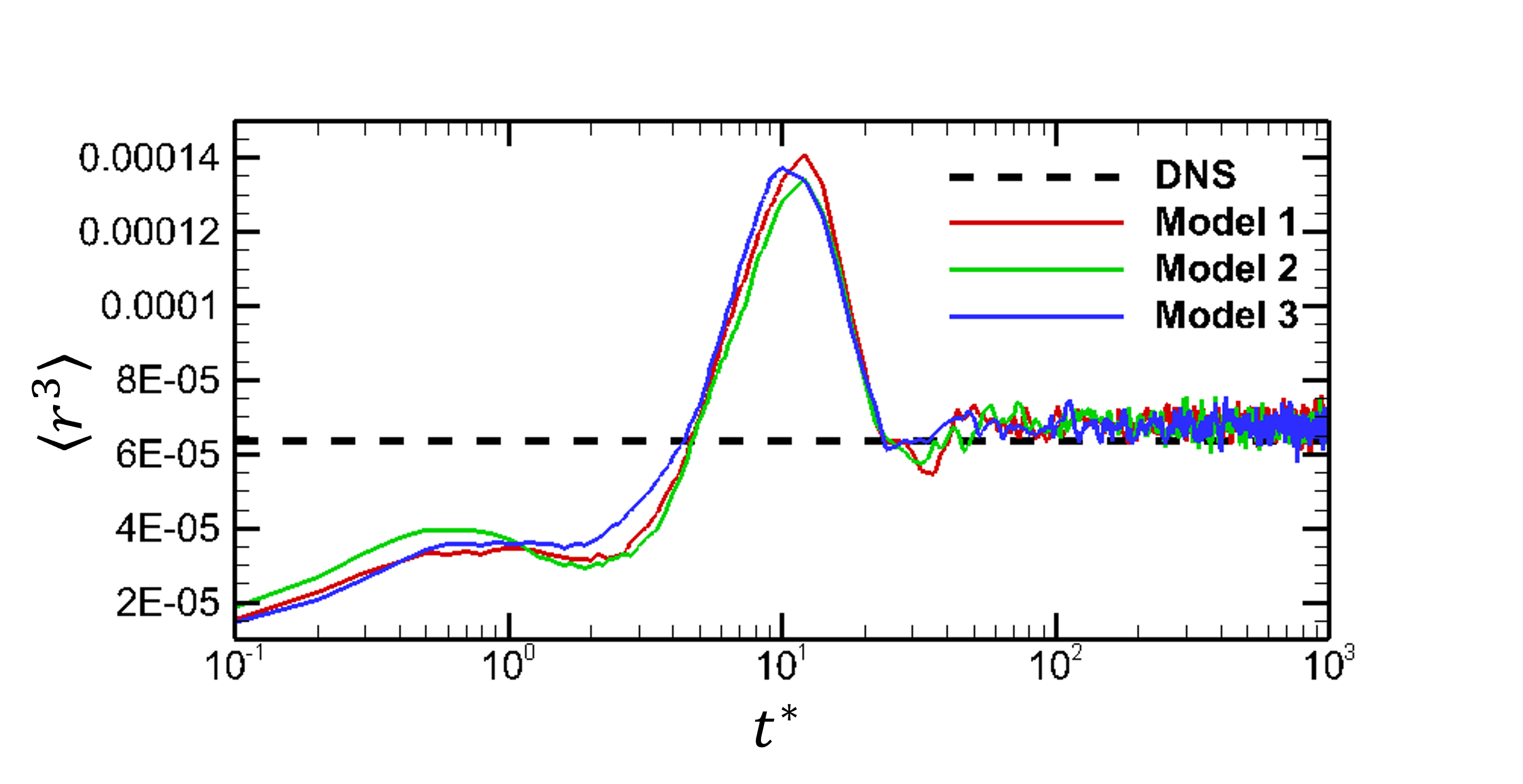}
        \caption{}
    \end{subfigure}
    \begin{subfigure}{0.46\textwidth}
        \centering
        \includegraphics[width=\textwidth]{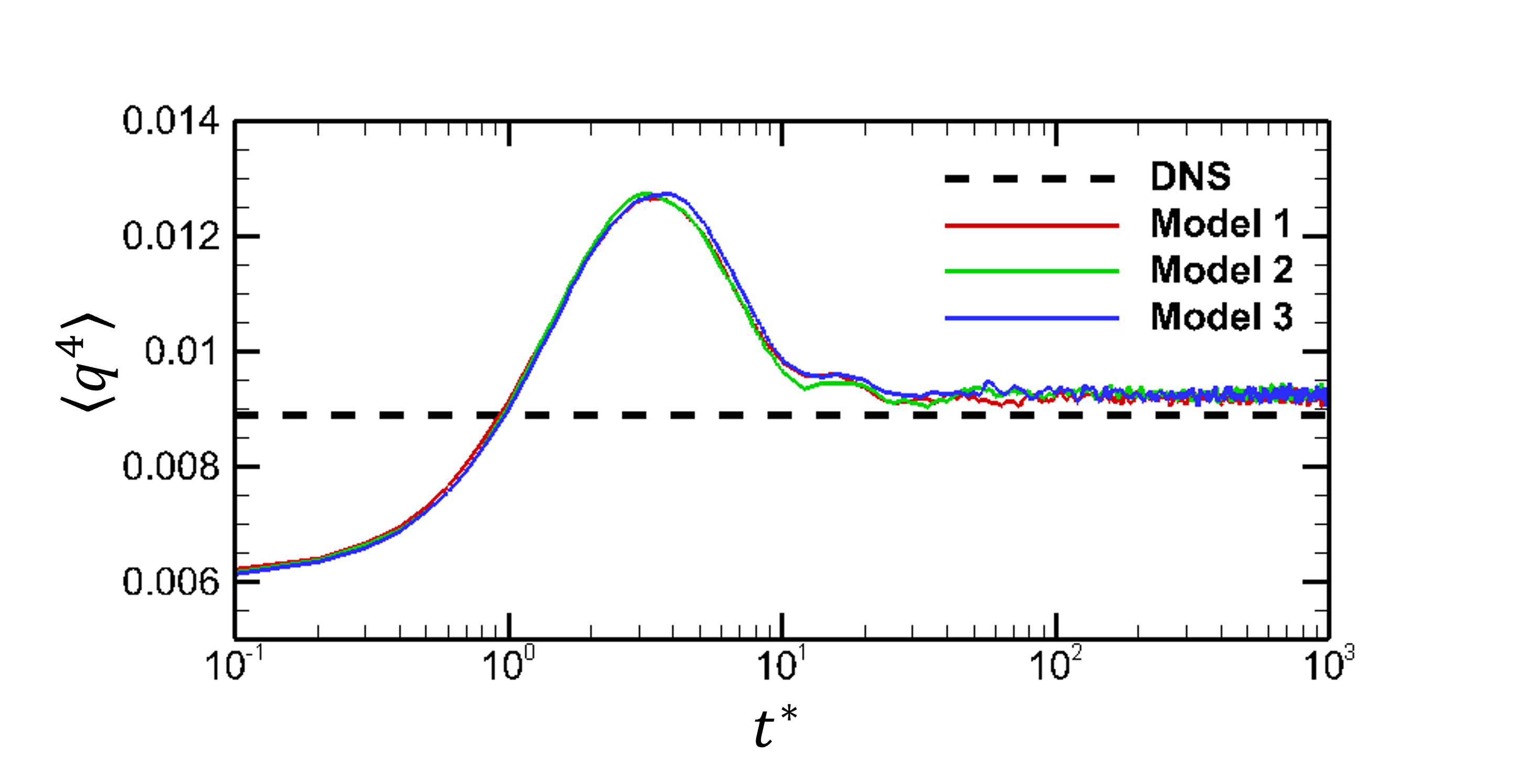}
        \caption{}
    \end{subfigure}
    \hfill 
    \begin{subfigure}{0.46\textwidth}
        \centering
        \includegraphics[width=\textwidth]{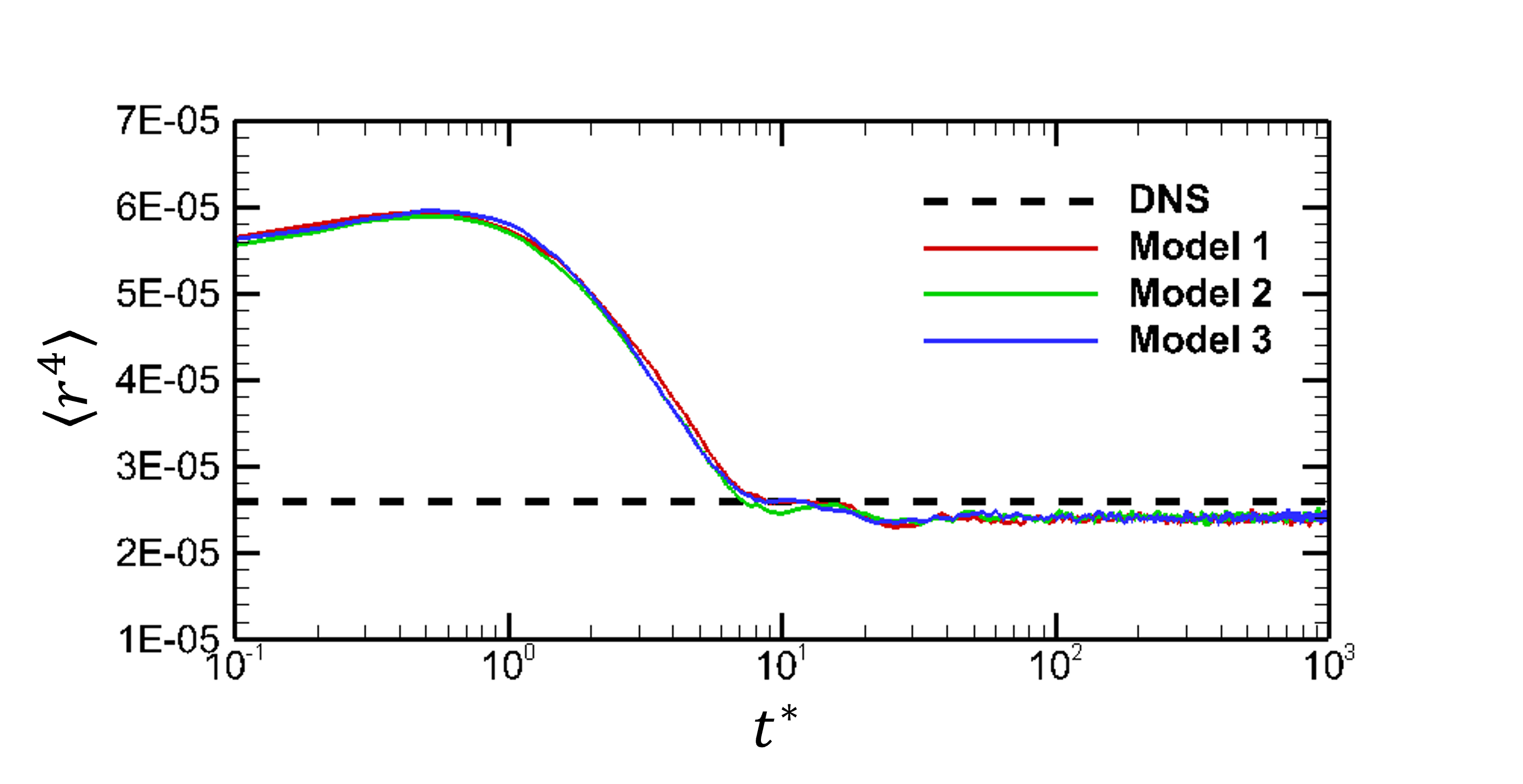}
        \caption{}
    \end{subfigure}
    \caption{\label{fig:Ch7:qrstats_evol_ts} Evolution of $q$ and $r$ statistics in global normalized time $^*$. Means: (a) $\langle q \rangle$ and (b) $\langle r \rangle$; second order moments: (c) $\langle q^2 \rangle$ and (d) $\langle r^2 \rangle$; third order moments: (e) $\langle q^3 \rangle$ and (f) $\langle r^3 \rangle$; fourth order moments: (g) $\langle q^4 \rangle$ and (h) $\langle r^4 \rangle$
    for the three models with different $\theta^*$ equations. 
    The dashed lines represent the DNS statistics. The time axis is in log-scale.}
\end{figure}

\begin{figure}
    \centering
    \begin{subfigure}{0.49\textwidth}
        \centering
        \includegraphics[width=\textwidth]{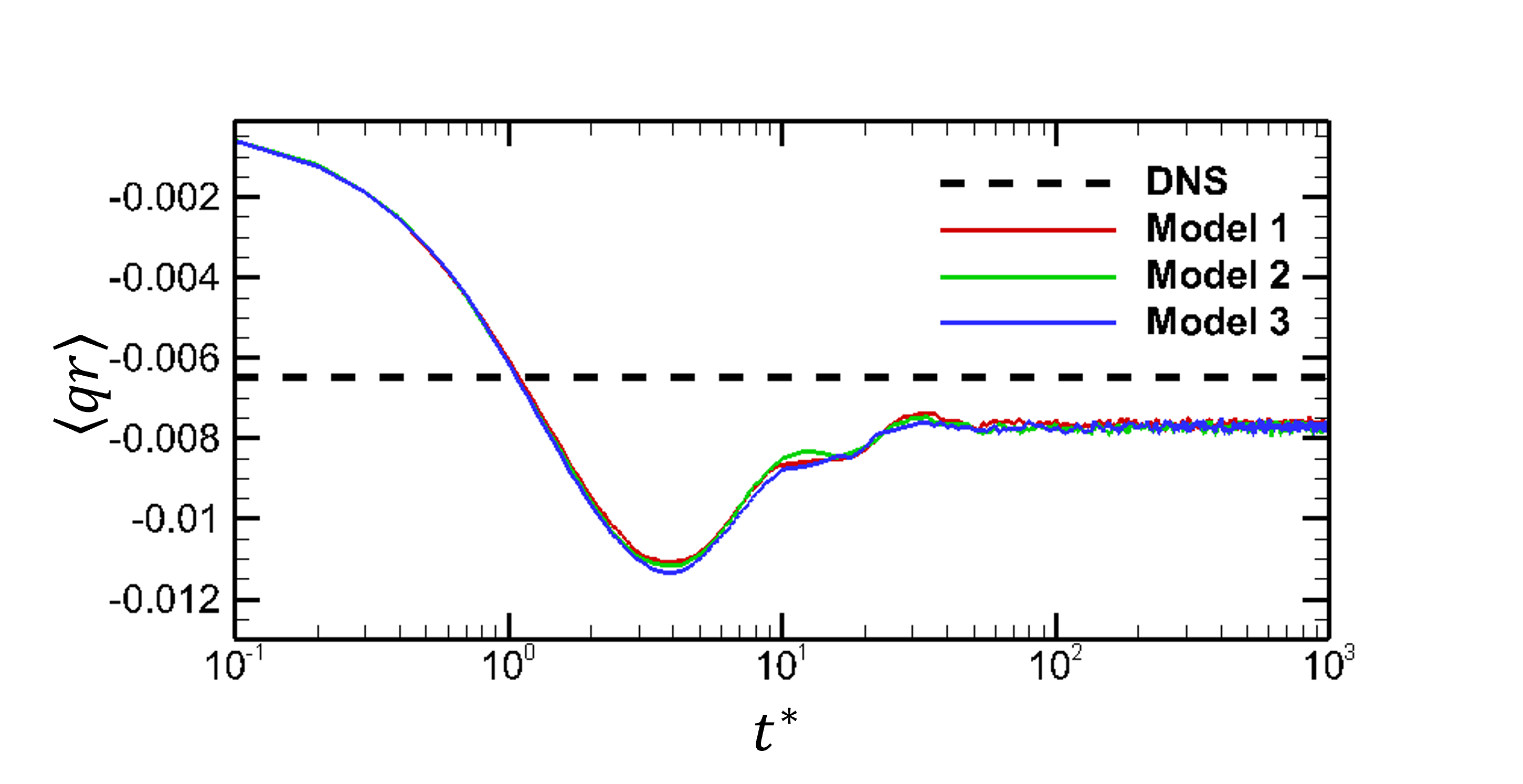}
        \caption{}
    \end{subfigure}
    \hfill 
    \begin{subfigure}{0.49\textwidth}
        \centering
        \includegraphics[width=\textwidth]{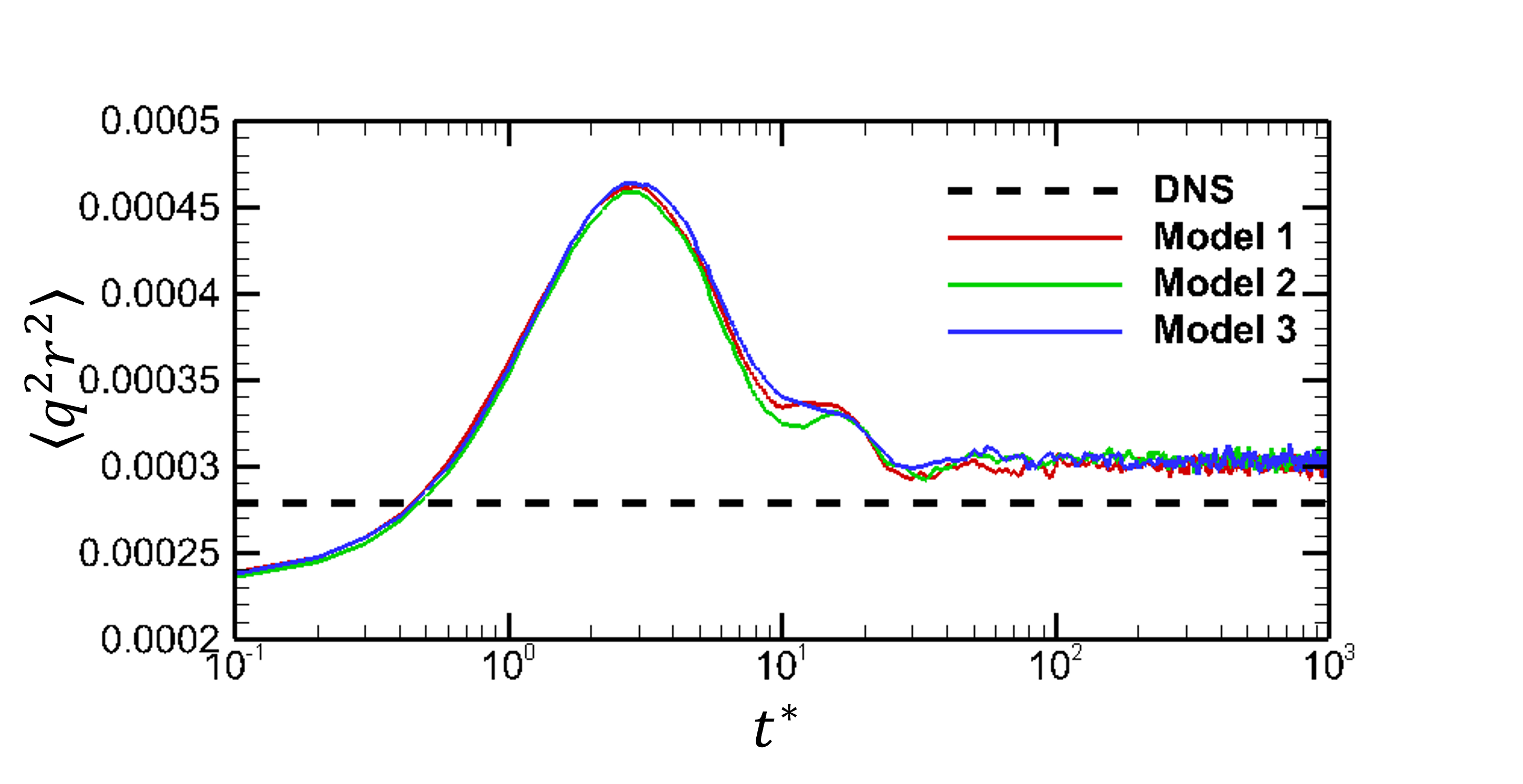}
        \caption{}
    \end{subfigure}
    \caption{\label{fig:Ch7:qrstats_evol_ts2} Evolution of the $q$ and $r$ moments: (a) $\langle qr \rangle $, and (b) $\langle q^2 r^2 \rangle $, in global normalized time $t^*$. The dashed lines represent the DNS statistics. The time axis is in log-scale.}
\end{figure}

Now, we compare the \textit{a posteriori} results of the normalized VG tensor of the model with that of DNS.
First, we study the moments of the second ($q$) and third ($r$) invariants of the tensor, which are important quantities as they determine the geometric shape of the local flow streamlines.
The evolution of up to fourth-order moments of $q$ and $r$ are plotted for each model in figure \ref{fig:Ch7:qrstats_evol_ts}.
It is first evident that all three models with the same $b_{ij}$-SDE but different $\theta^*$-SDEs produce nearly identical $q,r$ moment values. Thus, it appears that the variation in the $\theta^*$ model does not have a discernible impact on the $b_{ij}$ statistics of the models.
Starting from a randomly generated set of initial conditions, the $b_{ij}$-SDE drives the solution toward convergence to a statistically stationary state at $t \approx 72 \tau_\eta$ ($t^* \approx 60$). Therefore, the $b_{ij}$ model takes approximately $10$ times as long as the $\theta^*$ model to reach stationarity.
Up to at least fourth-order converged moments of both $q$ and $r$ are reasonably close to the DNS values.
Further, the time evolution of moments of correlation between $q$ and $r$, i.e. $\langle qr \rangle $, and $\langle q^2 r^2 \rangle $, are plotted in figure \ref{fig:Ch7:qrstats_evol_ts2}.
These moments show a slightly larger deviation from the DNS values as compared to all the other moments.


\begin{figure}
    \centering
    \begin{subfigure}{0.32\textwidth}
        \centering
        \includegraphics[width=\textwidth]{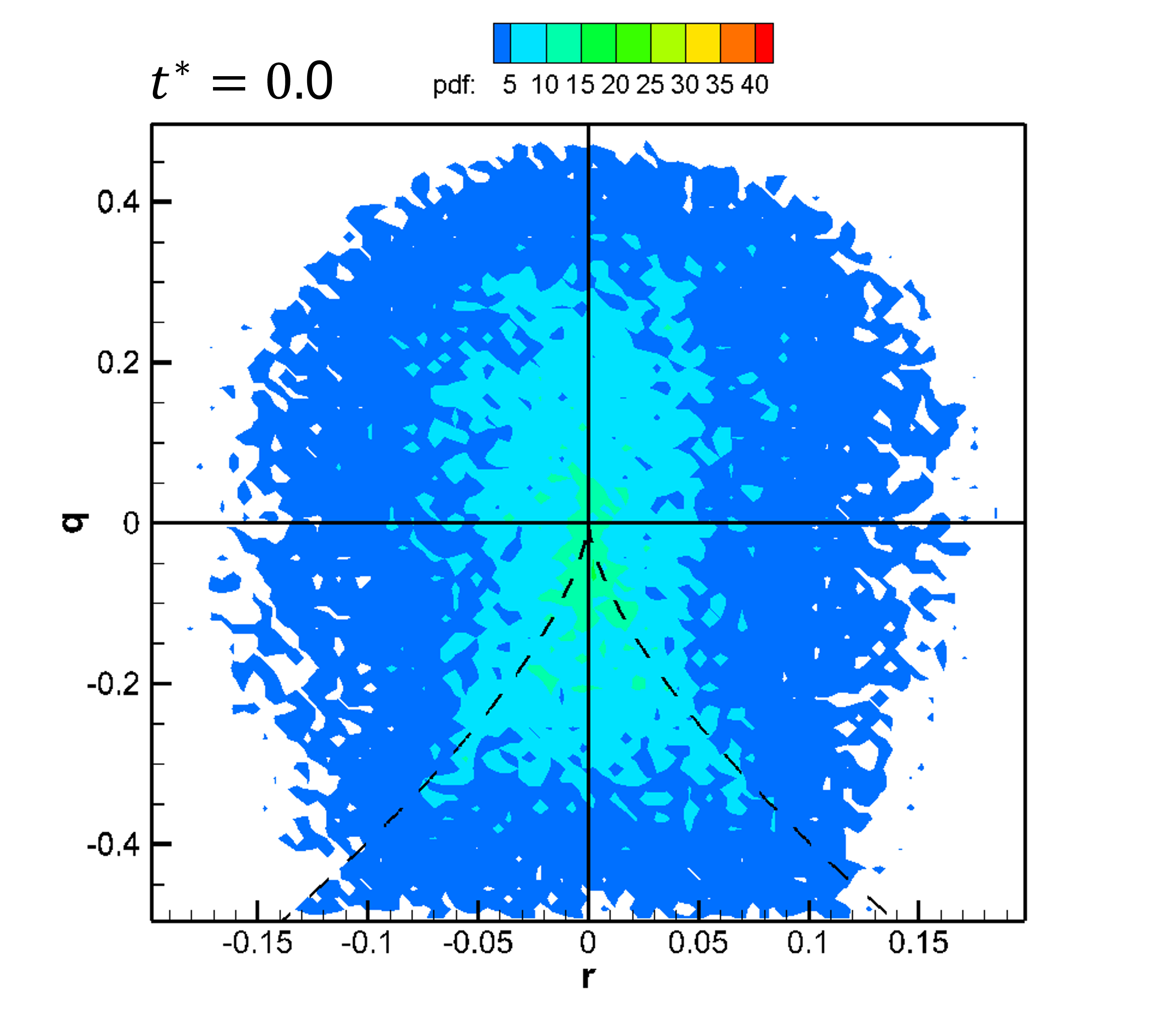}
        \caption{}
    \end{subfigure}
    \begin{subfigure}{0.32\textwidth}
        \centering
        \includegraphics[width=\textwidth]{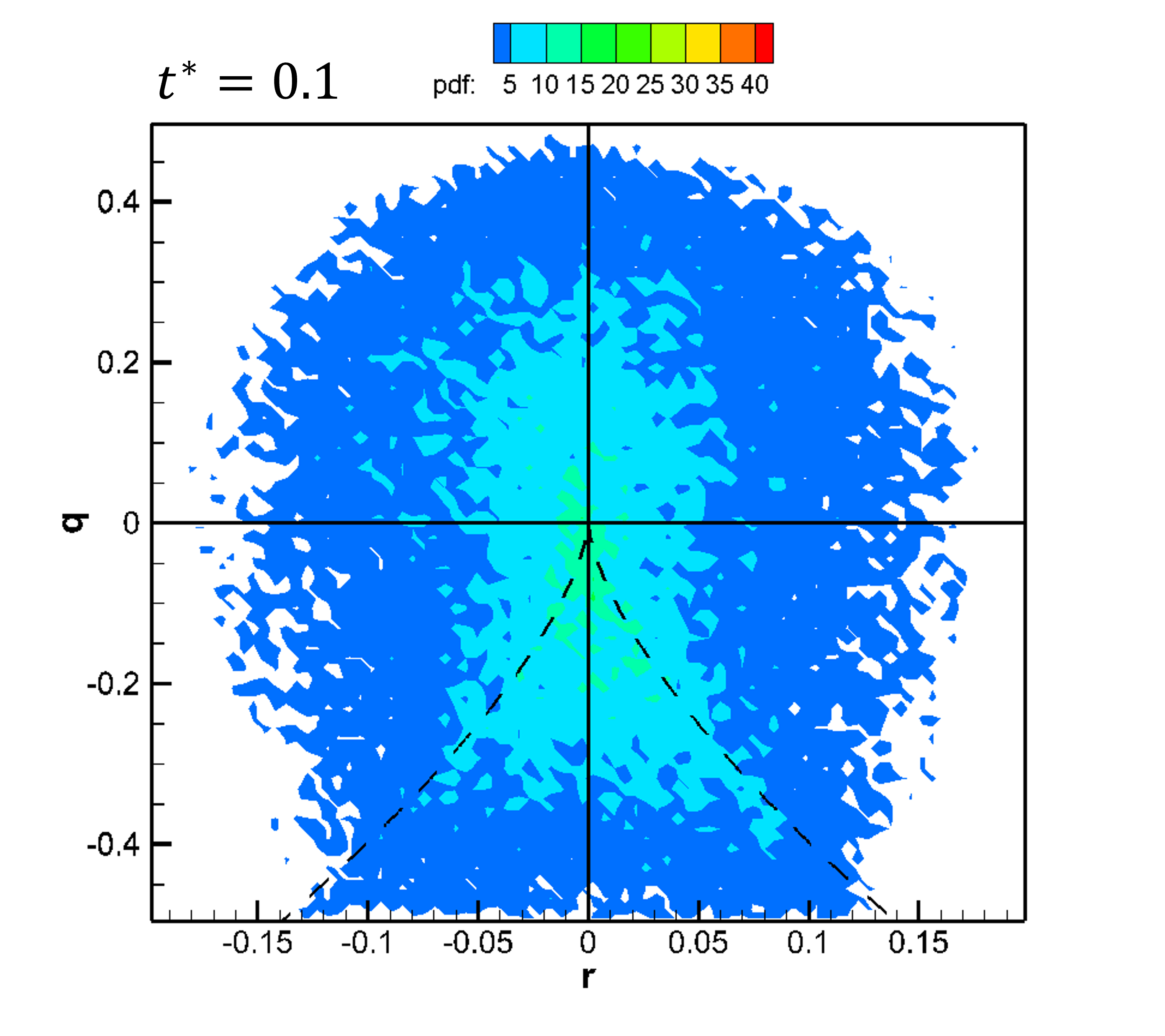}
        \caption{}
    \end{subfigure}
    \begin{subfigure}{0.32\textwidth}
        \centering
        \includegraphics[width=\textwidth]{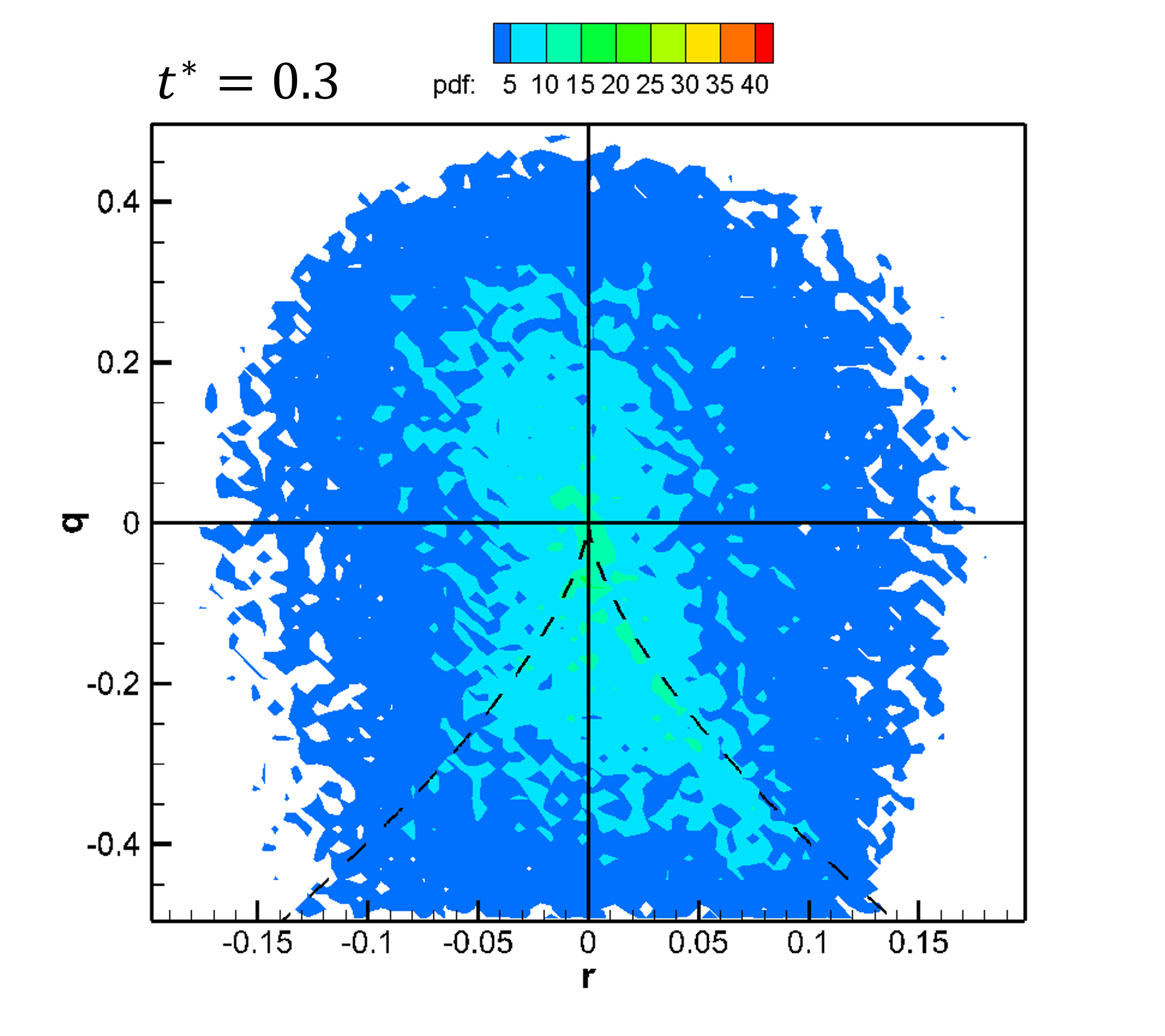}
        \caption{}
    \end{subfigure}
    \begin{subfigure}{0.32\textwidth}
        \centering
        \includegraphics[width=\textwidth]{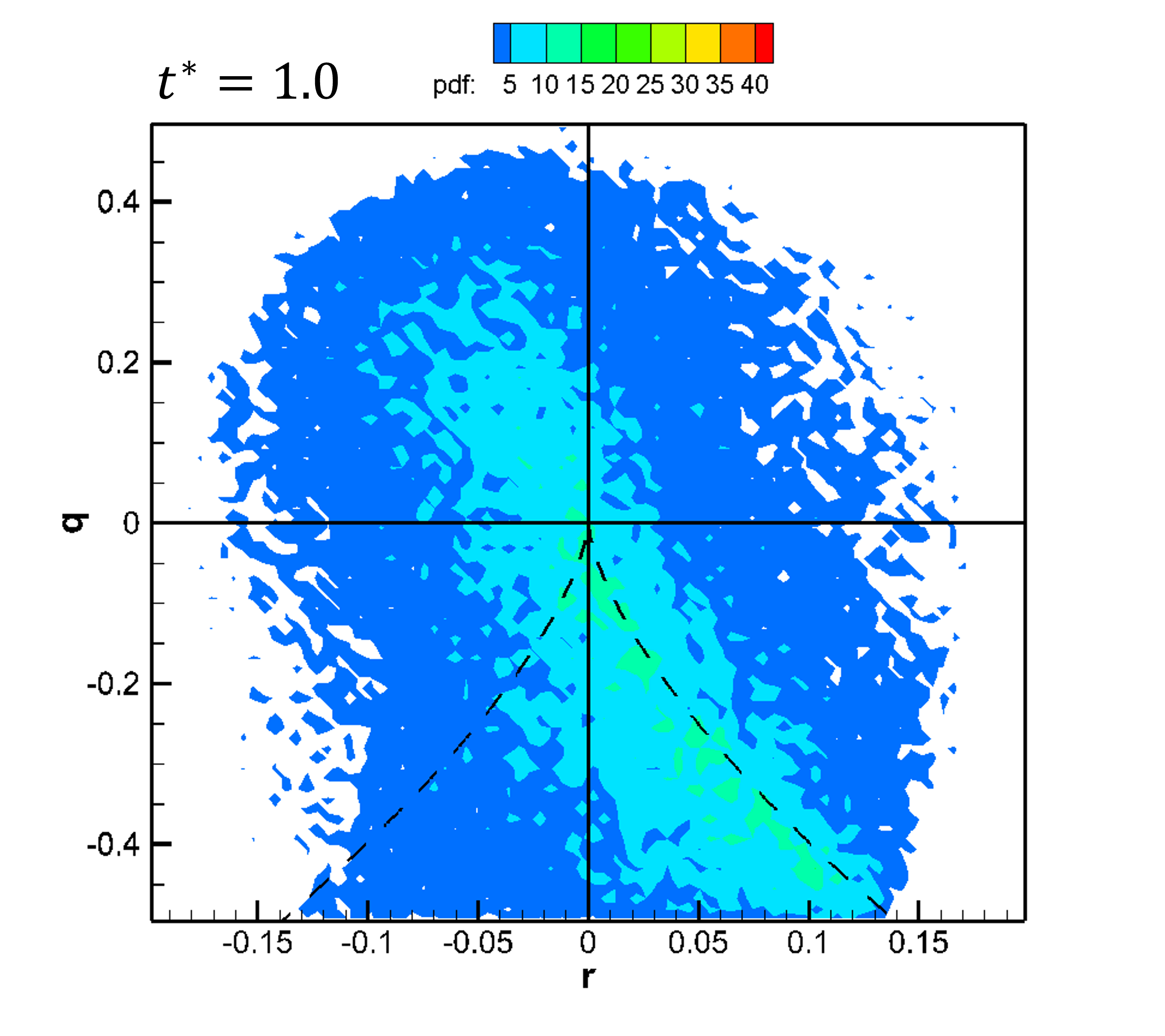}
        \caption{}
    \end{subfigure} 
    \begin{subfigure}{0.32\textwidth}
        \centering
        \includegraphics[width=\textwidth]{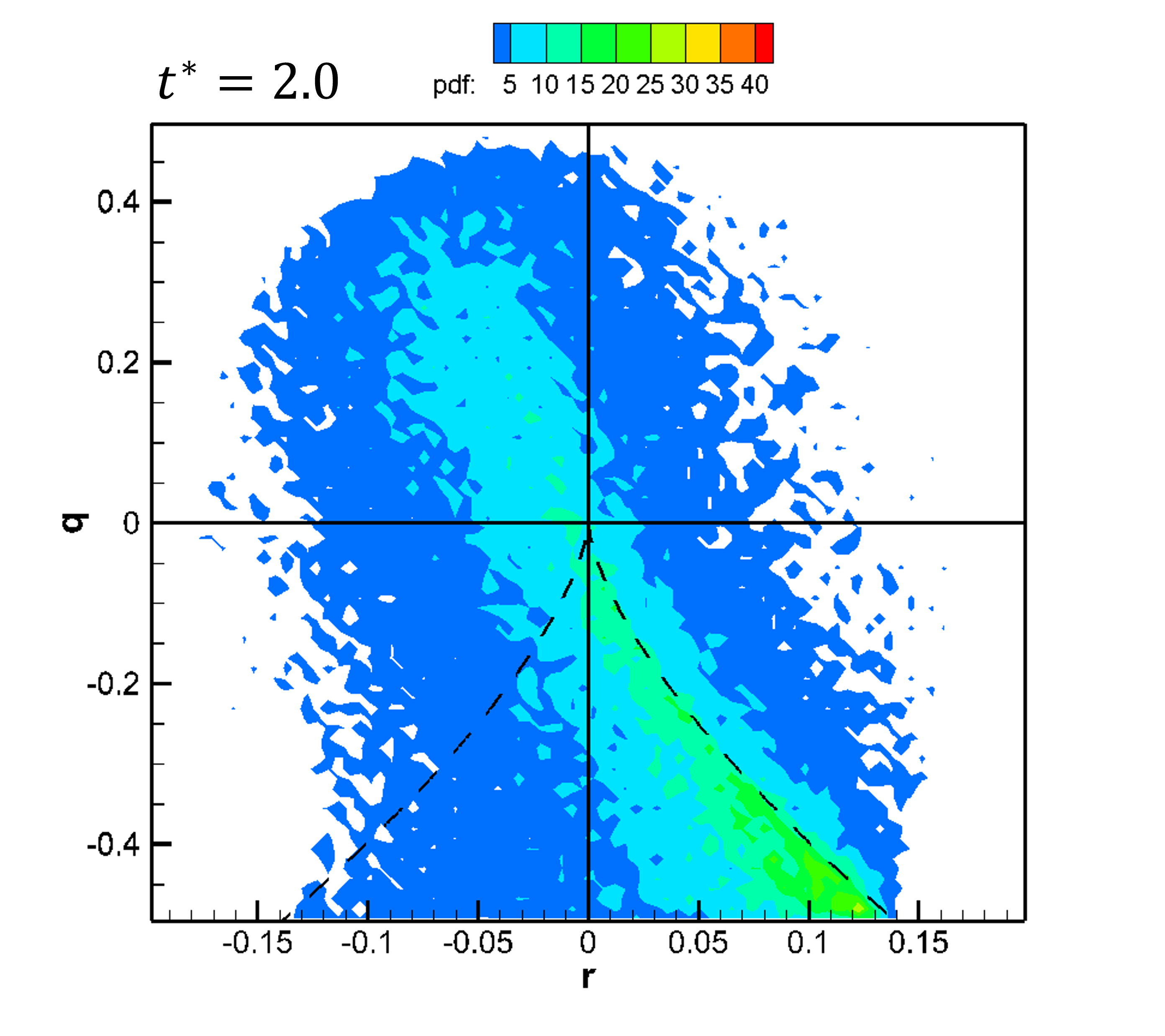}
        \caption{}
    \end{subfigure}
    \begin{subfigure}{0.32\textwidth}
        \centering
        \includegraphics[width=\textwidth]{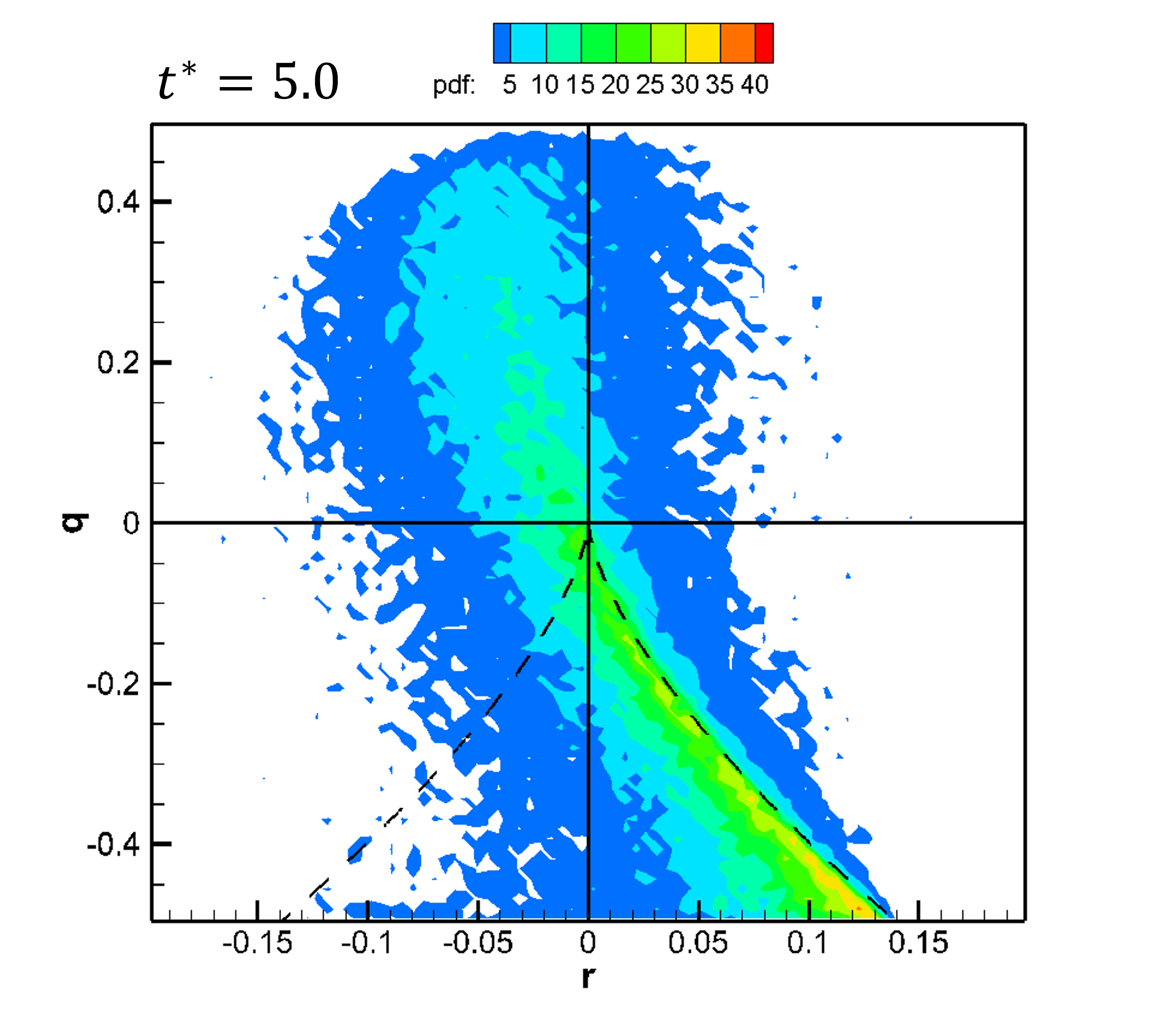}
        \caption{}
    \end{subfigure}
    \begin{subfigure}{0.32\textwidth}
        \centering
        \includegraphics[width=\textwidth]{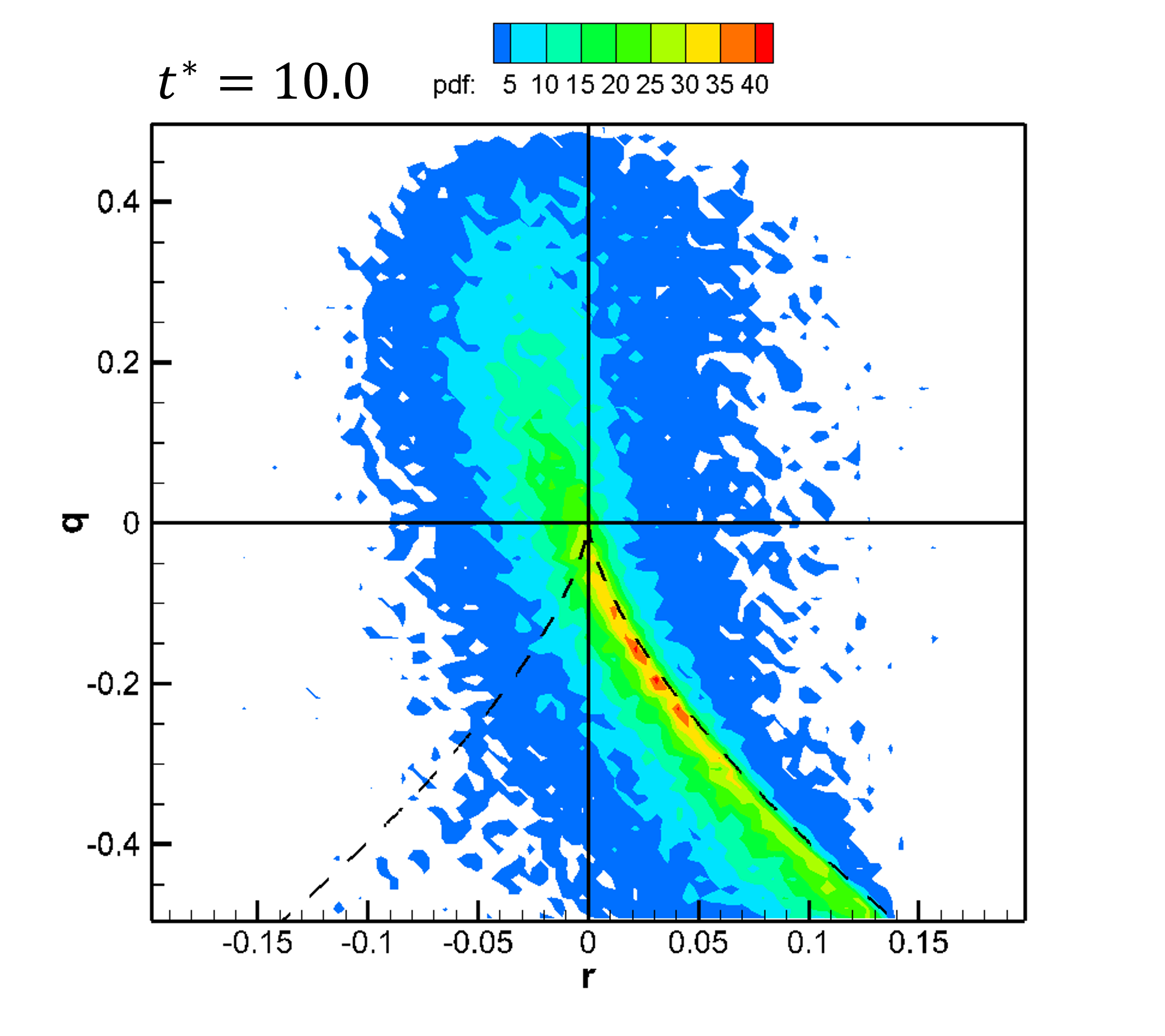}
        \caption{}
    \end{subfigure}
    \begin{subfigure}{0.32\textwidth}
        \centering
        \includegraphics[width=\textwidth]{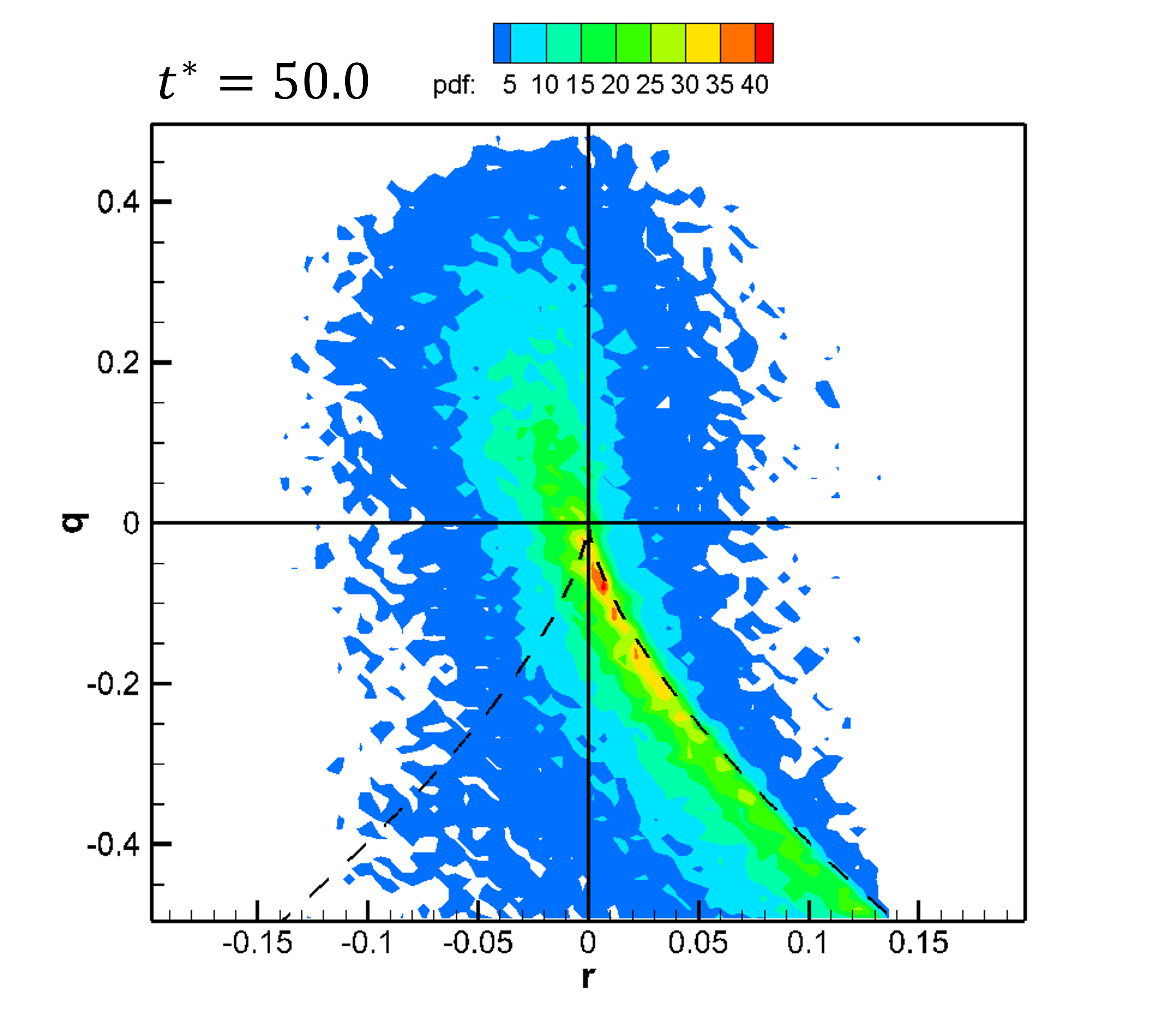}
        \caption{}
    \end{subfigure}
        \begin{subfigure}{0.32\textwidth}
        \centering
        \includegraphics[width=\textwidth]{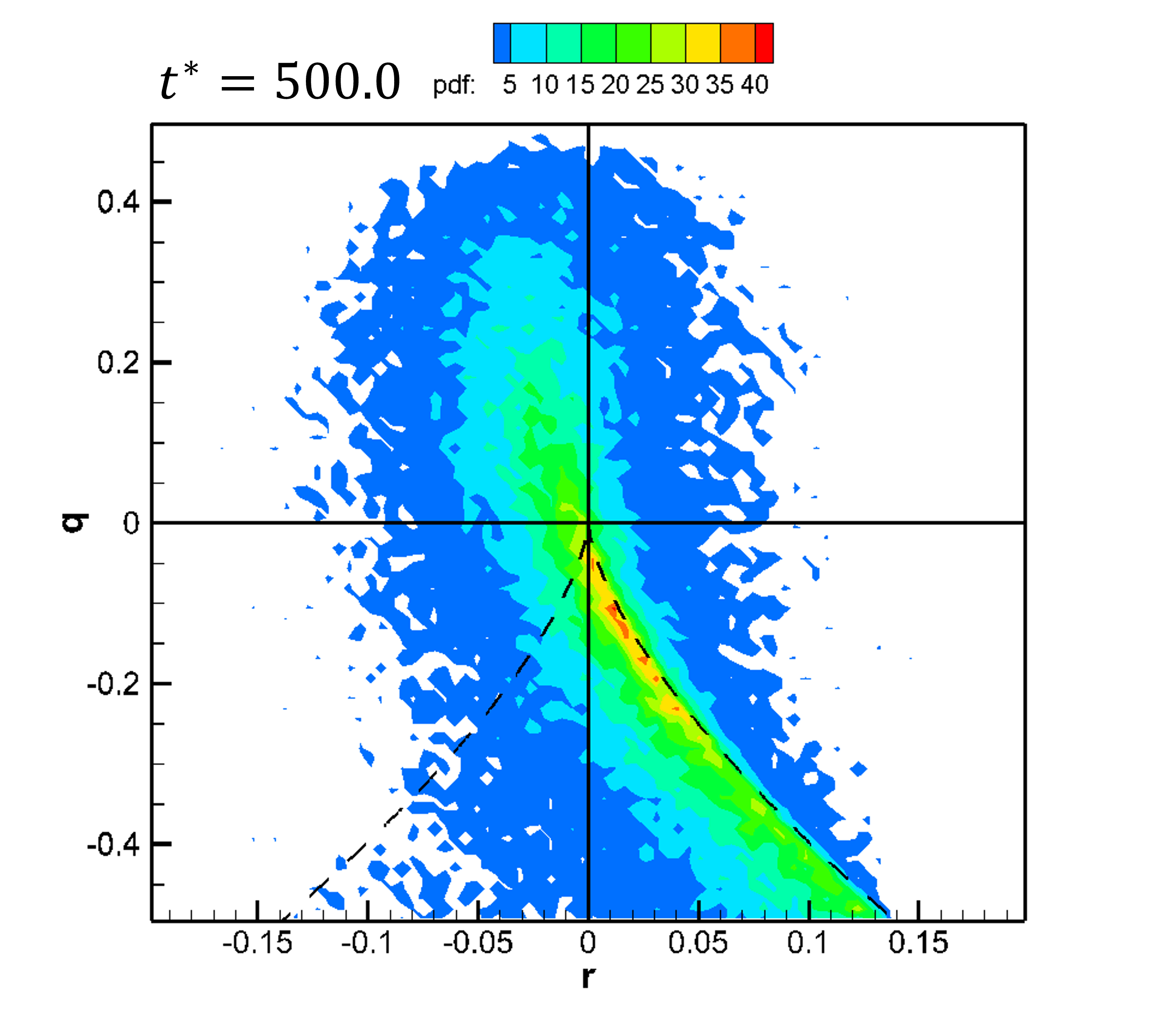}
        \caption{}
    \end{subfigure}
    \caption{\label{fig:Ch7:qrPDF_evol_ts} The evolution of the $q$-$r$ joint PDF during numerical propagation of model 3 at different global normalized time: (a) $t^* = 0.0$, (b) $t^* = 0.1$, (c) $t^* = 0.3$, (d) $t^* = 1.0$, (e) $t^* = 2.0$, (f) $t^* = 5.0$, (g) $t^* = 10.0$, (h) $t^* = 50.0$, (i) $t^* = 500.0$. The dashed lines represent the lines of zero-discriminant $(d = q^3 + (27/4)r^2) = 0$.}
\end{figure}

The evolution of $q$-$r$ joint PDF is now investigated for the propagation of model 3. The solutions of the other two models show similar trends and are, therefore, not presented separately.
The $q$-$r$ joint PDF is plotted with ensembles of only $40000$ particles at different times ($t^*$) in figure \ref{fig:Ch7:qrPDF_evol_ts}.  
The joint PDF of the initial field ($t^*=0$) is symmetric in $r$, as expected from a joint Gaussian distribution. 
As time progresses, the modeled dynamics cause the PDF to skew towards the right zero-discriminant line. The PDF contours shrink in size and steepen in magnitude as more and more particles accumulate along the right zero-discriminant line. 
This finally results in the characteristic teardrop-like shape, which becomes nearly invariant beyond $t^* \approx 60$. 
In this manner, our model reproduces the teardrop-shaped $q$-$r$ joint PDF, one of the key signatures of small-scale turbulence. 


\begin{figure}
    \centering
    \begin{subfigure}{0.45\textwidth}
        \centering
        \includegraphics[width=\textwidth]{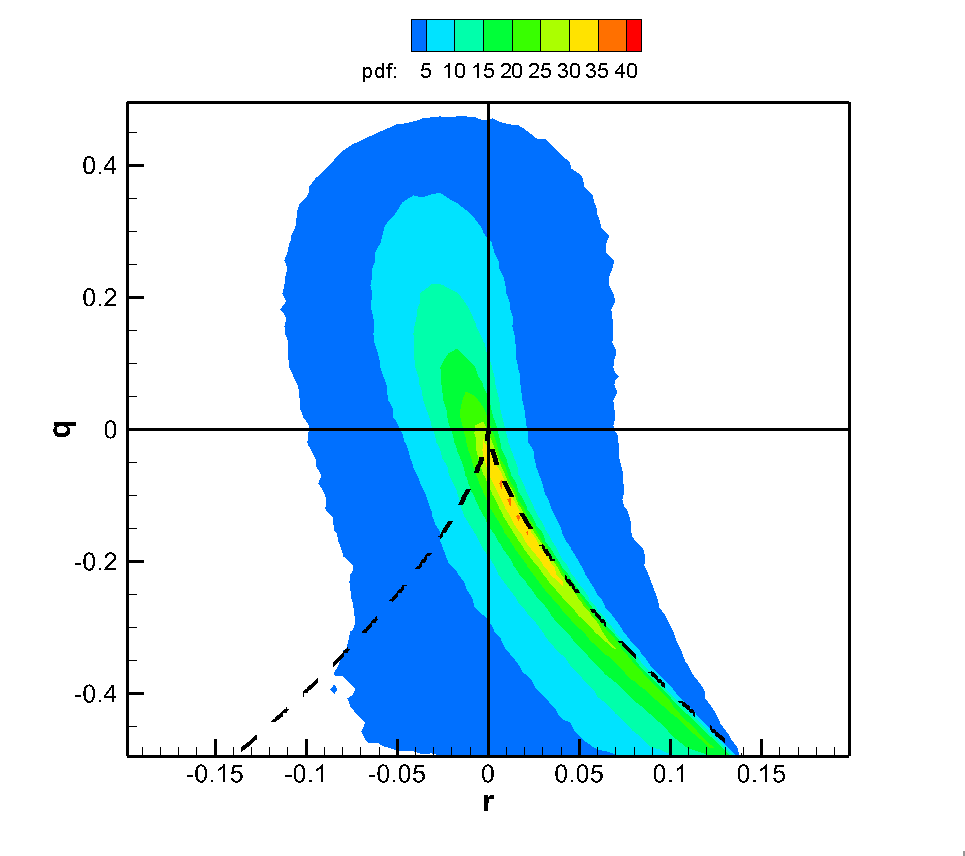}
        \caption{}
    \end{subfigure}
    \hfill 
    \begin{subfigure}{0.45\textwidth}
        \centering
        \includegraphics[width=\textwidth]{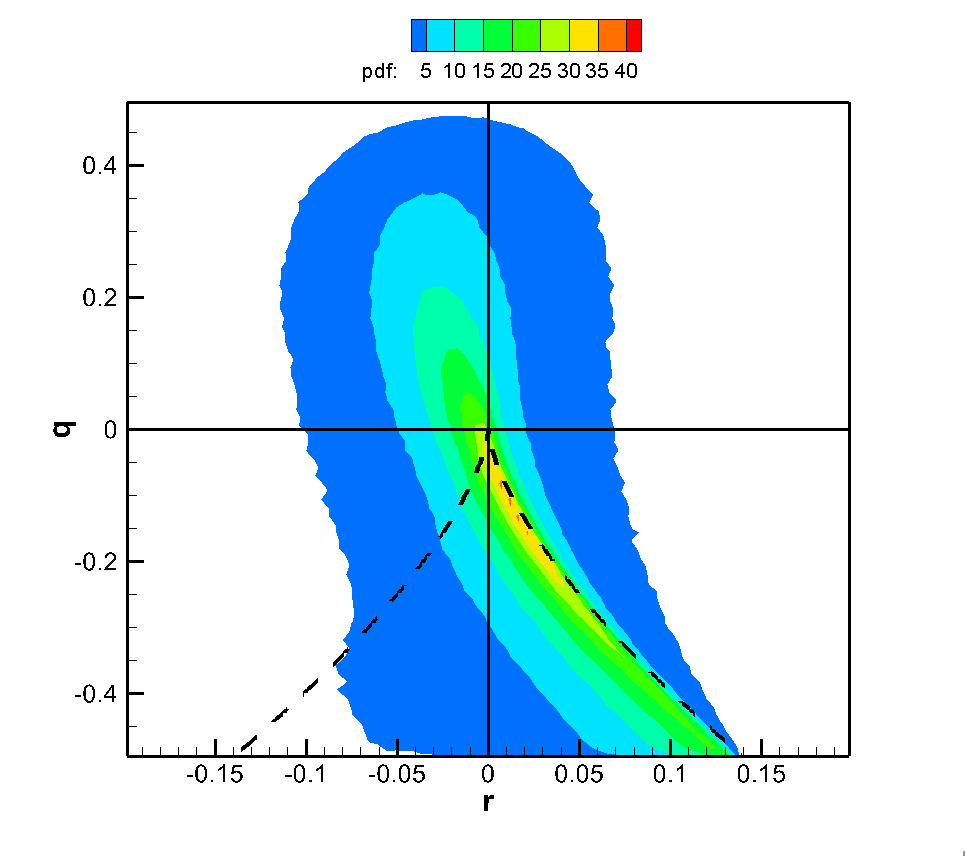}
        \caption{}
    \end{subfigure}
    \begin{subfigure}{0.45\textwidth}
        \centering
        \includegraphics[width=\textwidth]{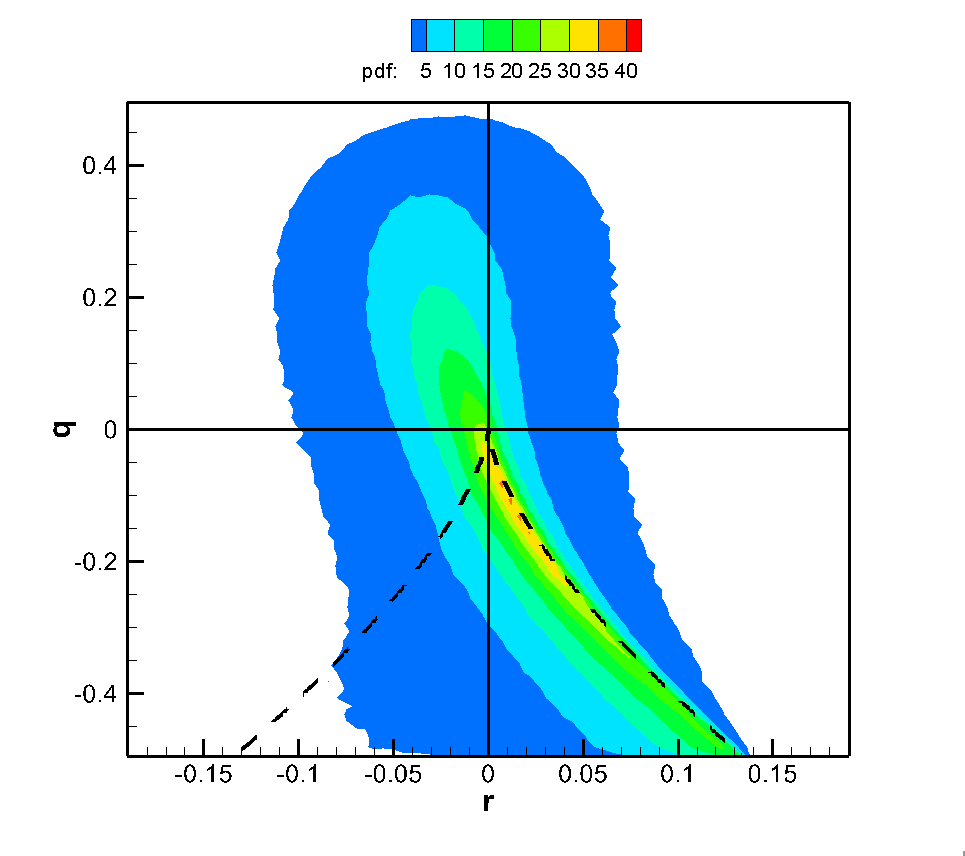}
        \caption{}
    \end{subfigure}
    \hfill 
    \begin{subfigure}{0.45\textwidth}
        \centering
        \includegraphics[width=\textwidth]{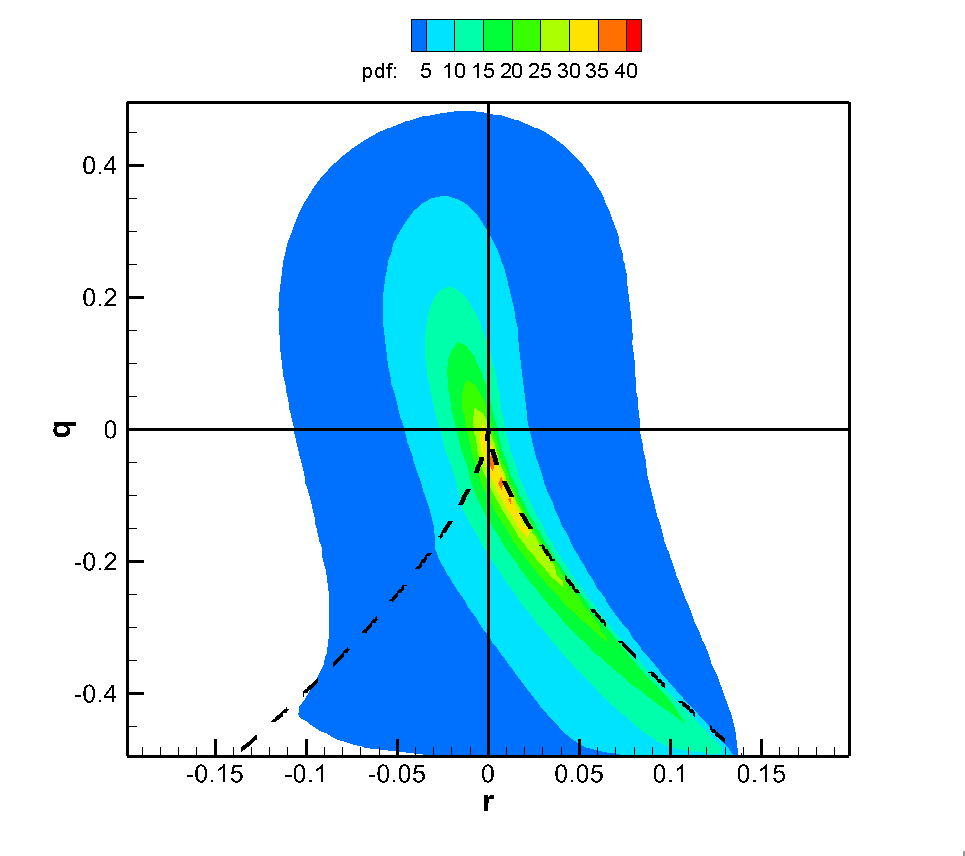}
        \caption{}
    \end{subfigure}    
    \caption{\label{fig:Ch7:qrPDF} Joint PDFs of $q$-$r$ obtained from the: solutions of (a) model 1, (b) model 2, and (c) model 3, and (d) DNS data. The dashed lines represent the zero-discriminant lines.}
\end{figure}

The converged $q$-$r$ joint PDF, averaged over multiple time realizations in the stationary state of the models' solutions, are plotted in figure \ref{fig:Ch7:qrPDF}(a-c) for the three models.
It is clear that all three models produce nearly identical $q$-$r$ joint PDFs showing excellent resemblance to that obtained from DNS data (figure \ref{fig:Ch7:qrPDF}d).
A closer comparison with the DNS joint PDF shows that the model's joint PDF has a slightly thinner tail in the strain-dominated bottom half of the teardrop and is slightly wider in the rotation-dominated top half.
Overall, the model is able to reproduce the joint PDF of $q$-$r$, one of the key features of velocity gradient geometry in turbulence with reasonable accuracy and without any distortion such as those commonly observed in previously proposed velocity gradient models \citep{johnson2016closure,pereira2016diss}.


\begin{figure}
    \centering
    \includegraphics[width=0.6\textwidth]{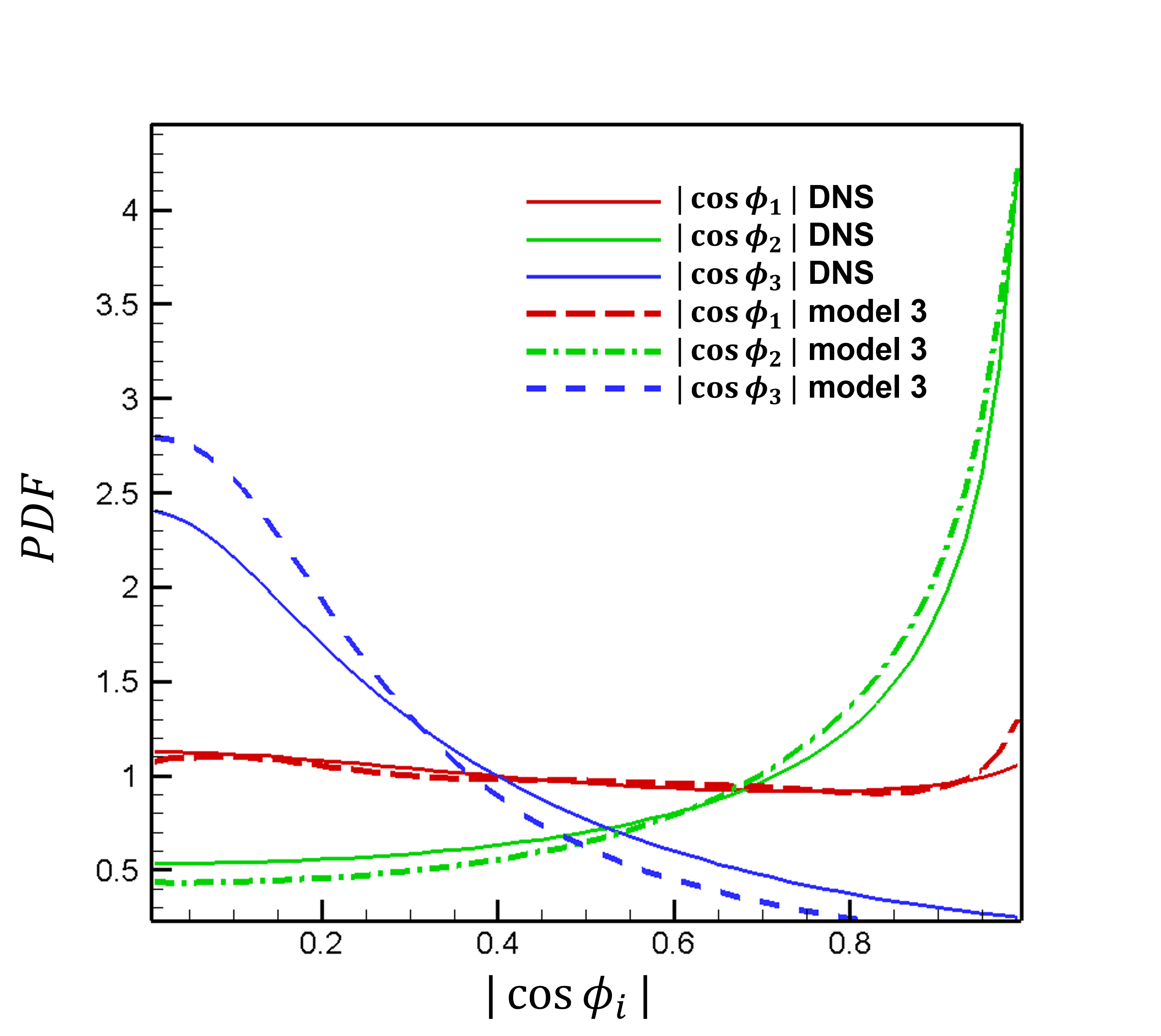}
    \caption{\label{fig:Ch7:cosPDF} PDFs of absolute values of cosine of angles between vorticity vector and strain-rate eigenvectors (1 - most expansive, 2 - intermediate, 3 - most compressive). The solid lines are the PDFs obtained from DNS data.}
\end{figure}

The alignment of vorticity with strain-rate eigendirections is another key feature of small-scale turbulence. 
From equation (\ref{eq:Ch7:b_sw}) in the eigen reference frame of the strain-rate tensor, one can show that the cosine of the angles of alignment between the vorticity vector and the three strain-rate eigenvectors are given by,
\begin{equation}
    \cos{\phi_i} = \frac{\omega_i}{|\bm{\omega}|} \;\;\; \forall \; i = 1,2,3.
\end{equation}
The angles $\phi_1,\phi_2,\phi_3 $ represent the angles of alignment of vorticity with the most expansive, intermediate, and most compressive strain-rate eigenvectors, respectively.  
In figure \ref{fig:Ch7:cosPDF}, the converged PDFs of the absolute values of the cosine of alignment angles are plotted for model 3 in comparison with that of DNS. 
Similar to other $b_{ij}$ statistics, the alignment PDFs produced by the other two models are nearly identical to that of model 3 and are therefore not displayed separately.
It is evident that the model is able to capture these PDFs with reasonable accuracy. 
It reproduces the preferential alignment of vorticity with the intermediate strainrate eigendirections reasonably well but slightly over-predicts the tendency of the vorticity vector to be perpendicular to the compressive strain-rate eigenvector. The PDF of the alignment with the most expansive strain-rate eigenvector is also captured very well by the model.

So far, we have established that in terms of the $\theta^*$ statistics, model 3 performs the best showing a slight advantage over model 1, while model 2 shows the highest deviation from the DNS statistics.
Further, the $b_{ij}$ model performs remarkably well in reproducing the $b_{ij}$ statistics accurately, which does not vary with the $\theta^*$ model.
This is somewhat surprising since even though the $b_{ij}$-SDE in local time $t'$ does not have an explicit dependence on $\theta^*$, the real-time evolution of $b_{ij}$ indirectly depends on the local VG magnitude $A$ ($\sim e^{\theta^*}$).
Yet, the $b_{ij}$ statistics of the model appear to be unchanged with the variation of $\theta^*$ model equation.

\subsection{VG tensor \label{sec:Ch7:resultsAij}}

\begin{figure}
    \centering
    \begin{subfigure}{0.58\textwidth}
        \centering
        \includegraphics[width=\textwidth]{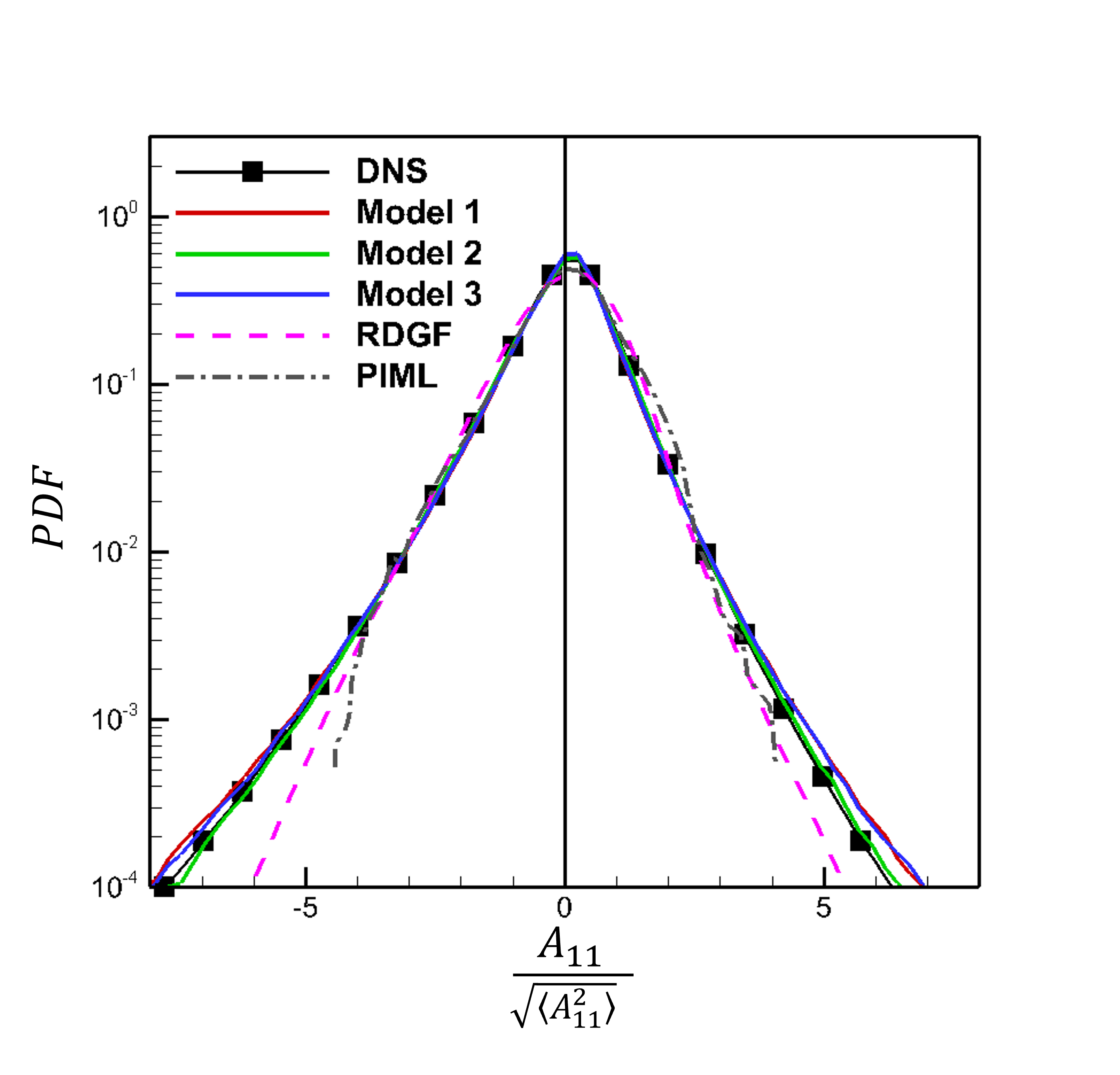}
        \caption{}
    \end{subfigure}
    \begin{subfigure}{0.58\textwidth}
        \centering
        \includegraphics[width=\textwidth]{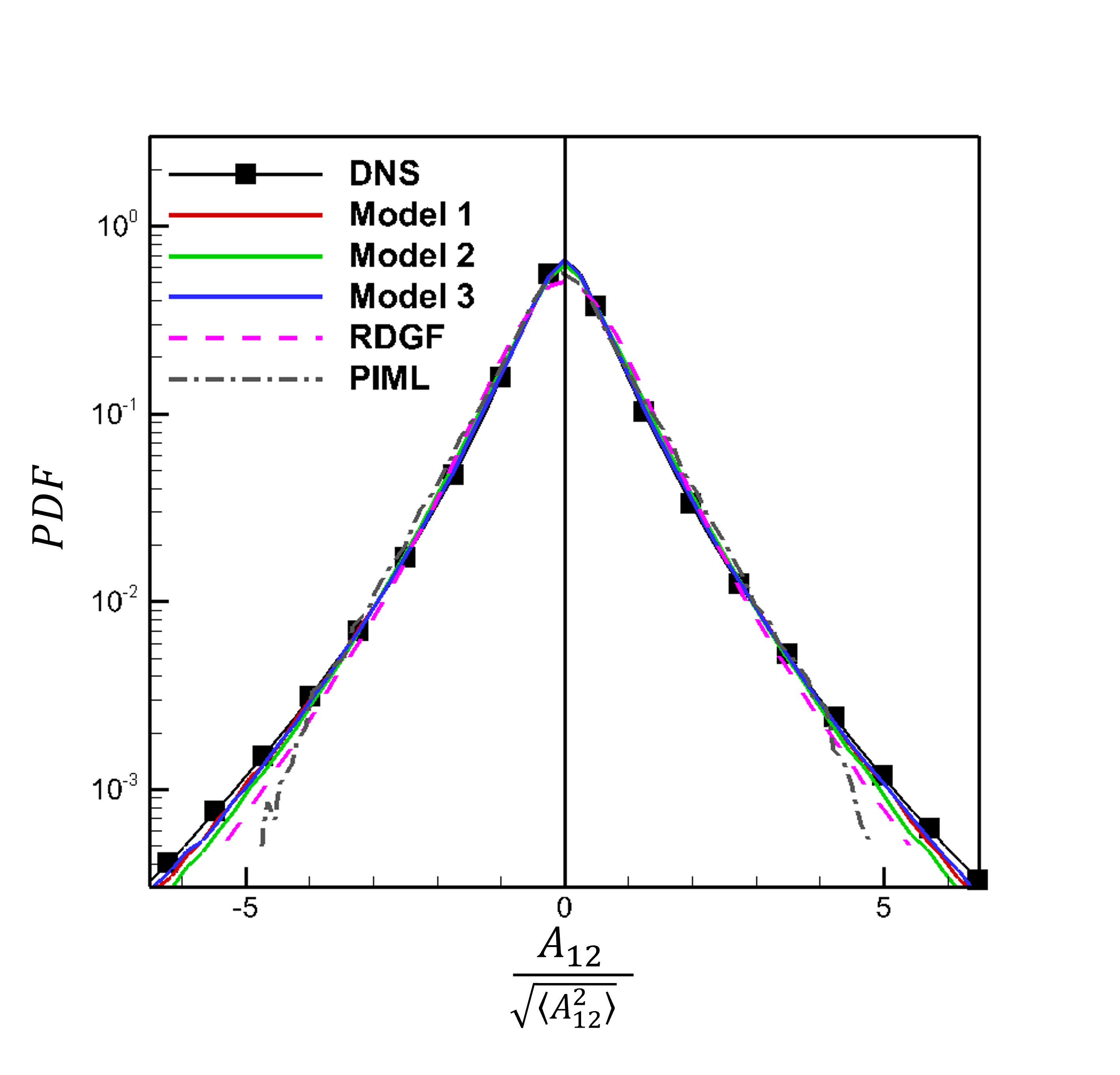}
        \caption{}
    \end{subfigure}
    \caption{\label{fig:Ch7:AijPDF} PDFs of: (a) longitudinal component of velocity gradient tensor, $A_{11}/\sqrt{\langle A^2_{11} \rangle} $, and (b) transverse component of velocity gradient tensor, $A_{12}/\sqrt{\langle A^2_{12} \rangle} $, in log-linear scale obtained from the solutions of the three models. The solid line marked with symbols represent the PDFs obtained from DNS data. The dashed and dash-dotted lines represent the PDFs obtained from previous models - RDGF \citep{johnson2016closure} and PIML \citep{tian2021physics}, respectively.}
\end{figure}

After examining the statistical results of $b_{ij}$ and $A$ individually, we now test the models' performance in capturing the overall velocity gradient ($A_{ij}$) statistics.
First, the PDFs of the longitudinal ($A_{11}$) and transverse ($A_{12}$) velocity gradients, normalized by their global root-mean-square values, are examined for all the three models in figure \ref{fig:Ch7:AijPDF}.
For comparison, we have also plotted the corresponding PDFs obtained from the DNS data and two of the recent velocity gradient models that have shown improved results compared to the other models in the literature - (i) recent deformation of Gaussian field (RDGF) model by \cite{johnson2016closure} and (ii) physics-informed machine learning (PIML) model by \cite{tian2021physics}.
Our models are able to reproduce the skewed $A_{11}$-PDF and the symmetric $A_{12}$-PDF, as observed in DNS, with reasonable accuracy.
They show significant improvement in capturing the PDFs of both $A_{11}$ and $A_{12}$ compared to both RDGF and PIML models. 
On closer observation, it is evident that while the PDFs are captured nearly perfectly in the densely populated part by all three models, there are smaller differences near the tails of the PDFs. 
Models 1 and 3 predict a slightly heavier-tailed distribution of $A_{11}$ than DNS, while model 2 produces a more accurate $A_{11}$-PDF.
On the other hand, model 3 appears to capture the $A_{12}$-PDF tails slightly more accurately than the other two.


\begin{table}
  \begin{center}
\def~{\hphantom{0}}
  \begin{tabular}{l|ccc|ccc|ccc}
      $~~$ & \multicolumn{3}{c|}{$A$} &  \multicolumn{3}{c|}{$A_{11}$} & \multicolumn{3}{c}{$A_{12}$} 
      \\[3pt]
      {$~~$} &
      \multirow{2}{*}{$\frac{\langle A^{3} \rangle}{\langle A^{2} \rangle^{3/2}}$} & 
        \multirow{2}{*}{$\frac{\langle A^{4} \rangle}{\langle A^{2} \rangle^{2}}$} & 
        \multirow{2}{*}{$\frac{\langle A^{6} \rangle}{\langle A^{2} \rangle^{3}}$} & 
        \multirow{2}{*}{$\frac{\langle A^{3}_{11} \rangle}{\langle A^{2}_{11} \rangle^{3/2}}$} & 
        \multirow{2}{*}{$\frac{\langle A^{4}_{11} \rangle}{\langle A^{2}_{11} \rangle^{2}}$} & 
        \multirow{2}{*}{$\frac{\langle A^{6}_{11} \rangle}{\langle A^{2}_{11} \rangle^{3}}$} & 
        \multirow{2}{*}{$\frac{\langle A^{3}_{12} \rangle}{\langle A^{2}_{12} \rangle^{3/2}}$} & 
        \multirow{2}{*}{$\frac{\langle A^{4}_{12} \rangle}{\langle A^{2}_{12} \rangle^{2}}$} & 
        \multirow{2}{*}{$\frac{\langle A^{6}_{12} \rangle}{\langle A^{2}_{12} \rangle^{3}}$} \\ 
        $~~$ &  &  &  &  &  &  &  &  &\\
        \hline
       DNS  $~$ & $~\bm{1.73}~$ & $~\bm{4.30}~$ & $~\bm{71.0}~$ & $~\bm{-0.59}~$ 
        & $~\bm{7.90}~$ & $~\bm{259}~$ & $~\bm{0.0}~$ & $~\bm{12.14}~$ & $~\bm{760}~$ \\
       Model 1 $~$ & $~1.68~$ & $~4.00~$ & $~60.5~$ & $~\bm{-0.58}~$ 
        & $~10.2~$ & $~607~$ & $~{0.0}~$ & $~10.3~$ & $~511~$ \\
	    Model 2 $~$ & $~1.58~$ & $~3.38~$ & $~39.7~$ & $~-0.53~$ 
        & $~\bm{8.06}~$ & $~\bm{298}~$ & $~{0.0}~$ & $~8.89~$ & $~342~$ \\
	    Model 3 $~$ & $~\bm{1.70}~$ & $~\bm{4.14}~$ & $~\bm{78.0}~$ & $~-0.55~$ 
        & $~9.65~$ & $~507~$ & $~{0.0}~$ & $~\bm{11.04}~$ & $~\bm{707}~$ \\
	    RDGF  $~$ & $~$--$~$ & $~$--$~$ & $~$--$~$ & $~-0.45~$ 
        & $~4.7~$ & $~$--$~$ & $~{0.0}~$ & $~6.8~$ & $~$--$~$ \\
  \end{tabular}
  \caption{Third, fourth and sixth order moments of VG magnitude ($A=\sqrt{A_{ij}A_{ij}}$), longitudinal VG component ($A_{11}$), transverse VG component ($A_{12}$) from DNS data, model 1, model 2, model 3 and RDGF model of \cite{johnson2016closure}. For each moment, the DNS value and the model's value closest to DNS are written in bold type font.}
  \label{tab:Ch7:Amoments}
  \end{center}
\end{table}

In order to determine the finer differences in the PDFs, we examine the higher order moments of the velocity gradient magnitude, $A$, as well as the velocity gradient components, $A_{11}$ and $A_{12}$. These moment values for all three models of this work, DNS data, and the RDGF model of \cite{johnson2016closure} are presented in table \ref{tab:Ch7:Amoments}.
The moment value produced by a model that is closest to the DNS is marked in bold type font.
It is evident that the moments of magnitude $A$ are best captured by model 3. 
The skewness, kurtosis, and $6^{th}$ order moment of the longitudinal component $A_{11}$ are reproduced best in model 2, although model 3 is not far behind and is slightly better than model 1.
The skewness of the transverse component $A_{12}$ is correctly captured as zero by all the models, maintaining a symmetrical probability distribution in each case.
The kurtosis and $6^{th}$ order moment of $A_{12}$ are also captured most closely by model 3. 
Overall, model 3 provides the most accurate representation of the probability distributions and moments of the velocity gradient tensor.


\begin{figure}
    \centering
    \begin{subfigure}{0.47\textwidth}
        \centering
        \includegraphics[width=\textwidth]{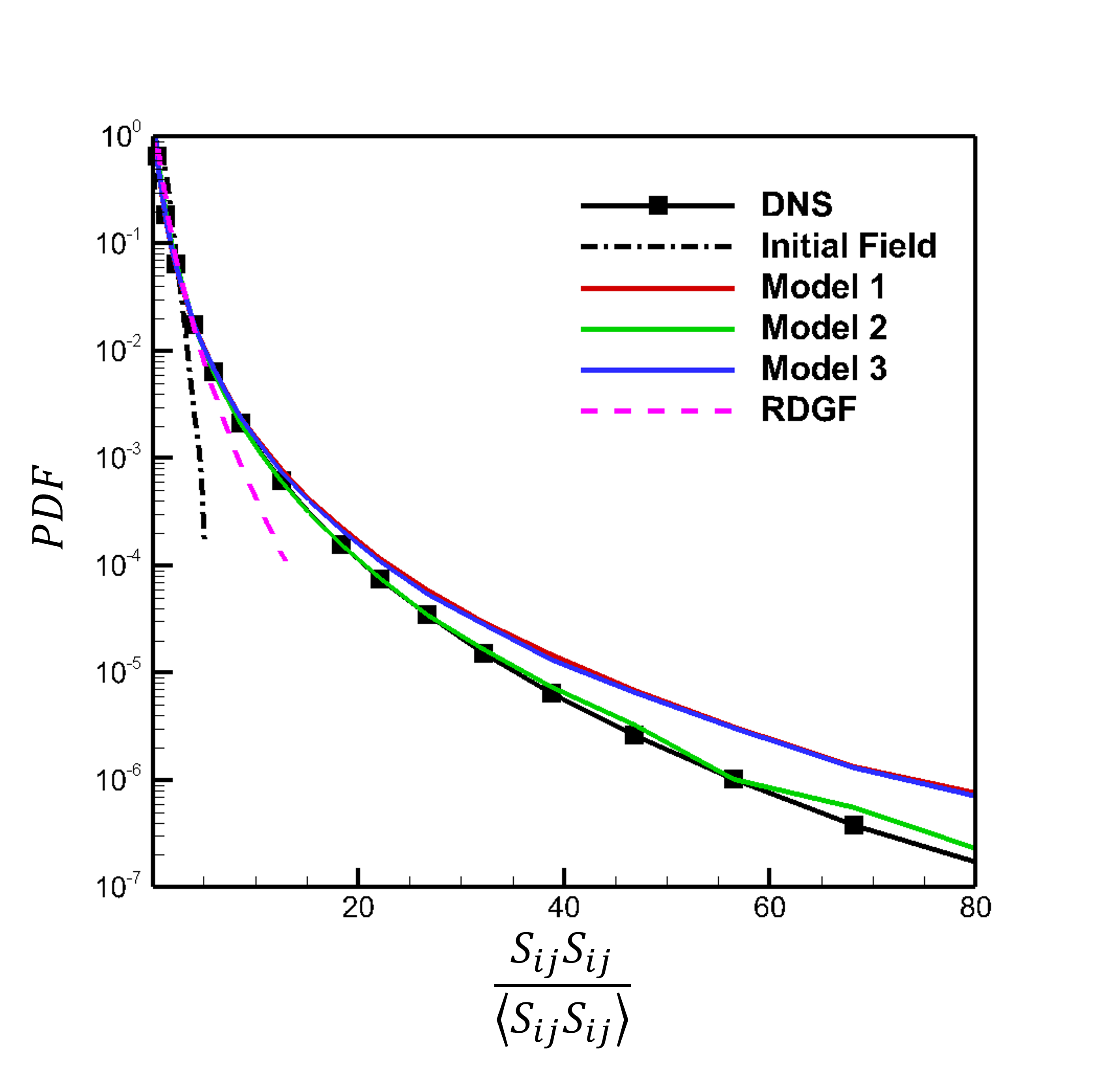}
        \caption{}
    \end{subfigure}
    \hfill
    \begin{subfigure}{0.47\textwidth}
        \centering
        \includegraphics[width=\textwidth]{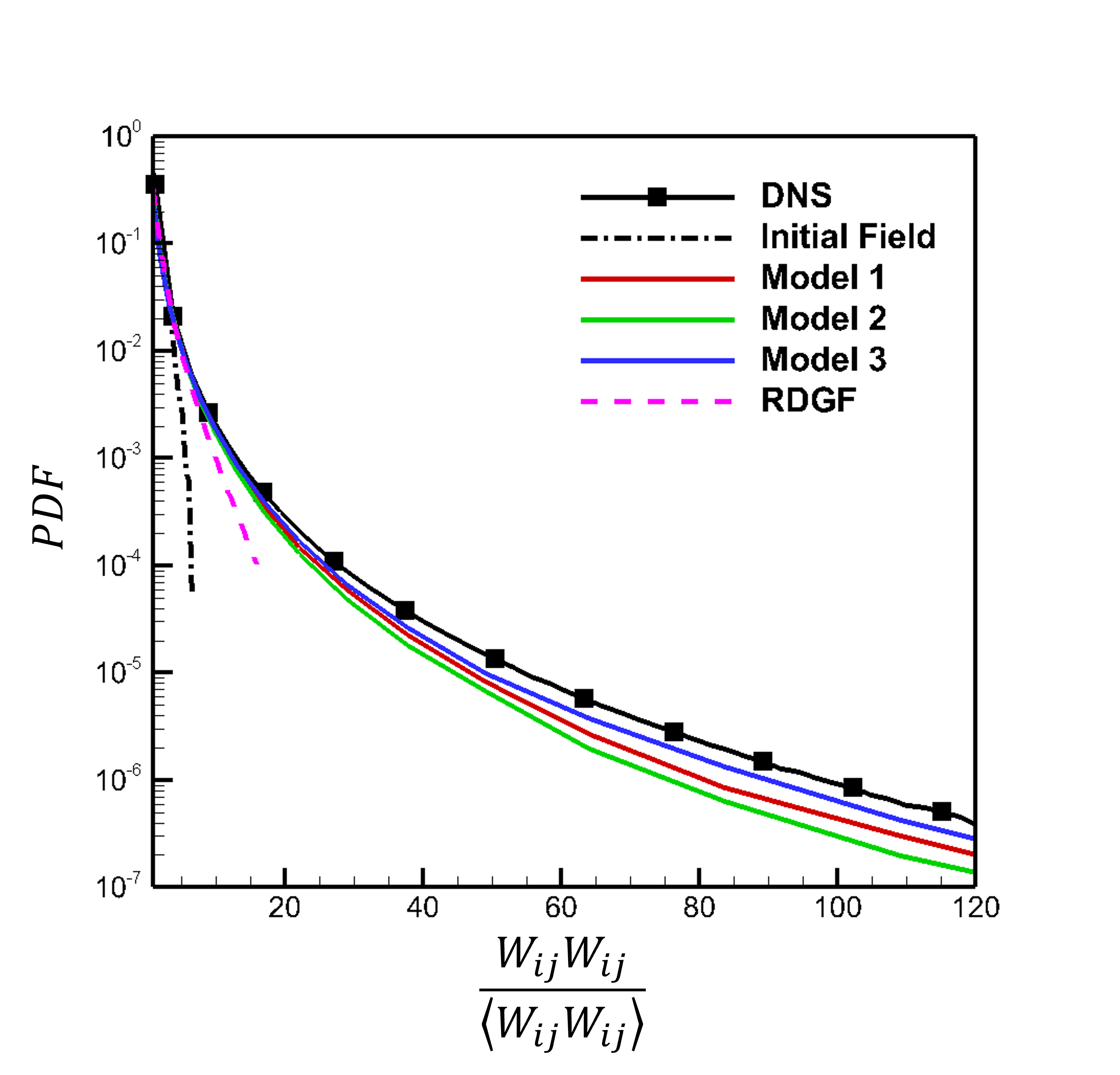}
        \caption{}
    \end{subfigure}
    \begin{subfigure}{0.47\textwidth}
        \centering
        \includegraphics[width=\textwidth]{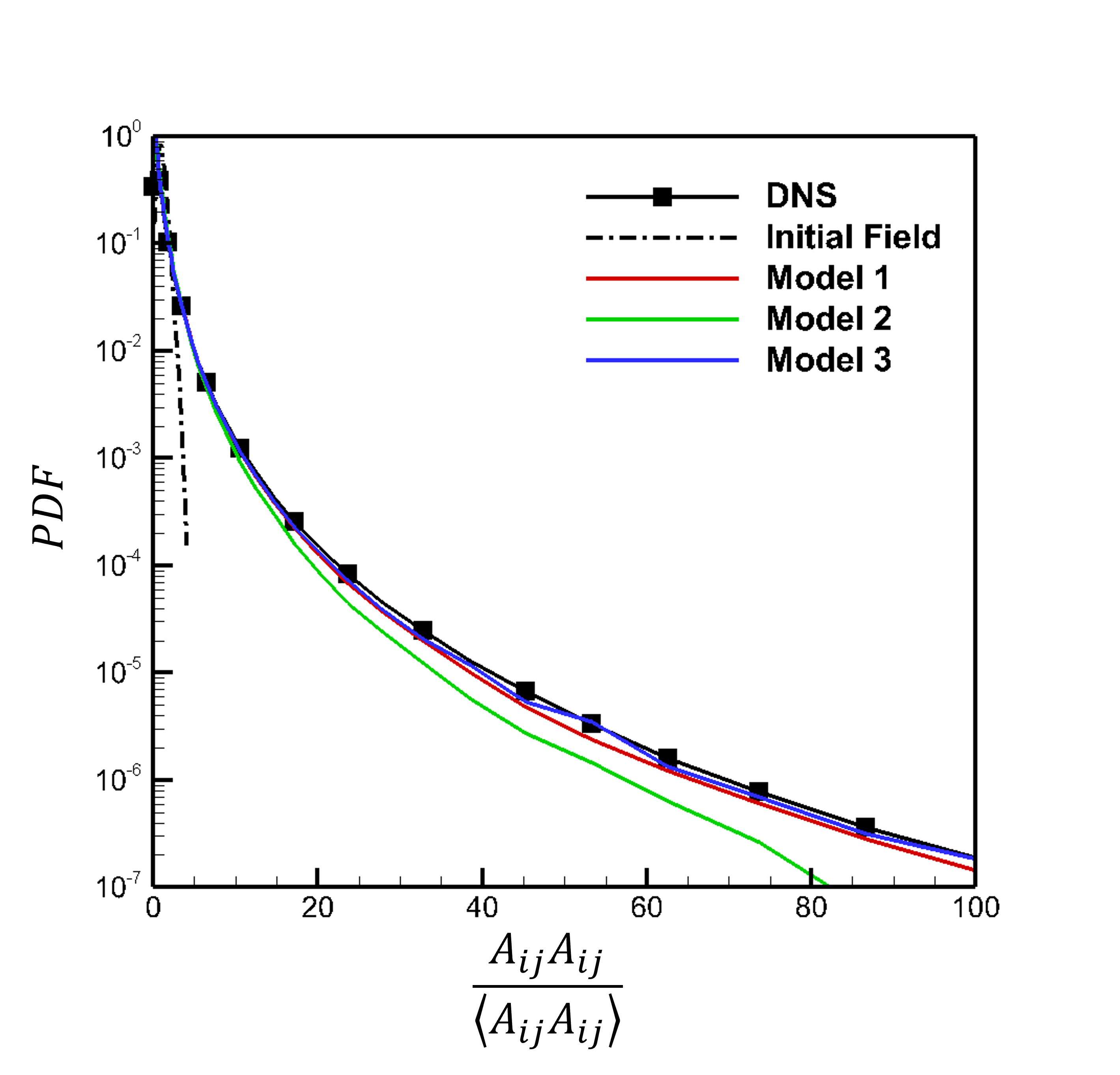}
        \caption{}
    \end{subfigure}
    \caption{\label{fig:Ch7:S2W2A2PDF} PDFs of: (a) dissipation rate, $S_{ij}S_{ij}/\sqrt{\langle S_{ij}S_{ij} \rangle} $, (b) enstrophy, $W_{ij}W_{ij}/\sqrt{\langle W_{ij}W_{ij} \rangle} $, and (c) pseudodissipation rate, $A_{ij}A_{ij}/\sqrt{\langle A_{ij}A_{ij} \rangle} $, in log-linear scale obtained from the solutions of the three models. The black solid line marked with symbols represent the PDFs obtained from DNS data; black dash-dotted line marks the PDFs for the initial field used in the model's simulations; dashed line represent the PDFs from the RDGF model of \cite{johnson2016closure}.}
\end{figure}

Finally, the PDFs of the dissipation rate ($ \nu S_{ij}S_{ij}$), enstrophy ($ \nu W_{ij}W_{ij}$), and pseudodissipation rate ($ \nu A_{ij}A_{ij}$), normalized by their global means, are computed from the converged stationary state solution of all the three models and plotted in figure \ref{fig:Ch7:S2W2A2PDF}. 
The PDFs obtained from the DNS data and those available from the RDGF model \citep{johnson2016closure} are also illustrated for comparison, along with the PDFs for the initial Gaussian field used in our model's simulations. 
It is interesting to note that the model is able to start from this Gaussian field and develop a turbulent flow field solution closely resembling that of DNS with the characteristic PDF-tails at extreme values.
It is clear that all three models reproduce the heavy-tailed probability distributions of both dissipation and enstrophy more accurately than the RDGF model.
Model 2 provides the most accurate representation of the dissipation PDF while models 1 and 3 over-predict the probability of occurrence of large dissipation rates near the tails of the PDF.
Enstrophy, which is more intermittent in nature than dissipation rate \citep{yeung2018effects,buaria2019extreme}, is captured best by model 3.
Models 1 and 2 under-predict the probability density of enstrophy near the extreme tails.
Taking the sum of the dissipation rate and enstrophy results in the pseudodissipation rate, which is reproduced quite accurately by model 3, even near the extreme tails.  
Overall, the results of model 3 constitute the closest representation of the velocity gradient statistics in turbulent flows.

\section{Conclusion \label{sec:Ch7:conc}}

A stochastic model for the Lagrangian evolution of velocity gradient (VG) tensor in an incompressible turbulent flow is presented. 
The bounded and well-behaved dynamics of the normalized velocity gradient tensor ($b_{ij}$) is modeled separately from the intermittent velocity gradient magnitude ($A$). 
The main nonlocal flow physics of pressure and viscous processes are well-behaved and amenable to modeling in the bounded framework of $b_{ij}$. Additionally, we can reduce the $b_{ij}$ space into a four-dimensional compact space. Thus, the closure modeling of these important nonlocal effects is performed using a simple but effective lookup table approach within the four-dimensional bounded state-space of $b_{ij}$.
On the other hand, the intermittent magnitude of the velocity gradient tensor is modeled as a modified lognormal process, in which DNS-data-based conditional variance is incorporated to better capture the intermittency.
The $b_{ij}$ model is generalizable to turbulent flows at different Reynolds numbers, while only the model for the magnitude requires Reynolds number-dependent parameters.

Numerical simulation of the Lagrangian model takes an initially random field and drives it toward a statistically stationary solution closely resembling DNS small-scale behavior. 
The model performs remarkably well in capturing the Eulerian PDFs and higher order moments of $b_{ij}$. 
Further, it is able to reproduce the characteristic teardrop shape of the joint probability distribution of the $b_{ij}$ invariants ($q,r$) without any discernible distortion, commonly observed in previous models. 
The vorticity-strainrate alignment angles are also captured with reasonable accuracy.
The model also reproduces up to sixth-order moments of velocity gradients and the heavy-tailed PDFs of velocity gradient magnitude, enstrophy, and dissipation rate with much improved accuracy over previous models in the literature.
This suggests that the presented model not only reproduces the small-scale geometric features of turbulence but is also able to capture its intermittent nature better than the previous velocity-gradient models. 



The main nonlinearities and nonlocal flow physics of the velocity-gradient dynamics are reproduced quite accurately by the $b_{ij}$ stochastic model based on direct tabulated data, without using machine learning. The intermittent behavior of the velocity gradients observed in DNS is also reproduced with improved accuracy by using a modified lognormal model of the VG magnitude. In future work, the magnitude will be modeled as a multifractal process to fully capture the extreme tails of the PDFs and the higher-order moments with greater accuracy.

The authors would like to acknowledge Dr. Diego Donzis of Texas A\&M University for providing part of the DNS data used in this work. 
The authors would also like to acknowledge Texas A\&M High Performance Research Computing, whose resources were used in this work.
Part of this work was presented at the American Physical Society (APS) Division of Fluid Dynamics meeting of 2020 and was a part of the PhD dissertation of R.D. at Texas A\&M University published in December 2021.

\appendix


\section{Relevant properties of It\^{o} process \label{sec:appA}}

\noindent 
\textit{It\^{o}'s lemma for scalar variables:} For a stochastic differential equation (SDE) of a scalar ($x$) of the form
\begin{equation}
    d x = f(x) dt + g(x) \; dW,
\end{equation}
the SDE for a function of the variable, $\varphi = \varphi({x})$, is given by
\begin{equation}
    d (\varphi(x)) = \bigg( \frac{\partial \varphi}{\partial t} + f(x) \frac{\partial \varphi}{\partial x} + \frac{1}{2}g^2(x) \frac{\partial^2 \varphi}{\partial x^2} \bigg) dt + g(x) \frac{\partial \varphi}{\partial x}\;dW 
\end{equation}

\noindent
\textit{It\^{o}'s lemma for tensorial variables:} For a system of SDEs of a tensor, $X_{ij}$, of the form
\begin{equation}
    d X_{ij} = F_{ij} (\bm{X}) dt + G_{ijkl} (\bm{X}) \; dW_{kl},
\end{equation}
the SDE for a function of the tensor, $\phi = \phi(\bm{X})$, is given by
\begin{equation}
    d \phi = \bigg( \frac{\partial \phi}{\partial t} + F_{ij} \frac{\partial \phi}{\partial X_{ij}} + \frac{1}{2} G_{ijkl} G_{pqkl} \frac{\partial^2 \phi}{\partial X_{ij}X_{pq}} \bigg) dt + G_{ijkl} \frac{\partial \phi}{\partial X_{ij}}\;dW_{kl} 
\end{equation}

\noindent
\textit{It\^{o}'s product rule:} For SDEs of two scalar variables, $x_1$ and $x_2$, given by
\begin{eqnarray}
    & dx_1 = f_1(x_1) dt + g_1(x_1) \; dW \;, \\
    & dx_2 = f_2(x_2) dt + g_2(x_2) \; dW 
\end{eqnarray}
the SDE of the product of the two variables is
\begin{equation}
    d(x_1 x_2) = x_1 d(x_2) + d(x_1) x_2 + d(x_1) d(x_2)
\end{equation}

\section{Derivation of $b_{ij}$ SDE from $A_{ij}$ SDE \label{sec:appB}}

The system of SDEs for the velocity gradient tensor $A_{ij}$ is given by
\begin{eqnarray}
    & d A_{ij} = M_{ij} dt + K_{ijkl}\; dW_{kl} \nonumber \\ 
    & \text{where,} \;\;  \langle dW_{ij} \rangle = 0 \;\;\; \text{and} \;\; \langle dW_{ij}dW_{kl} \rangle = \delta_{ik} \delta_{jl} dt
    \label{eq:appD:AijSDE}
\end{eqnarray}
Applying It\^{o}'s lemma we can obtain the SDE of the Frobenius norm of the tensor, $\phi = A^2 = A_{ij}A_{ij}$, as
\begin{equation}
    d(\phi) = (2A_{ij}M_{ij} + K_{ijkl}K_{ijkl}) dt + 2 A_{ij} K_{ijkl}\;dW_{kl}
\end{equation}
neglecting terms of the order of $\mathcal{O}(dt^n) \; \forall \; n>1$.
Then, the SDE of the VG magnitude, $A = \sqrt{A^2} = \sqrt{\phi}$, is obtained using It\^{o}'s lemma:
\begin{equation}
    d(A) = \bigg( \frac{A_{ij}M_{ij}}{A} + \frac{K_{ijkl}K_{ijkl}}{2A} - \frac{A_{ij}K_{ijkl}A_{mn}K_{mnkl}}{2A^3} \bigg) dt + \frac{A_{ij}K_{ijkl}}{A}\;dW_{kl}
\end{equation}
Next, the SDE of its reciprocal s obtained using It\^{o}'s lemma
\begin{equation}
    d\bigg(\frac{1}{A}\bigg) = - \frac{1}{A^2} \Bigg[ \bigg( \frac{A_{ij}M_{ij}}{A} + \frac{K_{ijkl}K_{ijkl}}{2A} - \frac{3}{2}\frac{A_{ij}K_{ijkl}A_{mn}K_{mnkl}}{A^3} \bigg) dt + \frac{A_{ij}K_{ijkl}}{A}\;dW_{kl} \Bigg]
    \label{eq:appD:1overA}
\end{equation}
Finally, applying It\^{o}'s product rule to determine the SDE for the normalized VG tensor, $b_{ij} \equiv \frac{A_{ij}}{A}$,
\begin{equation}
    d b_{ij} = d \;\bigg(\frac{1}{A} \;.\; A_{ij} \bigg) = \frac{1}{A} \; dA_{ij} + A_{ij} \; d\bigg(\frac{1}{A}\bigg) + dA_{ij}\;d\bigg( \frac{1}{A} \bigg) 
    \label{eq:appD:bijSDE}
\end{equation}
and using equations (\ref{eq:appD:AijSDE}) and (\ref{eq:appD:1overA}), we obtain
\begin{eqnarray}
    d b_{ij} =  \bigg(  \frac{M_{ij}}{A} - \frac{b_{ij}b_{kl}M_{kl}}{A} 
                       -\frac{b_{ij}K_{pqkl}K_{pqkl}}{2A^{2}} 
      - \frac{b_{pq}K_{pqkl}K_{ijkl}}{A^{2}} \nonumber \\ 
      + \frac{3}{2} \frac{b_{ij}b_{pq}K_{pqkl}b_{mn}K_{mnkl}}{A^{2}} \bigg) dt +
      \bigg(  \frac{K_{ijkl}}{A} - \frac{b_{ij}b_{pq}K_{pqkl}}{A} \bigg) \; dW_{kl}.
    \label{eq:appD:bijSDE2}
\end{eqnarray}
Rearranging, we can write the final form of the $b_{ij}$-SDE as follows
\begin{eqnarray}
    d b_{ij} = \bigg( \frac{M_{ij}}{A^2} - b_{ij}b_{kl}\frac{M_{kl}}{A^2}
                       -\frac{1}{2}b_{ij}\frac{K_{pqkl}}{A^{3/2}}\frac{K_{pqkl}}{A^{3/2}} 
      - b_{pq}\frac{K_{pqkl}}{A^{3/2}}\frac{K_{ijkl}}{A^{3/2}} \nonumber \\ 
      + \frac{3}{2} b_{ij}b_{pq}\frac{K_{pqkl}}{A^{3/2}}b_{mn}\frac{K_{mnkl}}{A^{3/2}} \bigg) dt' + 
      \bigg(  \frac{K_{ijkl}}{A^{3/2}} - b_{ij}b_{pq}\frac{K_{pqkl}}{A^{3/2}} \bigg) \; dW'_{kl}
\end{eqnarray}
where, $dt' = Adt $ and $ dW'_{ij} = A^{1/2} dW_{ij}$. 
Note that all the terms on the RHS of the $b_{ij}$ SDE are non-dimensional, including $dt'$, $dW'_{kl}$, $M_{ij}/A^2$ and  $K_{ijkl}/A^{3/2}$.
This equation can also be written as
\begin{eqnarray}
    & d b_{ij} = (\mu_{ij} + \gamma_{ij}) dt' + D_{ijkl} \;dW'_{kl} \;\;\; \text{where} \nonumber \\
    & \mu_{ij} = \frac{M_{ij}}{A^2} - b_{ij}b_{kl}\frac{M_{kl}}{A^2}  \;\;, \;\; 
    D_{ijkl} = \frac{K_{ijkl}}{A^{3/2}} - b_{ij}b_{pq}\frac{K_{pqkl}}{A^{3/2}} \;\;, \nonumber \\
    & \gamma_{ij} = -\frac{1}{2}b_{ij}\frac{K_{pqkl}}{A^{3/2}}\frac{K_{pqkl}}{A^{3/2}} 
      - b_{pq}\frac{K_{pqkl}}{A^{3/2}}\frac{K_{ijkl}}{A^{3/2}} + \frac{3}{2}b_{ij}b_{pq}\frac{K_{pqkl}}{A^{3/2}}b_{mn}\frac{K_{mnkl}}{A^{3/2}} 
    \label{eq:appD:bijSDEterms}
\end{eqnarray}
where $\mu_{ij}$ is the mean drift coefficient tensory, $\gamma_{ij}$ is the additional drift coefficient tensor and $D_{ijkl}$ is the diffusion coefficient tensor.

\section{Incompressibility constraint \label{sec:appC}}

To prove that the system of SDEs of $b_{ij}$ in equation (\ref{eq:Ch7:bijSDEterms}) satisfies the incompressibility constraint, we take the trace on both sides of the $b_{ij}$-SDE:
\begin{equation}
    db_{ii} = (\mu_{ii} + \gamma_{ii}) dt' + D_{iikl} \; dW'_{kl}
    \label{eq:appD:1}
\end{equation}
Now, since $b_{ii} = 0$, we have
\begin{equation}
    \mu_{ii} = \frac{M_{ii}}{A^2} - b_{ii}b_{kl}\frac{M_{kl}}{A^2}  = 0.
    \label{eq:appD:2}
\end{equation} 
Further, since $K_{iikl} = 0$ by construction, it can be easily showed that
\begin{eqnarray}
    & \gamma_{ii} = -\frac{1}{2}b_{ii}\frac{K_{pqkl}}{A^{3/2}}\frac{K_{pqkl}}{A^{3/2}} 
      - b_{pq}\frac{K_{pqkl}}{A^{3/2}}\frac{K_{iikl}}{A^{3/2}} + \frac{3}{2}b_{ii}b_{pq}\frac{K_{pqkl}}{A^{3/2}}b_{mn}\frac{K_{mnkl}}{A^{3/2}} = 0 \nonumber \\ 
    & D_{iikl} = \frac{K_{iikl}}{A^{3/2}} - b_{ii}b_{pq}\frac{K_{pqkl}}{A^{3/2}}  = 0
    \label{eq:appD:3}
\end{eqnarray} 
Therefore, from equations (\ref{eq:appD:1}-\ref{eq:appD:3}), we have
\begin{equation}
    d b_{ii} = 0 
\end{equation}


\section{Normalization constraint \label{sec:appD}}

Next, we prove that the $b_{ij}$ SDE maintains the Frobenius norm of unity. For this, we first derive the SDE for the Frobenius norm of $b_{ij}$, using It\^{o}'s product rule: 
\begin{equation}
    d(b_{ij} b_{ij}) = (2 b_{ij} \mu_{ij} + 2 b_{ij} \gamma_{ij} + D_{ijkl} D_{ijkl}) dt' + 2 b_{ij} D_{ijkl} \; dW'_{kl}
    \label{eq:appD:dbijbij}
\end{equation}
Now, the first term is zero by construction since
\begin{equation}
    2 b_{ij} \mu_{ij} = 2 b_{ij} \bigg( \frac{M_{ij}}{A^2} - b_{ij}b_{kl}\frac{M_{kl}}{A^2} \bigg) = 2 \frac{ b_{kl}M_{kl} }{A^2} - 2 b_{ij}b_{ij}\frac{b_{kl}M_{kl}}{A^2} =  0
\end{equation}
provided $b_{ij}b_{ij} = 1$.
The second term can be expanded as follows:
\begin{eqnarray}
    2 b_{ij} \gamma_{ij} & = & 2 b_{ij} \bigg( -\frac{1}{2}b_{ij}\frac{K_{pqkl}}{A^{3/2}}\frac{K_{pqkl}}{A^{3/2}} 
      - b_{pq}\frac{K_{pqkl}}{A^{3/2}}\frac{K_{ijkl}}{A^{3/2}} + \frac{3}{2}b_{ij}b_{pq}\frac{K_{pqkl}}{A^{3/2}}b_{mn}\frac{K_{mnkl}}{A^{3/2}}  \bigg) \nonumber \\
      & = & - b_{ij}b_{ij} \frac{K_{pqkl}}{A^{3/2}} \frac{K_{pqkl}}{A^{3/2}} - 2 b_{ij} \frac{K_{ijkl}}{A^{3/2}} b_{pq} \frac{K_{pqkl}}{A^{3/2}} + 3 b_{ij}b_{ij} b_{pq} \frac{K_{pqkl}}{A^{3/2}} b_{mn} \frac{K_{mnkl}}{A^{3/2}} \nonumber \\
      & = & - \bigg( \frac{K_{pqkl}}{A^{3/2}} \frac{K_{pqkl}}{A^{3/2}} - b_{ij} \frac{K_{ijkl}}{A^{3/2}} b_{pq} \frac{K_{pqkl}}{A^{3/2}}  \bigg)
\end{eqnarray}
since $b_{ij}b_{ij} = 1$.
And the third term can be expanded as
\begin{eqnarray}
    D_{ijkl} D_{ijkl} & = & \bigg( \frac{K_{ijkl}}{A^{3/2}} - b_{ij}b_{pq}\frac{K_{pqkl}}{A^{3/2}} \bigg) \bigg( \frac{K_{ijkl}}{A^{3/2}} - b_{ij}b_{pq}\frac{K_{pqkl}}{A^{3/2}} \bigg) \nonumber \\
    & = & \frac{K_{ijkl}}{A^{3/2}} \frac{K_{ijkl}}{A^{3/2}} - b_{ij} \frac{K_{ijkl}}{A^{3/2}} b_{pq} \frac{K_{pqkl}}{A^{3/2}} 
\end{eqnarray}
Therefore, the second and third terms cancel each other out.
Finally, the diffusion term is also zero due to the form of the diffusion coefficient as:
\begin{equation}
    2 b_{ij} D_{ijkl} = 2 b_{ij} \bigg( \frac{K_{ijkl}}{A^{3/2}} - b_{ij}b_{pq}\frac{K_{pqkl}}{A^{3/2}} \bigg) = 2 b_{ij}\frac{K_{ijkl}}{A^{3/2}} - 2 b_{ij} b_{ij} b_{pq} \frac{K_{pqkl}}{A^{3/2}} = 0
\end{equation}
Thus, it is proved that for the given form of $\mu_{ij}$, $\gamma_{ij}$ and $D_{ijkl}$, the equation (\ref{eq:appD:dbijbij}) simplifies to
\begin{equation}
    d (b_{ij}b_{ij} ) = 0.
\end{equation}
In other words, the form of the $b_{ij}$-SDE (equation \ref{eq:appD:bijSDEterms}) automatically ensures that $b_{ij}b_{ij}$ remains unity at all times provided it is initially unity.

\section{Galilean invariance \label{sec:appE}}

Now we demonstrate that the approach of closure modeling of the normalized anisotropic pressure Hessian ($\bm{h}$) and viscous Laplacian ($\bm{\tau}$) tensors satisfies Galilean invariance.
The tensor $\bm{h}$ is modeled as
\begin{equation}
    \bm{h} = \bm{Q} \; \bm{\tilde{h}} \; \bm{Q}^T
    \label{eq:appD:A2}
\end{equation}
where $\bm{\tilde{h}}$ is the pressure Hessian tensor in the principal frame of strain-rate tensor ($\bm{s}$).
This $\bm{\tilde{h}}$ is obtained from data-driven closure as a function of $\bm{\tilde{b}}$, also in principal reference frame.
Thus,
\begin{equation}
    \bm{Q} = [\bm{E_1} \;\; \bm{E_2} \;\; \bm{E_3}]
\end{equation}
where $\bm{E_i}$ are the right eigenvectors of $\bm{s}$ corresponding to its eigenvalues $a_i$ and $\bm{E_i}$ constitute the columns of the rotation matrix $\bm{Q}$.
Let us rotate the coordinate frame of the observer by certain angles, using a rotation matrix $\bm{R}$. Let the tensors and vectors in new reference frame be marked by $\;'\;$. 
Then the tensor $\bm{s}$ becomes 
\begin{eqnarray}
    \bm{s'} = \bm{R} \; \bm{s} \; \bm{R}^T 
\end{eqnarray}
and its eigenvectors also rotate by the same angles since
\begin{eqnarray}
    & \bm{s} \bm{E_i} = a_i \bm{E_i} \;\; \implies \; 
    \bm{R}^T \bm{s'} \bm{R}  \bm{E_i} = a_i \bm{E_i} \nonumber \\
    & \implies \bm{s'} \bm{R} \bm{E_i} = a_i \bm{R} \bm{E_i} \;\;
    \implies \; \bm{s'} \bm{E'_i}= a_i \bm{E'_i} \;\; \text{where} \;\;
    \bm{E'_i} = \bm{R} \bm{E_i}
\end{eqnarray}
Since $\bm{E'_i}$ constitute the columns of the rotated tensor $\bm{Q'}$, we can say 
\begin{eqnarray}
    \bm{Q'} = \bm{R} \; \bm{Q}
    \label{eq:appD:A1}
\end{eqnarray}
Therefore, using equations (\ref{eq:appD:A2}) and (\ref{eq:appD:A1}), the pressure Hessian tensor in the new reference frame becomes,
\begin{eqnarray}
    \bm{h'} = \bm{Q'} \;\bm{\tilde{h}}\; \bm{Q'}^T
    = \bm{R} \; \bm{Q} \; \bm{\tilde{h}} \; \bm{Q}^T \bm{R}^T
    = \bm{R} \; \bm{h} \; \bm{R}^T
    \label{eq:appD:A3}
\end{eqnarray}
Note that $\bm{\tilde{h}} = \bm{\tilde{h}}(q,r,a_2,\omega_2)$, all four of which are either frame invariant or specifically defined in the principal reference frame and therefore $\bm{\tilde{h}}$ is unaltered by frame rotation.
It is evident from equation (\ref{eq:appD:A3}) that the new tensor $\bm{h'}$ also rotates by the same angles with respect to the old $\bm{h}$ as the new frame rotates with respect to the old frame. 
This proves that the model for pressure Hessian tensor $\bm{h}$ is Galilean invariant. The same proof applies to the viscous Laplacian tensor $\bm{\tau}$. 

Aside from the mean pressure and viscous terms discussed above, all the other terms in the $b_{ij}$ stochastic differential equation are functions of $b_{ij}$ itself and it can be shown that they are also Galilean invariant by construction.

\section{Numerical schemes for stochastic differential equations \label{sec:appF}}


In this work, the numerical scheme used to propagate the $b_{ij}$-SDE in computational time $t'$ is a second order weak predictor-corrector scheme given by:
\begin{eqnarray}
    b'_{ij} &=& b^{(n)}_{ij} + \mu_{ij}(\bm{b}^{(n)})\Delta t' + \gamma_{ij}(\bm{b}^{(n)})\Delta t' + D_{ijkl}(\bm{b}^{(n)}) \; \xi_{kl}\sqrt{\Delta t'} \\
    b^{(n+1)}_{ij} &=&  b^{(n)}_{ij} + \frac{1}{2} \big[ \mu_{ij}(\bm{b}^{(n)}) + \mu_{ij}(\bm{b}') \big] \Delta t' 
    + \frac{1}{2} \big[ \gamma_{ij}(\bm{b}^{(n)}) + \gamma_{ij}(\bm{b}')  \big] \Delta t' \nonumber \\
    & & \;\;\;\;\;\;\; + \frac{1}{2} \big[ D_{ijkl}(\bm{b}^{(n)}) + D_{ijkl}(\bm{b}') \big] \; \xi_{kl}\sqrt{\Delta t'} 
\end{eqnarray}
where each component of $\xi_{ij}$ is an independent standardized Gaussian random variable. 
The $\theta^*$-SDE can be written in the computational timescale $t'$ as follows:
\begin{eqnarray}
    d \theta^* &=& - \theta^* dt^* + \beta(q,r) \; d W^*  \nonumber \\
              &=&  - \theta^* \frac{\langle A \rangle}{A} dt' + \beta(q,r) \sqrt{ \frac{\langle A \rangle}{A} } \; d W' 
\end{eqnarray}
where, $\beta(q,r)$ represents the different diffusion coefficients discussed in section \ref{sec:Ch7:modelths}.
The $\theta^*$-SDE is also propagated using the second order weak predictor-corrector scheme:
\begin{eqnarray}
    & {\theta^*}' = {\theta^*}^{(n)} - {\theta^*}^{(n)} \frac{\langle A \rangle}{A} \Delta t' + \beta(q^{(n)},r^{(n)}) \xi \sqrt{ \frac{\langle A \rangle}{A}  } \sqrt{\Delta t'} \\
    & {\theta^*}^{(n+1)} = {\theta^*}^{(n)} - \frac{1}{2} \big[ {\theta^*}^{(n)} + {\theta^*}' \big] \frac{\langle A \rangle}{A} \Delta t' 
    + \frac{1}{2} \big[ \beta(q^{(n)},r^{(n)}) + \beta(q',r') \big] \xi \sqrt{ \frac{\langle A \rangle}{A}  } \sqrt{\Delta t'}
\end{eqnarray}
where $q^{(n)},r^{(n)}$ represent the second and third invariants of the $\bm{b}^{(n)}$ tensor and $q',r'$ represent the second and third invariants of the $\bm{b}'$ tensor. Here, the VG magnitude $A = e^{(\sigma_\theta \theta^* + \langle \theta \rangle)}$, for constant values of $\langle \theta \rangle, \sigma_\theta$ from DNS.

\section{Direct numerical simulation data \label{sec:appG}}

In this work, DNS data of forced isotropic turbulent flows of Taylor Reynolds number, $Re_\lambda =u'\lambda/\nu$, ranging from $1$ to $588$ have been used. Here, $u'$ is the root-mean-square velocity and $\lambda$ is the Taylor microscale of the flow, and $\nu$ is the kinematic viscosity of the fluid. The simulations are spatially well-resolved with $k_{max} \eta > 1.3$, where $k_{max}$ is the highest resolved wave number and $\eta$ is the Kolmogorov length scale. These datasets are obtained from the following two sources: (1) Johns Hopkins Turbulence Database \citep{perlman2007data,li2008public}; the data has been widely used in the literature for investigating velocity gradient statistics \citep{johnson2016large,elsinga2017scaling,danish2018multiscale} as well as its Lagrangian dynamics \citep{yu2010lagrangian,yu2010scaling} in turbulence, and (2) Donzis research group at Texas A\&M University; in the past the data has been used to study small-scale dynamics, intermittency, and anomalous scaling \citep{donzis2008dissipation,donzis2010short,yakhot2017emergence}.
Further details about the DNS data are provided in table \ref{tab:table1}.

\begin{table}
  \begin{center}
\def~{\hphantom{0}}
  \begin{tabular}{lccc}
 $Re_\lambda$ & Grid points & $k_{max}\eta$ & Source \\ \hline
 $~~1$ & $256^3$ & $105.6$ & \cite{yakhot2017emergence} \\
 $~~6$ & $256^3$ & $34.8$  & \cite{yakhot2017emergence} \\
 $~~9$ & $256^3$ & $26.6$ & \cite{yakhot2017emergence}\\
 $~14$ & $256^3$ & $19.87$ & \cite{yakhot2017emergence}\\
 $~18$ & $256^3$ & $15.59$ & \cite{yakhot2017emergence}\\
 $~25$ & $256^3$ & $11.51$ & \cite{yakhot2017emergence}\\
 $~35$ & $64^3$ & $1.45$ & \cite{yakhot2017emergence}\\
 $~86$ & $256^3$ & $2.83$ & \cite{donzis2008dissipation}\\
 $225$ & $512^3$ & $1.34$ & \cite{donzis2008dissipation}\\
 $385$ & $1024^3$ & $1.41$ & \cite{donzis2008dissipation}\\
 $427$ & $1024^3$ & $1.32$ & \cite{li2008public}\\
 $588$ & $2048^3$ & $1.39$ & \cite{donzis2008dissipation}\\
  \end{tabular}
  \caption{\label{tab:table1}Details of forced isotropic incompressible turbulence data.}
  \end{center}
\end{table}


\bibliographystyle{jfm}


\end{document}